\documentclass[twocolumn]{aastex62}

\usepackage{lineno}
\received{\today}
\submitjournal{ApJ}

\shorttitle{A Wideband Polarization Study of Hydra A}
\shortauthors{Baidoo et al.}
\usepackage{amsmath}

\begin{document}

\title{A Wideband Polarization Observation of Hydra A with the Jansky Very Large Array}

\correspondingauthor{Lerato Baidoo}
\email{mll.sebokolodi@gmail.com}

\author[0000-0003-0520-0696]{Lerato Baidoo}
\affiliation{National Radio Astronomy
  Observatory, 1003 Lopezville Rd, Socorro, NM 87801}
\affiliation{Center for Radio Astronomy Techniques and Technologies,
  Department of Physics and Electronics, Rhodes University, PO Box 94,
  Makhanda 6140, South Africa}
\affiliation{Dunlap Institute for Astronomy and
  Astrophysics, University of Toronto, Toronto, ON M5S 3H4, Canada}

\author[0000-0001-7097-8360]{Richard A. Perley} \affiliation{National Radio Astronomy
  Observatory, 1003 Lopezville Rd, Socorro, NM 87801}
\affiliation{Center for Radio Astronomy Techniques and Technologies,
  Department of Physics and Electronics, Rhodes University, PO Box 94,
  Makhanda 6140, South Africa}

\author[0009-0003-5578-0614]{Jean Eilek} \affiliation{Physics Department, New Mexico Tech,
  Socorro, NM 87801} \affiliation{Adjunct Astronomer at the National
  Radio Astronomy Observatory, 1003 Lopezville Rd, Socorro, NM 87801}

\author{Oleg Smirnov}
\affiliation{Center for Radio Astronomy Techniques and Technologies,
  Department of Physics and Electronics, Rhodes University, PO Box 94,
  Makhanda 6140, South Africa}
\affiliation{South African Radio Astronomy Observatory, SARAO, 2 Fir
  Street, Black River Park, Observatory, 7925, South Africa}

\author[0000-0003-1997-0771]{Valentina Vacca} \affiliation{INAF, Observatorio Astronomico
  di Caligiari, via della Scienza 5, I-09047 Selargius (CA), Italy}

\author[0000-0001-5246-1624]{Torsten En{\ss}lin} \affiliation{Max-Planck-Institute for
  Astrophysics, Karl-Schwarzschild-Str. 1, 85748 Garching, Germany}

%% Note that the \and command from previous versions of AASTeX is now
%% depreciated in this version as it is no longer necessary. AASTeX 
%% automatically takes care of all commas and "and"s between authors names.

%% AASTeX 6.2 has the new \collaboration and \nocollaboration commands to
%% provide the collaboration status of a group of authors. These commands 
%% can be used either before or after the list of corresponding authors. The
%% argument for \collaboration is the collaboration identifier. Authors are
%% encouraged to surround collaboration identifiers with ()s. The 
%% \nocollaboration command takes no argument and exists to indicate that
%% the nearby authors are not part of surrounding collaborations.

%% Mark off the abstract in the ``abstract'' environment. 
\begin{abstract}
We present results of a wideband high-resolution polarization study of
Hydra A, one of the most luminous FR I radio galaxies known and
amongst the most well-studied.  The radio emission from this source
displays extremely large Faraday rotation measures ($\mathrm{RM}$s), ranging from
$-12300$ rad m$^{-2}$ to $5000$ rad m$^{-2}$, the majority of
which are believed to originate from magnetized thermal gas external
to the radio tails.  The radio emission from both tails strongly
depolarizes with decreasing frequency.  The depolarization, as a
function of wavelength, is commonly non-monotonic, often showing
oscillatory behavior, with strongly non-linear rotation of the
polarization position angle with $\lambda^2$. A simple model, based on
the $\mathrm{RM}$ screen derived from the high frequency, high resolution data,
predicts the lower frequency depolarization remarkably well.  The
success of this model indicates the majority of the depolarization can
be attributed to fluctuations in the magnetic field on scales $< 1500$
pc, suggesting the presence of turbulent magnetic field/electron
density structures on sub-kpc scales within a Faraday rotating (FR)
medium. 
\end{abstract}

%% Keywords should appear after the \end{abstract} command. 
%% See the online documentation for the full list of available subject
%% keywords and the rules for their use.

%% From the front matter, we move on to the body of the paper.
%% Sections are demarcated by \section and \subsection, respectively.
%% Observe the use of the LaTeX \label
%% command after the \subsection to give a symbolic KEY to the
%% subsection for cross-referencing in a \ref command.
%% You can use LaTeX's \ref and \label commands to keep track of
%% cross-references to sections, equations, tables, and figures.
%% That way, if you change the order of any elements, LaTeX will
%% automatically renumber them.
%%
%% We recommend that authors also use the natbib \citep
%% and \citet commands to identify citations.  The citations are
%% tied to the reference list via symbolic KEYs. The KEY corresponds
%% to the KEY in the \bibitem in the reference list below. 
\section{Introduction} \label{sec:intro}
Hydra A (3C $218$) is a wide-tailed Fanaroff-Riley type I (FR I) radio
galaxy. With a spectral luminosity of $P_{178\text{MHz}} = 3\times
10^{26}$ W Hz$^{-1}$ sr$^{-1}$ \citep{TAYLOR1990}, it is one of the most luminous FRI galaxies known.
%Hydra A's luminosity is estimated to be
%factor of $10$ above the FRI - FRII break \citep{TAYLOR1990}.
The radio galaxy is hosted by a cD elliptical galaxy situated at
redshift $0.054$\footnote{Assuming H$_0=69.3$ km s$^{-1}$ Mpc$^{-1}$,
  $\Omega_{M}=0.288$, and $\Omega_{\Lambda}=0.712$, the projected linear scale
  of Hydra A is 1.1 kpc arcsec$^{-1}$.}  \citep{DWARAKANATH1995,
  OWEN1995, TAYLOR1996}. This galaxy is the dominant member of the
relatively poor Abell cluster A780 \citep{ABELL1958}.

\begin{figure*}
 \centering 
 \includegraphics[width=7in]{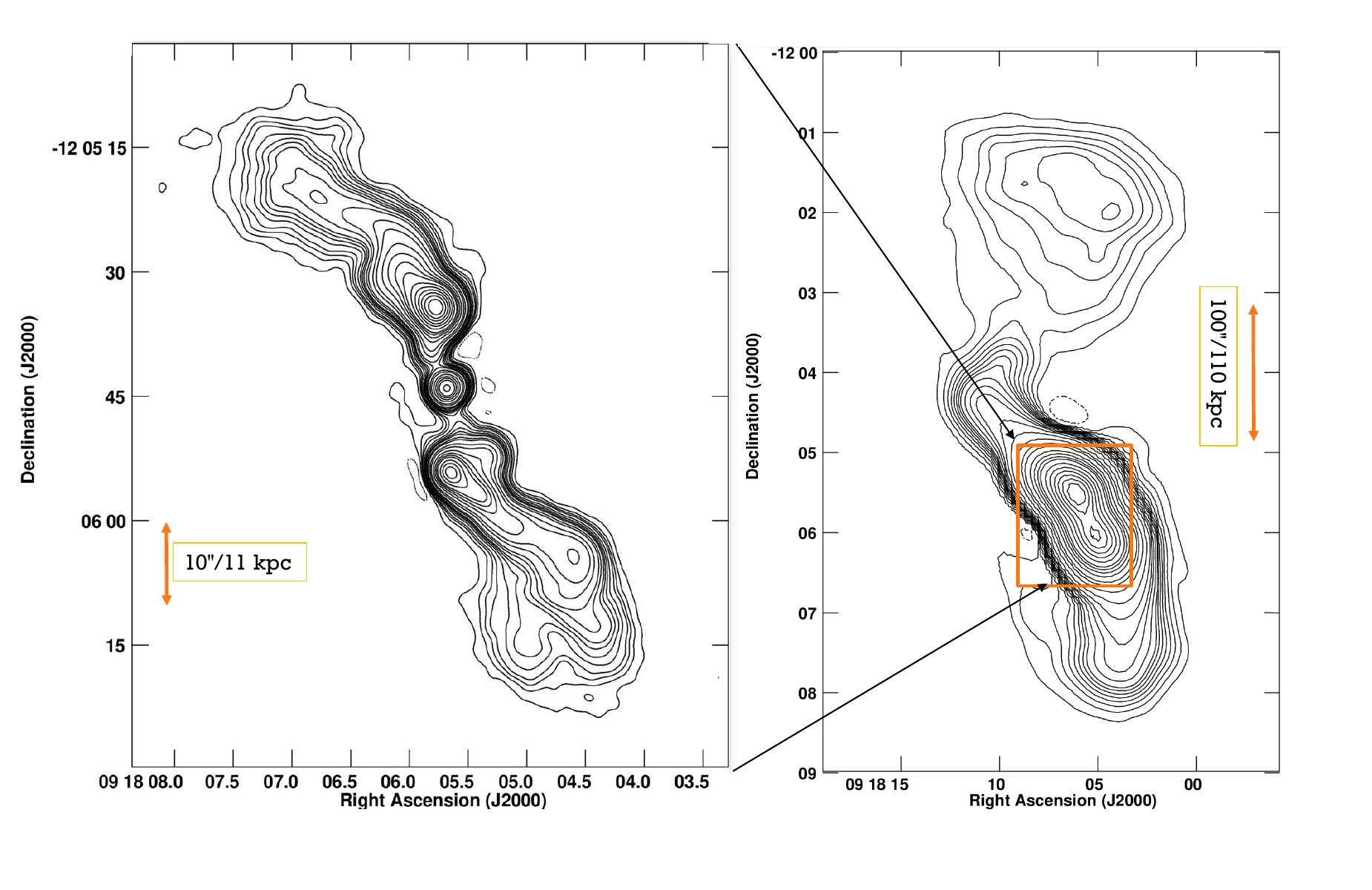}
 \caption{Total intensity contour maps of Hydra A, showing the inner
   tails (left) and jets at 11.1 GHz with 2.65$\arcsec$ resolution,
   and the low-brightness extended tails (right) at 1.04 GHz with 28$\arcsec$ resolution.  The colored box in the right panel shows
   the extent of the left panel image.  These images are from the VLA
   Calibrator Monitoring Program observations taken with the VLA in
   2013. Left plot: The peak brightness is 0.59 Jy beam$^{-1}$ and contour levels are 0.0059$\times$(-0.2, 0.2, 0.5, 0.75, 1, 1.5, 2, 2.5, 3, 4, 5, 7.5, 10, 15, 20, 25, 30, 40, 50, 60, 70, 80, 90) Jy beam$^{-1}$. Right plot: The peak brightness is 20.1 Jy beam$^{-1}$ and contour levels are 0.0201$\times$(-0.1, 0.1, 0.2, 0.3, 0.4, 0.5, 0.6, 0.8, 1, 1.25, 1.5, 2, 3, 5, 7.5, 10, 15, 20, 25, 30, 35, 40, 50, 60, 70, 80, 90) Jy beam$^{-1}$. \label{fig:1}}
 \end{figure*}

At low radio frequencies, Hydra A's detectable radio structure extends
$\sim 530$ kpc in the North-South direction, and $\sim 265$ kpc in the
East-West direction \citep{TAYLOR1990, LANE2004}.  The right panel of
Figure~\ref{fig:1} shows this large-scale structure at 1.04 GHz with 28$\arcsec$ resolution.  The left panel of this figure shows the
brighter inner tails and jet emission at 11.1 GHz with 2.65$\arcsec$
resolution, whose polarimetric structures are the subject of this
paper. The polarized emission from the outermost parts of the source
is too faint to be detected in our observations.

Hydra A is embedded in a cool core X-ray cluster of luminosity $L=4 \times 10^{44}$ ergs s$^{-1}$ between $0.5$ and $4.5$ keV
\citep{DAVID1990}.  The cluster gas temperature decreases from roughly
$ 4$ keV at $200$ kpc radius to $3$ keV at $10$ kpc
\citep{MCNAMARA2000,2001DAVID}, while the electron density, $n_e$,
within 10 kpc is $0.06$ cm$^{-3}$ \citep{MCNAMARA2000}, and decreases
with radius as $r^{-0.4}$ out to $30$ kpc, and as $r^{-1.6}$ out to
$100$ kpc \citep{2001DAVID}.

X-ray deficit regions are found coincident with the radio tails
\citep{MCNAMARA2000,2001DAVID,NULSEN2002}. These are most obvious on
X-ray maps superimposed by radio intensity at $330$ MHz, see Figures 1 and
2 of \citet{2009SIMIONESCU}, and Figure 2 of \citet{2005NULSEN}. Such
regions -- commonly referred to `cavities' -- are common in clusters
of galaxies hosting radio galaxies,
\citep[e.g][]{1993BOEHRINGER,1994CARILLI,2000FABIAN,2001BLANTON,2002HEINZ},
and are believed to result from exclusion of the thermal cluster gas
by the expanding bubble of synchrotron-emitting relativistic gas
originating from the nucleus.  Surrounding the cavities is a region of
enhanced X-ray surface brightness gas and pressure
\citep{2005NULSEN,2009SIMIONESCU}. For Hydra A, this enhanced region
extends $6\arcmin$ north and $4.3\arcmin$ east of the AGN
\citep{2005NULSEN}. This enhancement is interpreted as weak shocks of
Mach number $\sim 1.3$ due to the same energy outburst from the
 nucleus that created the current radio
source \citep{2009SIMIONESCU}. Similar weak shocks are also observed
in the Perseus cluster \citep{2003FABIAN}, M87 \citep{2002YOUNG} and
Cygnus A \citep{SNIOS2018}.

The radio emission from Hydra A displays extremely large Faraday
rotation measures ($\mathrm{RM}$), with the northern tail showing $\mathrm{RM}$s ranging
from $-1000$ and $+3300$ rad m$^{-2}$, and the southern tail showing
$\mathrm{RM}$s from a few $\times 1000$ rad m$^{-2}$ down to $-12000$ rad
m$^{-2}$ \citep{TAYLOR1990,TAYLOR1993}. The rotation measures are
predominantly positive in the northern tail, and predominantly
negative and patchy in the southern tail \citep{TAYLOR1993}.
Large rotation measure gradients are also observed in both tails, with
gradients of up to $\sim 1000$ rad m$^{-2}$ arcsec$^{-1}$ in the
northern tail, and much larger values in the southern tail
\citep{TAYLOR1993}.  Additionally, the southern tail is less polarized
compared to the northern tail, and depolarizes much more rapidly with
frequency \citep{TAYLOR1993}. The $\mathrm{RM}$ maps of Hydra A were also statistically analyzed by \citet{2005Vogt} and \citet{2011Kuchar} for the underlying magnetic power spectra, and the Kolmogorof like spectra were reported.

\citet{TAYLOR1990} attributed the observed depolarization in the tails
to the large transverse gradients in $\mathrm{RM}$ rotating the intrinsic
source polarization by more than 1 radian over the angular scale of
the resolution beam.

The large $\mathrm{RM}$ can be due to the ambient cluster
gas, or a boundary layer of compressed cluster gas surrounding the
tails, or a region of mixed synchrotron gas with external thermal gas
in the boundary layer. Using the most recent RM reconstructed map from \citet{2022Hutschenreuter}, we estimate the RM contribution from our Galaxy in the direction of Hydra A  to be $-1.038 \pm 9.639$ rad m$^{-2}$, derived over 5 deg radius -- suggesting that our Galaxy cannot account for the observed RMs and gradients. Finally, the
optical line emitting gas observed by
  \citet{1988BAUM} extends only to a radius of $15\arcsec$ from the
nucleus, and hence cannot account for the overall
large-scale high $\mathrm{RM}$s across the tails.

\citet{TAYLOR1993} argued against internal Faraday
  rotations based on three observables: the lack of correlation
  between the observed $\mathrm{RM}$s with the depolarization; the similarity
  in $\mathrm{RM}$s across the tails and jets would imply that the internal
  mixing occurs equally in these regions -- but this is unlikely since
  they are intrinsically different; and the absence of non-linearities
  in position angle vs. $\lambda^2$ which are expected for rotations
  $>90^{\circ}$. On the other hand, the asymmetries in the $\mathrm{RM}$
  distribution between the tails is explained using the
  Laing-Garrington effect originating from a medium of size $\lesssim
  2 \times$ the source \citep{1988LAING,
    1991GARRINGTON}. \citet{TAYLOR1993} estimated a source inclination
  to the sky plane of $ \leq 60^{\circ}$.

Most, if not all, depolarization mechanisms will not produce a
perfectly linear relation between the observed electric vector and
$\lambda^2$.  The physical details of the depolarization process will
be contained in these non-linearities, along with the detailed decline
of the polarization fraction.  The study by \citet{TAYLOR1993}
consisted of only five wavelengths spanning $2$ to $3.6$
cm. With such sparse sampling, subtle depolarization effects, such as
those due to turbulent small-scale magnetic fields or to boundary
layer effects will not be visible. Additionally, the depolarization effects, which are most
manifest at long wavelengths, may not be seen at such short
wavelengths. With the availability of the wideband Jansky Very Large
Array \citep[JVLA, ][]{2001PERLEY}, we have the capability to observe
Hydra A at lower frequencies than those available to
\citet{TAYLOR1990, TAYLOR1993}, -- notably, with the $2$ -- $4$ GHz
system -- and with complete frequency coverage, thus allowing a
detailed study of the depolarization characteristics of the source.

In this paper, we present a full polarization study of Hydra A using
the wideband ($2$ -- $12$ GHz) capabilities of the JVLA. The data have
unprecedented high sensitivity and high spectral resolution. Our goals
are to better determine the polarization structures of the jets and
tails, to accurately determine the depolarization characteristics, and
to determine the characteristics of the magnetic field structures
likely responsible for the extraordinary depolarization.

This paper is organized as follows: The observations and calibration
of the data are presented in Section \ref{sec:observations}, and the
imaging of the data in Section \ref{sec:imaging}. Section
\ref{sec:newdata} presents our polarization results, followed by the
results of a high-frequency high-resolution Faraday rotation study in
section \ref{sec:faradayrotation}. In section \ref{sec:beamtest} we
show the results of applying the high frequency, high resolution model
to the low frequency data.  A summary is presented in Section
\ref{sec:discussion}.

\section{Observations and Data Calibration}\label{sec:observations}

Hydra A was observed in all four VLA configurations under project
code 13B-088 at L (1 - 2 GHz), S (2 - 4 GHz), C (4 - 8
GHz), and X (8 - 12 GHz) band resulting in a total frequency
coverage of $1$ - $12$ GHz. The observing dates, durations and
configurations are shown in Table \ref{tablog}. All bands (L, S, C, and X) were observed together.

\begin{table}[!ht]
\centering
\caption{Observing log. \label{tablog}}
\begin{tabular}{c c c}
 \hline 
Array & Observation  & Duration \\
configuration & date & [hr] \\
\hline  
 B & 2013 Dec 14 & $ 6.0$ \\
 A & 2014 Feb 27 & $ 5.0$ \\
 A & 2014 Mar 07 & $ 5.0$ \\
 D & 2014 Jun 27 & $4.0$ \\
 C & 2014 Oct 19 & $ 4.0$ \\
 \hline
\end{tabular} 
\end{table}

The data were taken with a time resolution of 2 seconds in
A-configuration, and 3 seconds in other configurations.  The frequency
channelization varied with band and configuration: 2 MHz for S-, C-
and X-bands in B, C, and D configuration, 1 MHz for L-band in those
same configurations. For A configuration, 2 MHz was used for X-band, 1
MHz for C-band, 0.5 MHz for S-band, and 0.25 MHz for L-band.  These
data were subsequently resampled to the same spectral resolution as in
the B, C and D configurations using the AIPS program {\tt SPEC}.  This
resulted in a spectral resolution of 1 MHz in L-band, and 2 MHz in the
other bands. These values were chosen so as to minimize both the
bandwidth smearing and bandwidth depolarization.

All calibration and imaging of the data were done using the
Astronomical Image Processing System (AIPS) software
\citep{GREISEN1990,VANMOORSEL1996}.

The editing and calibration procedures were exactly the same as for
Cygnus A \citep[see][for details]{2020SEBOKOLODI}.  After calibration,
the data were averaged in frequency and time to: 1 MHz/12 seconds for
L-band, 2 MHz/12 seconds for S-Low (2 - 3 GHz), 4 MHz/12 seconds for
S-Hi (3 - 4 GHz), and 8 MHz/12 seconds for C- and X-band.

As the original external phase and amplitude calibration is not
sufficiently accurate to enable high-fidelity imaging,
self-calibration of the Hydra A data was performed to remove gain
drifts between the individual spectral windows due to time-variable
changes in bandpass shape, and between the data taken in the four
separate configurations.  For this purpose, we utilized the emission
from the bright, unresolved nucleus to put all the data on a common
flux density and positional scale. We achieved this by using the long spacings to phase/amplitude reference the shorter ones.  This was done by making an A-configuration-only image, and self-calibrating the data from that model.  Once this was done, we use the improved image to self-calibrate the ``B'' configuration, then adding those data to the ``A'' data, and making a model using both.  Then repeat this with C and D configurations, extending the short-spacing limit downwards each time to ensure stable solutions.

\section{Imaging and Data Products}\label{sec:imaging}
Following the self-calibration process, we made cube images of Stokes
$Q$, $U$, and $I$ using a single scale cleaning algorithm. The spatial
extent of these image cubes is $4$k by $4$k, sampled with
pixel size of $0.05\arcsec$.  The cubes were made at two standard
resolutions, namely: $1.50\arcsec \times 1.0\arcsec$ and $0.50 \arcsec
\times 0.35 \arcsec$.  The lower resolution includes data between 2 -
12 GHz, and the higher includes data between 6 - 12 GHz.  The L-band
data (1 -- 2 GHz) were not utilized in this study, as the
depolarization at these frequencies at our resolution of 2 -- 3
arcseconds was nearly total.

\begin{figure}
 \centering
 \includegraphics[width=1.12\linewidth]{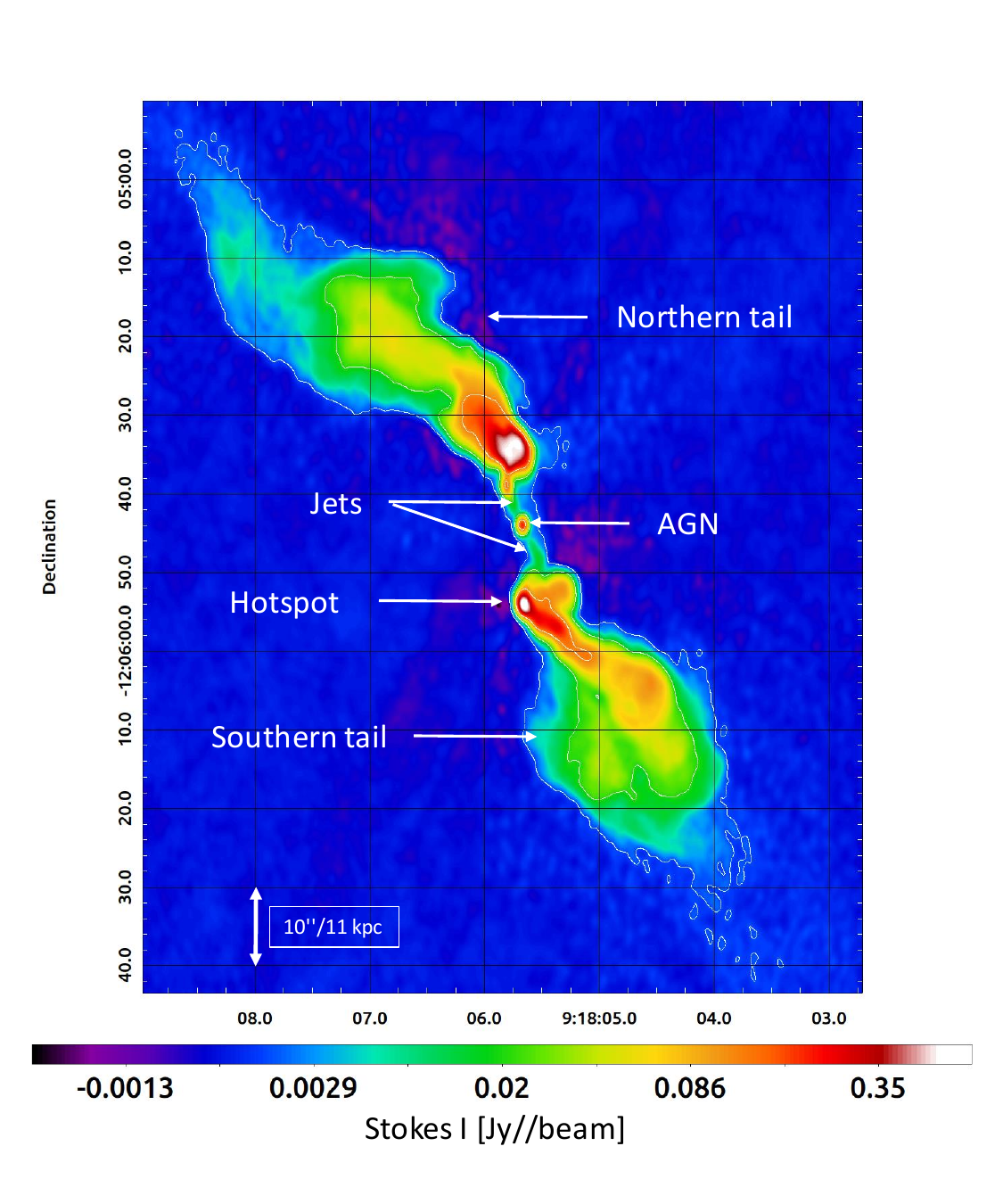}
 \caption{False-color total intensity map at 2.051 GHz with $1.5\arcsec \times 1.0 \arcsec$ resolution, showing the inner jets and tails more
   clearly.  This image utilizes data taken for this
   paper.  \label{fig:SColor}}
 \end{figure}

A false-color rendition of the inner regions of the source at 2.05
GHz, with $1.5\arcsec \times 1.0 \arcsec$ resolution is shown in
Figure~\ref{fig:SColor}.  This figure shows the `S' symmetry of the jets
and the inner tails.  The southern lobe shows a diffuse hotspot a few
kpc from the nucleus, while the northern lobe has no discernable
hotspot \citep{TAYLOR1990}.

To maximize image sensitivity, we combined spectral channels during imaging. However, bandwidth depolarization affects the number of channels we can average together, which depends on the channel frequency and the maximum $\mathrm{RM}$ present in the data. The number of channels
$N_{\mathrm{avg}}$ to average together was set by the requirement that the
rotation of the plane of polarized emission due to an $\mathrm{RM}$ of 12000 rad
m$^{-2}$ be less than 10 degrees.  Table \ref{tab2} shows the
channelization utilized for the different bands.

\begin{table}[!ht]
 \centering
 \caption{The number of frequency planes in each band utilized to
   avoid Faraday depolarization. \label{tab2}}
 \begin{tabular}{c c c c c}
 \hline 
 Band & $\nu$-interval & $\Delta \nu$ &  $N_{\mathrm{avg}}$ & $N_{\mathrm{planes}}$\\
  & [GHz] & [MHz] &  & \\ 
 \hline   
 %L & 1-2 & 1 & 1\\
 $\mathrm{S_{lo}}$ & 2-3 & 2 & 1 & 512\\
 $\mathrm{S_{hi}}$ & 3-4 & 4 & 1 & 256\\
 $\mathrm{C_{lo}}$ & 4-6 & 8 & 1 & 256\\
 $\mathrm{C_{hi}}$ & 6-8 & 16 & 2 & 128\\
 $\mathrm{X_{lo}}$ & 8-10 & 32 & 4 & 64\\
 $\mathrm{X_{hi}}$ & 10-12 & 64 & 8 & 32 \\ 
 \hline  
 \end{tabular}
\end{table}

We used single-scale CLEAN because of its speed.  However, the
accuracy of single-scale deconvolution by CLEAN degrades at high
frequencies and high resolutions, particularly for extended emission
such as the tails of Hydra A. The problem is particularly severe in
Stokes $I$, where the brightness of the extended regions is often broken
by CLEAN into synthesized beam-sized `islands' of emission.  To reduce
these effects, we fitted a smooth brightness profile for each spatial
position through the frequency axis of the cube, and used the smoothed
value to estimate the Stokes $I$ emission for each frequency plane.
This was done separately for the lower-resolution $2$ - $6$ GHz and
higher-resolution $6$ - $12$ GHz cubes to avoid significant errors resulting from steepining of spectra at higher frequencies \citep{2009Cotton}. The polarized emission cubes
($Q$ and $U$) do not suffer from this problem, as the high $\mathrm{RM}$ results in
these images being dominated by rapidly varying brightnesses which are
efficiently and effectively handled by CLEAN.

The off-source noise in the Stokes $Q$ and $U$ images at $0.50$\arcsec$\times$
$0.35$\arcsec ranges between $0.035$ mJy beam$^{-1}$ and $0.12$ mJy
beam$^{-1}$, and in Stokes $I$ between $0.04$ mJy beam$^{-1}$ and $0.3$
mJy beam$^{-1}$. At $1.50$\arcsec$\times$ $1.0$\arcsec, the off-source
noise ranges between $0.06$ mJy beam$^{-1}$ and 1.1 mJy beam$^{-1}$
for Stokes $Q$ and $U$ images, and $0.1$ mJy beam$^{-1}$ and $5$ mJy
beam$^{-1}$ for Stokes $I$ images.

%\subsection{Polarized Emission in $\lambda^2$-space}
From the Q and U images, we derived the polarized intensity image
$P=\sqrt{Q^2+U^2}$, corrected for Ricean bias, and the polarization
angle $\chi = 0.5\arctan{U/Q}$, and their associated errors as described
in \citet{2020SEBOKOLODI}. We also compute depolarization ratios by taking the ratio of two fractional polarization maps at two different frequencies or resolutions. We estimate the errors in the depolarization ratio map by computing the propagation of error of the two fractional polarization maps.

Additionally, we calculate the Faraday spectra for every line-of-sight
using the RM-synthesis technique as described in
\citet{2020SEBOKOLODI}.  For a more complete description, see the
original work by \citet{2005BRENTJENS}. The rotation measure transfer
function for our $2-12$ GHz data has width $\sim$180 rad m$^{-2}$,
while for the $6 - 12$ GHz data it is $\sim 2030$ rad m$^{-2}$. The
computed Faraday spectra were deconvolved using the AIPS task
`TARS'.
% We assumed uniform weighting for all channels and defined
% $\lambda^2_0$ as a weighted mean of all our $\lambda^2$.

\section{Polarization Results}\label{sec:newdata}
In this section, we analyze the polarization data of Hydra A in
detail.  In particular, we determine how the polarization changes
with frequency and with resolution. This is essential since the
depolarization can occur due to differential rotation of the emitting
gas, due to magnetized thermal gas along the line-of-sight (``true''
or ``internal'' depolarization), or within our synthesized beam due to unresolved
polarization structures (``beam'' depolarization). Distinguishing
between these phenomena is difficult, but is necessary to have a clear
understanding of both phenomena if we are to fully understand the
underlying physics associated with radio galaxies.

\subsection{Polarization as a Function of Frequency}
\subsubsection{Fractional Polarization Maps }
\begin{figure*}
 \center
   \begin{minipage}[b]{0.45\linewidth}%[3]
    \centering
    \includegraphics[width=0.95\linewidth]{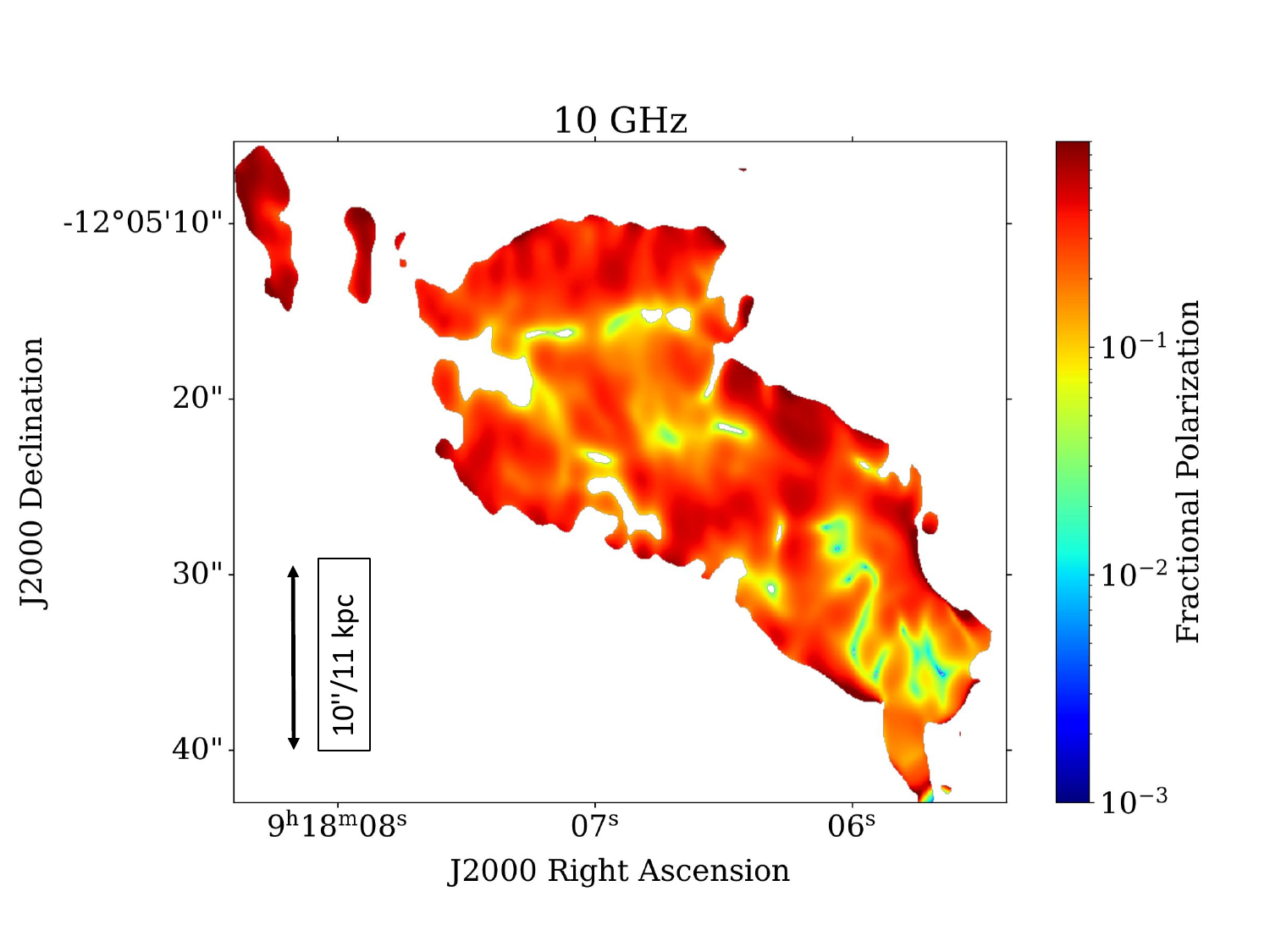}
  \end{minipage}
     \begin{minipage}[b]{0.45\linewidth}%[5]
    \centering
    \includegraphics[width=0.95\linewidth]{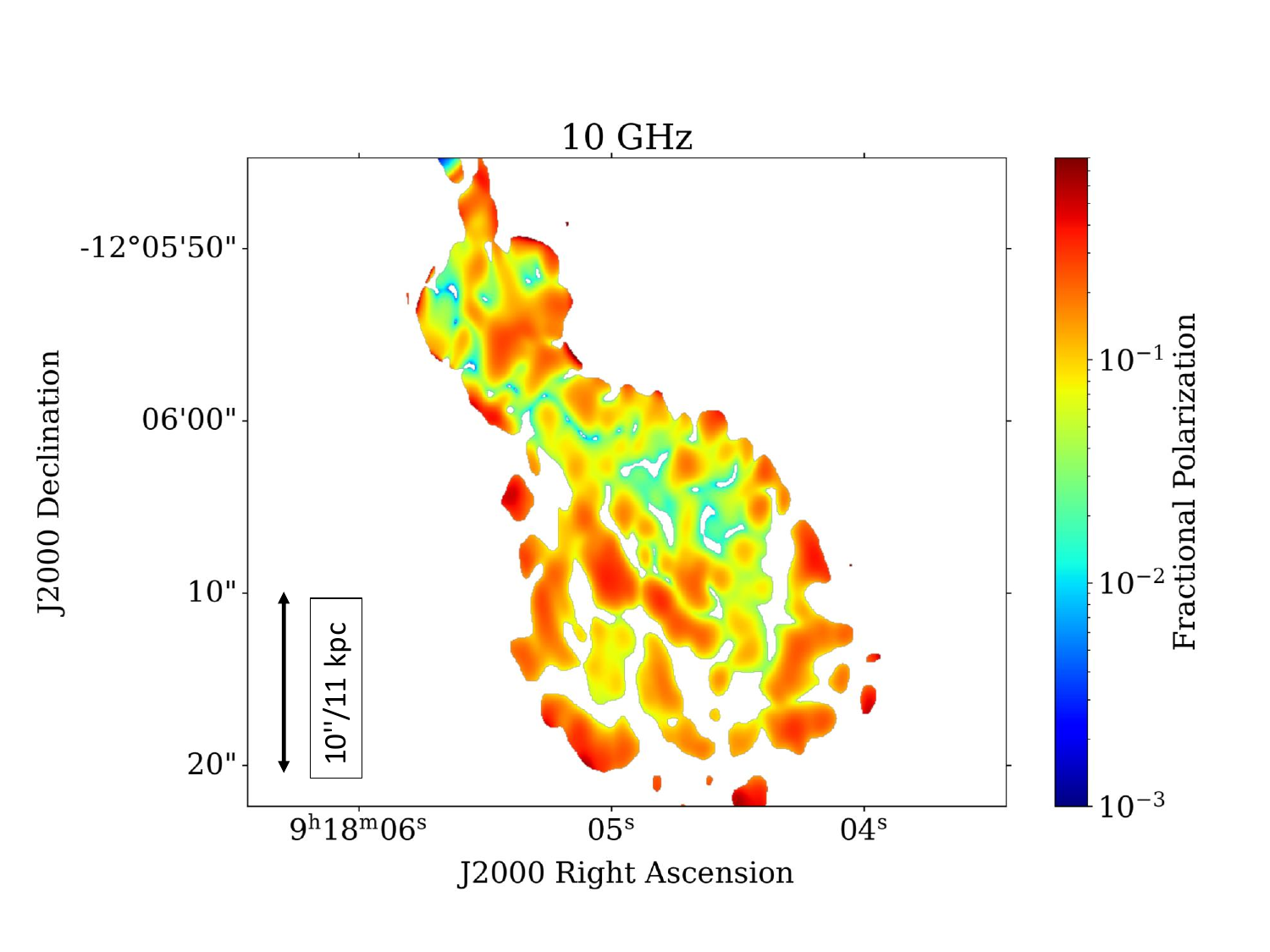}
  \end{minipage}
  
     \begin{minipage}[b]{0.45\linewidth}%[3]
    \centering
    \includegraphics[width=0.95\linewidth]{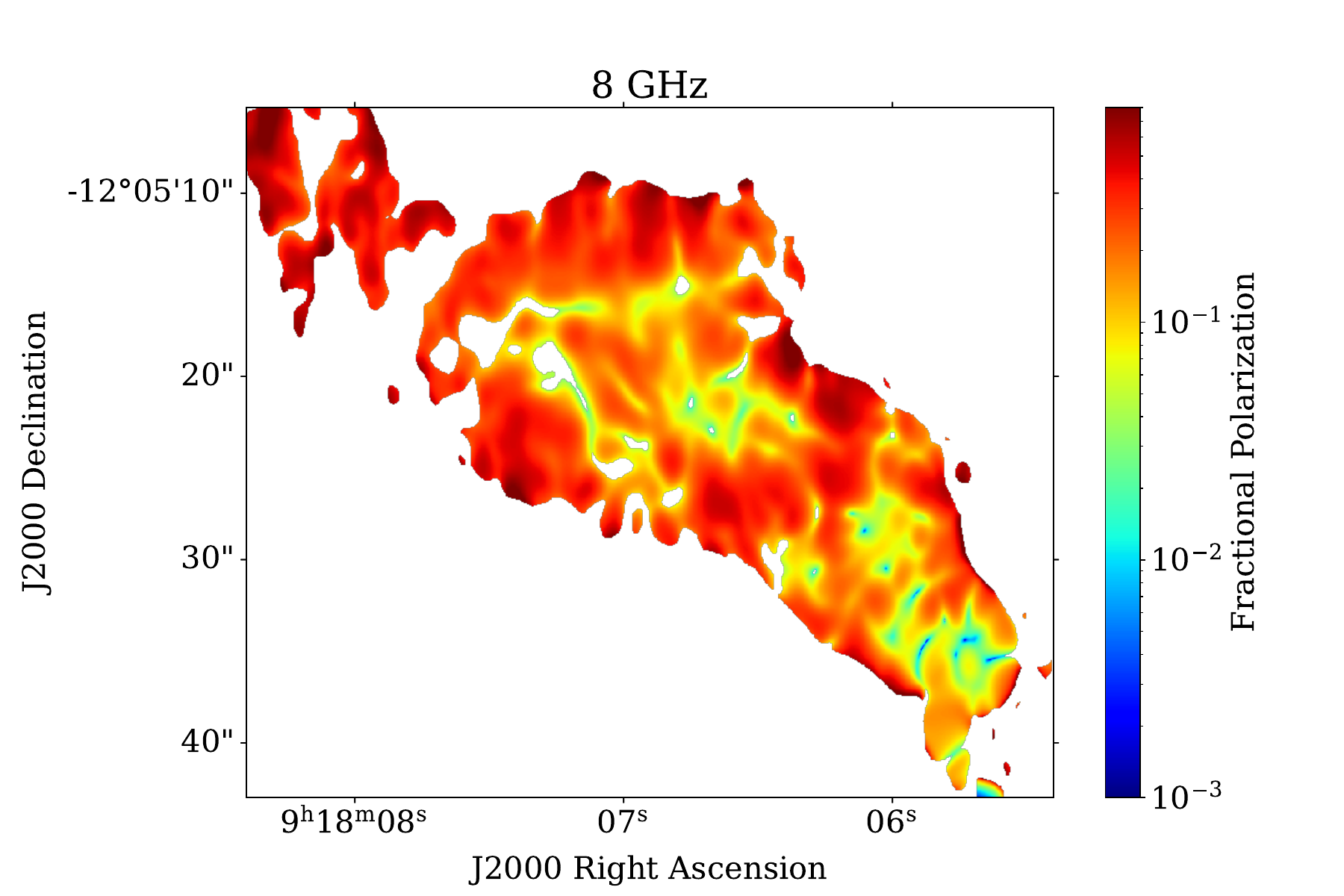}
  \end{minipage}
       \begin{minipage}[b]{0.45\linewidth}%[3]
    \centering
    \includegraphics[width=0.95\linewidth]{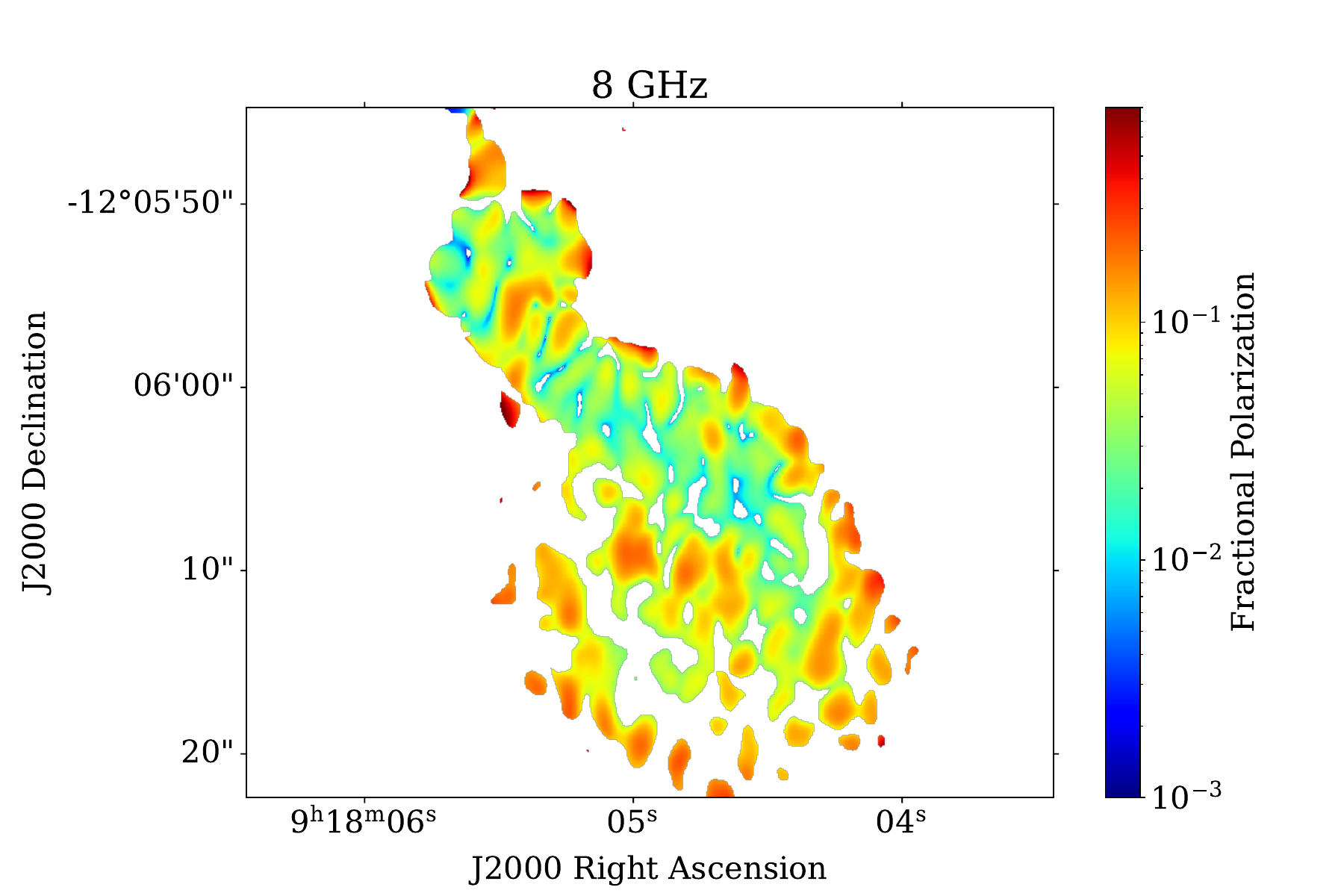}
  \end{minipage}
       \begin{minipage}[b]{0.45\linewidth}%[3]
    \centering
    \includegraphics[width=0.95\linewidth]{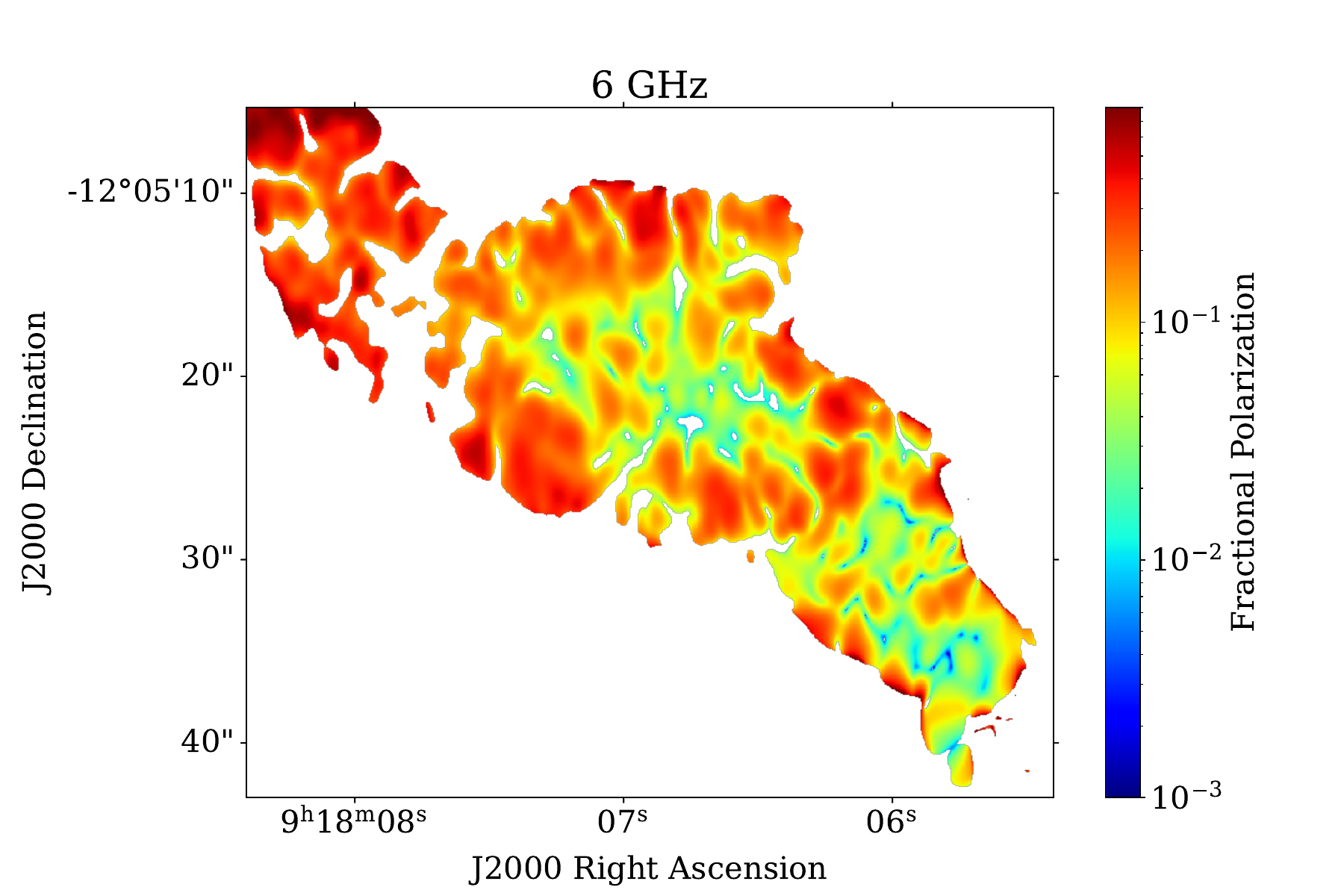}
  \end{minipage}
       \begin{minipage}[b]{0.45\linewidth}%[3]
    \centering
    \includegraphics[width=0.95\linewidth]{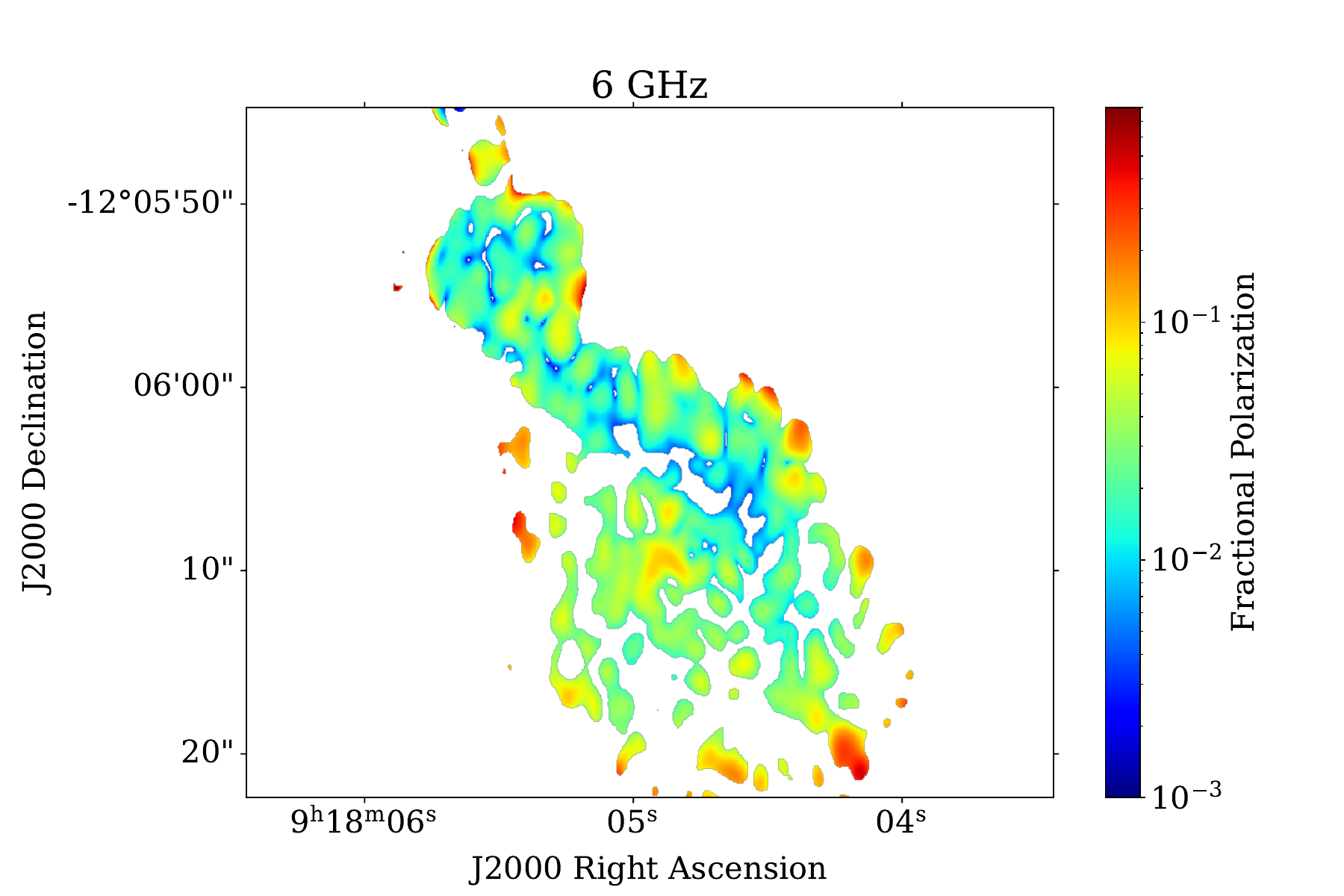}
  \end{minipage}
       \begin{minipage}[b]{0.45\linewidth}%[3]
    \centering
    \includegraphics[width=0.95\linewidth]{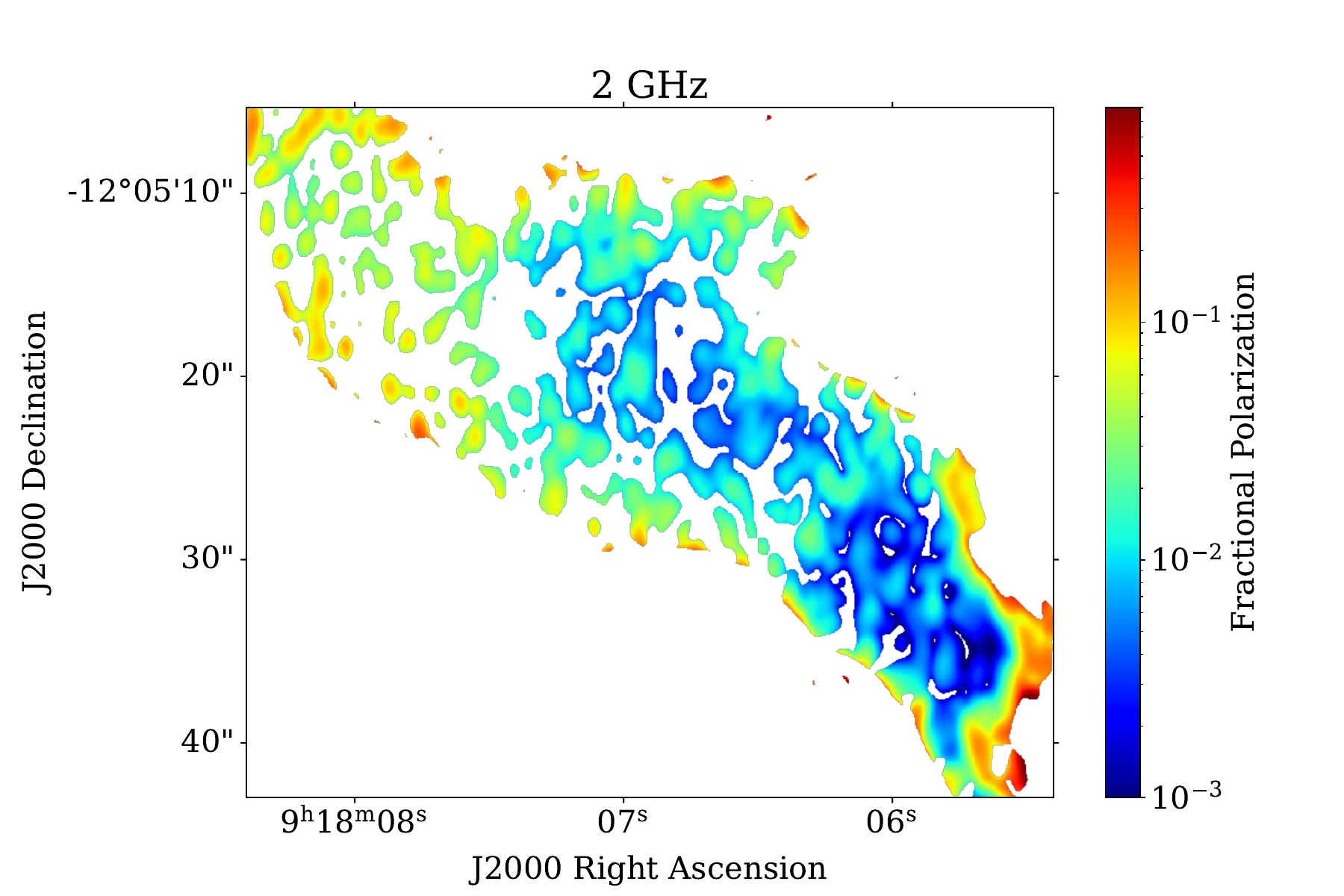}
  \end{minipage}
       \begin{minipage}[b]{0.45\linewidth}%[3]
    \centering
    \includegraphics[width=0.95\linewidth]{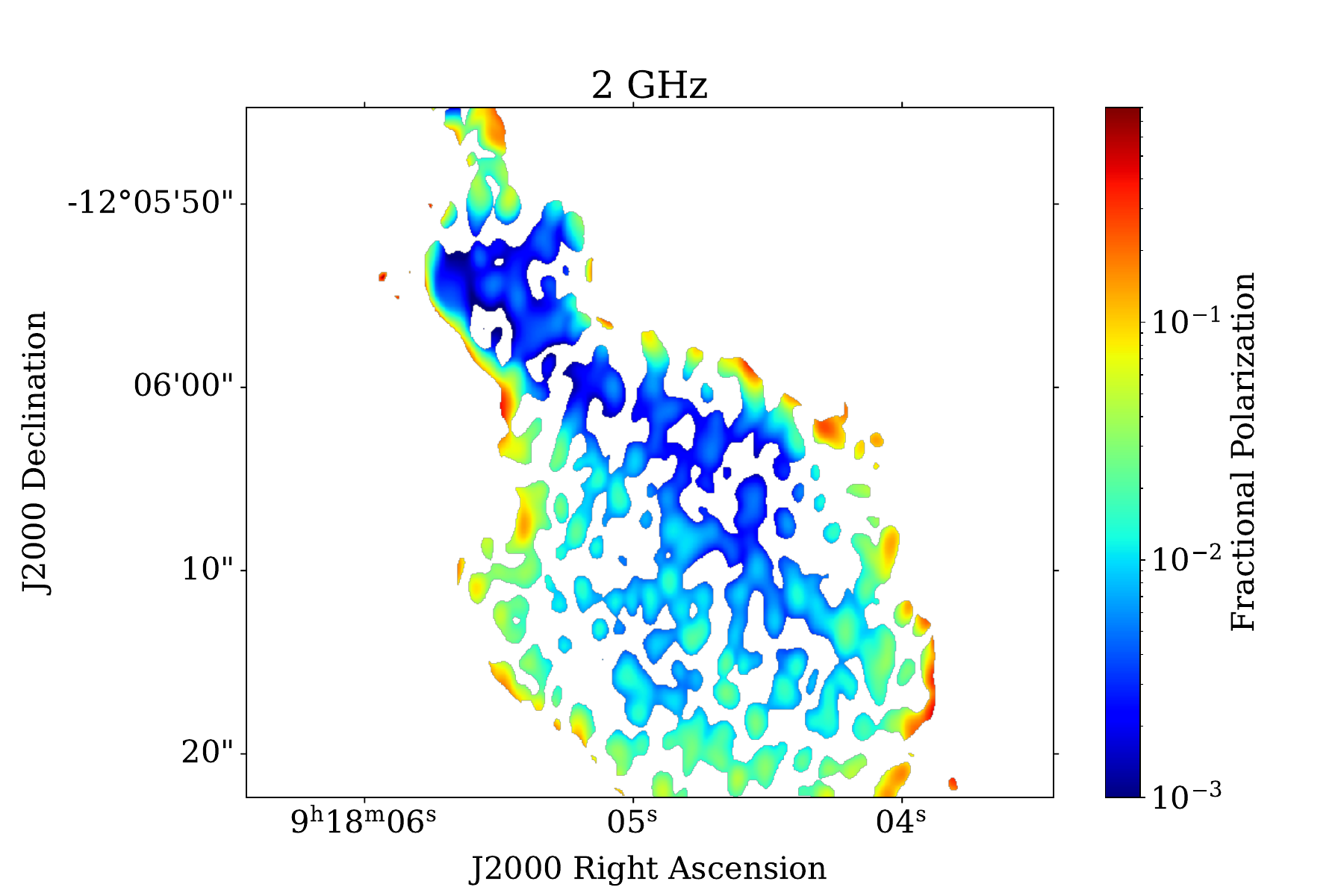}
  \end{minipage}
  \caption{The fractional polarization $p$ across the tails of Hydra A
    at selected frequencies at a resolution of $1.5\arcsec \times 1
    \arcsec$. Pixels shown have SNR in $p/\sigma_p > 60\%$, where
    $\sigma_p$ is the estimated rms noise in the polarization.
    Left: Northern tail. Right: Southern tail. Both tails depolarize
    significantly with decreasing frequency. The northern tail is
    relatively less depolarized compared to the southern
    tail. The physical scale shown in the top panel is the same for the rest of the panels. \label{fig:freqmaps}}
  \end{figure*}

Figure \ref{fig:freqmaps} shows the fractional polarization maps
across the tails of Hydra A at $1.5\arcsec \times 1.0 \arcsec$ resolution. The
northern tail is more polarized than the southern tail at all
frequencies. The fractional polarization of both tails decreases with
decreasing frequency. The inner regions of the tails close to the
nucleus depolarize more rapidly than the outer regions. Similarly to
Cygnus A \citep{2020SEBOKOLODI}, the notable structural features seen
in the Stokes `$I$'
 image, such as the hotspot and jets, are not
discernable in the fractional polarization maps.  The fractional
polarization is relatively smooth across the northern tail, and
patchier across the southern tail. However, the fractional
polarization of both tails becomes clumpy at low frequencies.

\subsubsection{Frequency-Dependent Depolarization Ratio (FDR)}
\begin{figure}[!ht]
 \center
   \begin{minipage}[b]{1\linewidth}%[3]
    \centering
    \includegraphics[width=1.0\linewidth]{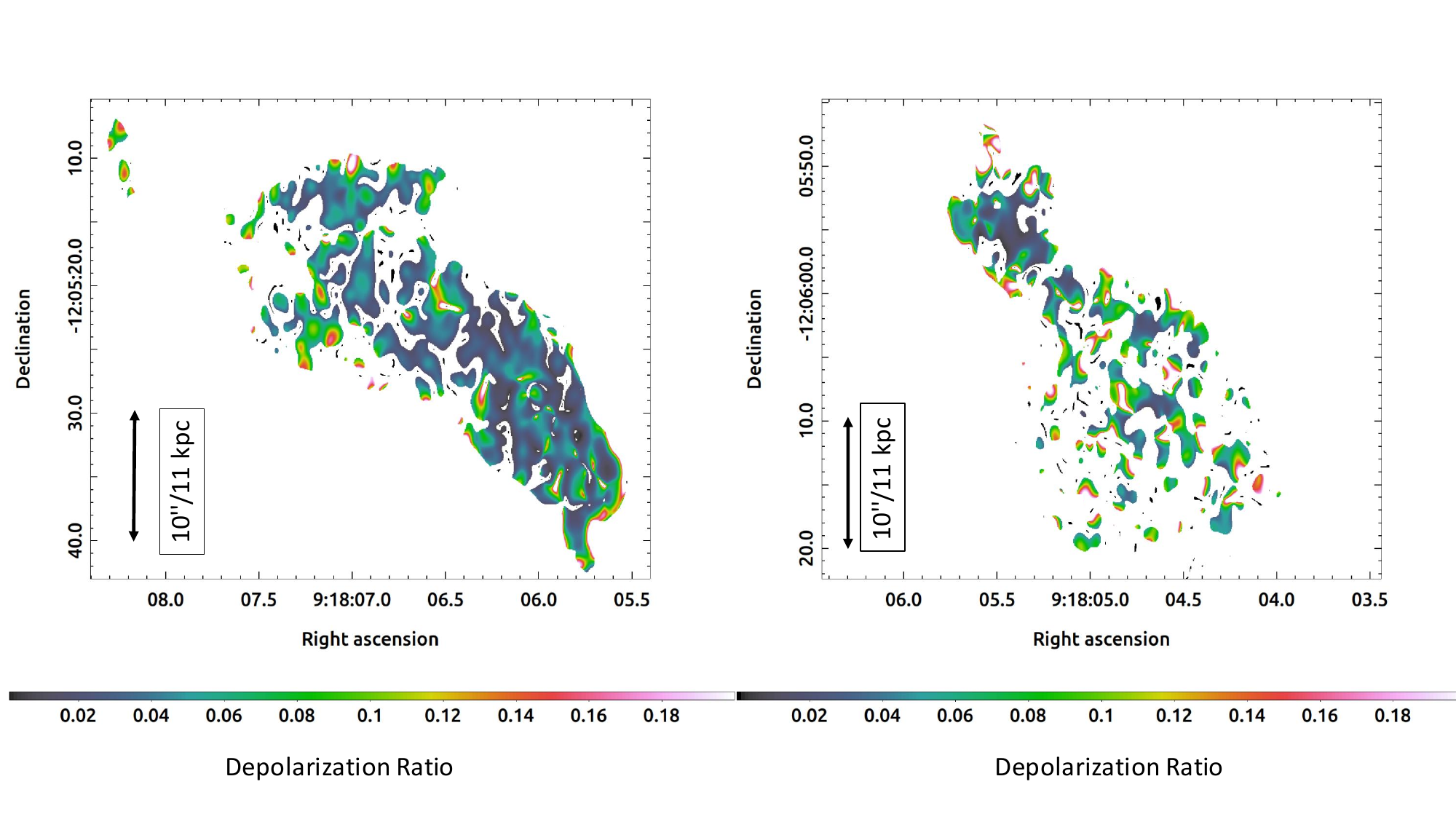}
  \end{minipage}
     \begin{minipage}[b]{1\linewidth}%[3]
    \centering
    \includegraphics[width=1.0\linewidth]{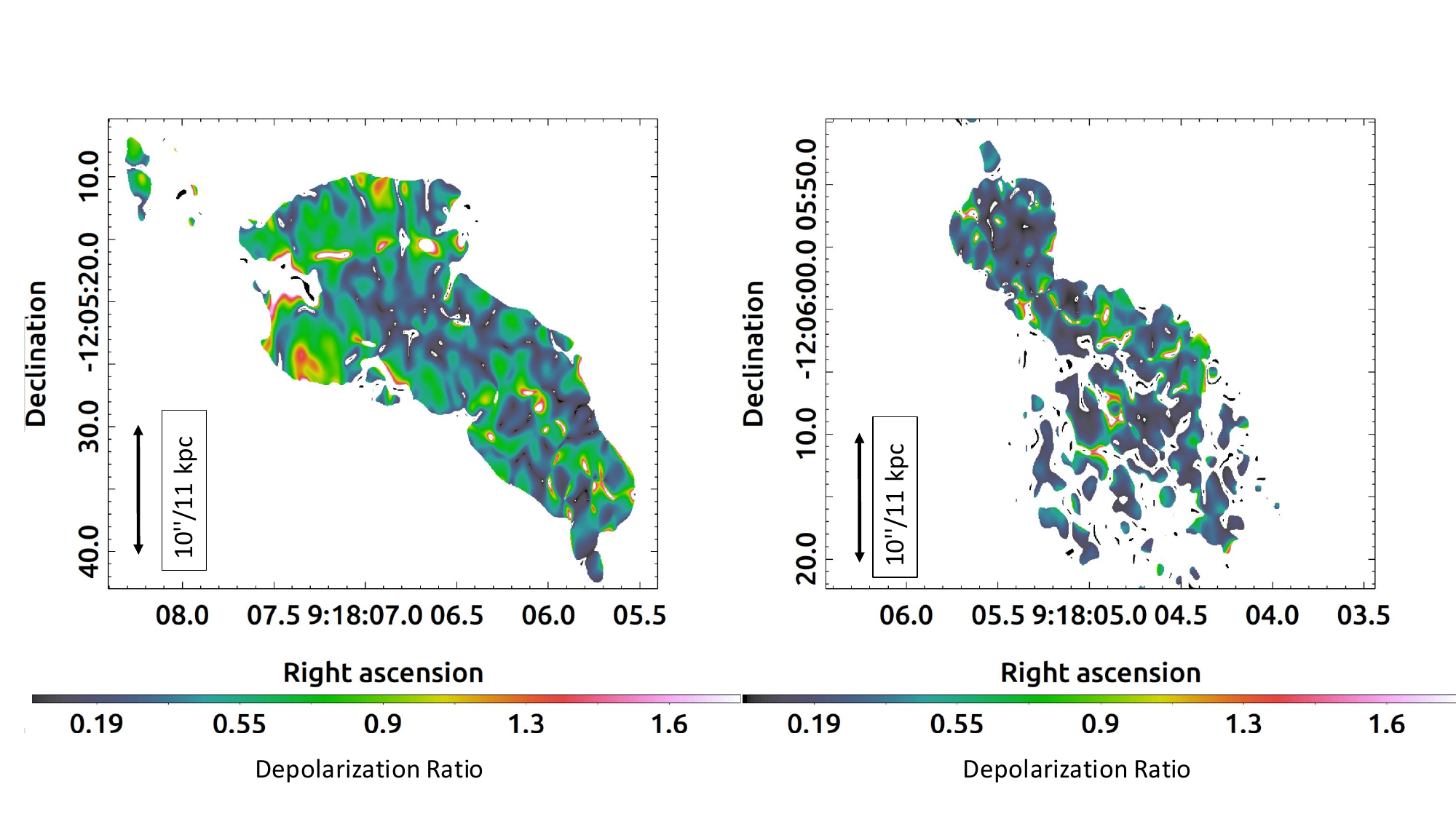}
  \end{minipage}
\caption{Depolarization ratio maps across the tails at $1.5\arcsec \times 1.0 \arcsec$ resolution. Top panel: FDR1 -- the ratio of the 2 GHz
  polarization image to the 10 GHz polarization image. Bottom panel:
  FDR2 -- the ratio of the 6 GHz polarization image to the 10 GHz
  polarization image. Left panel: Northern tail. Right panel: Southern tail. Only pixels with fractional
error of less than $80\%$ are shown. Note the range of color scales in the upper and lower panels are different, due to the stronger depolarization at 2 GHz. Attempting to match the color scales hides the important variations.
  \label{fig:fdr-distribuion}}
\end{figure}

\begin{figure}
 \center
   \begin{minipage}[b]{1\linewidth}%[3]
    \centering
    \includegraphics[width=1.05\linewidth]{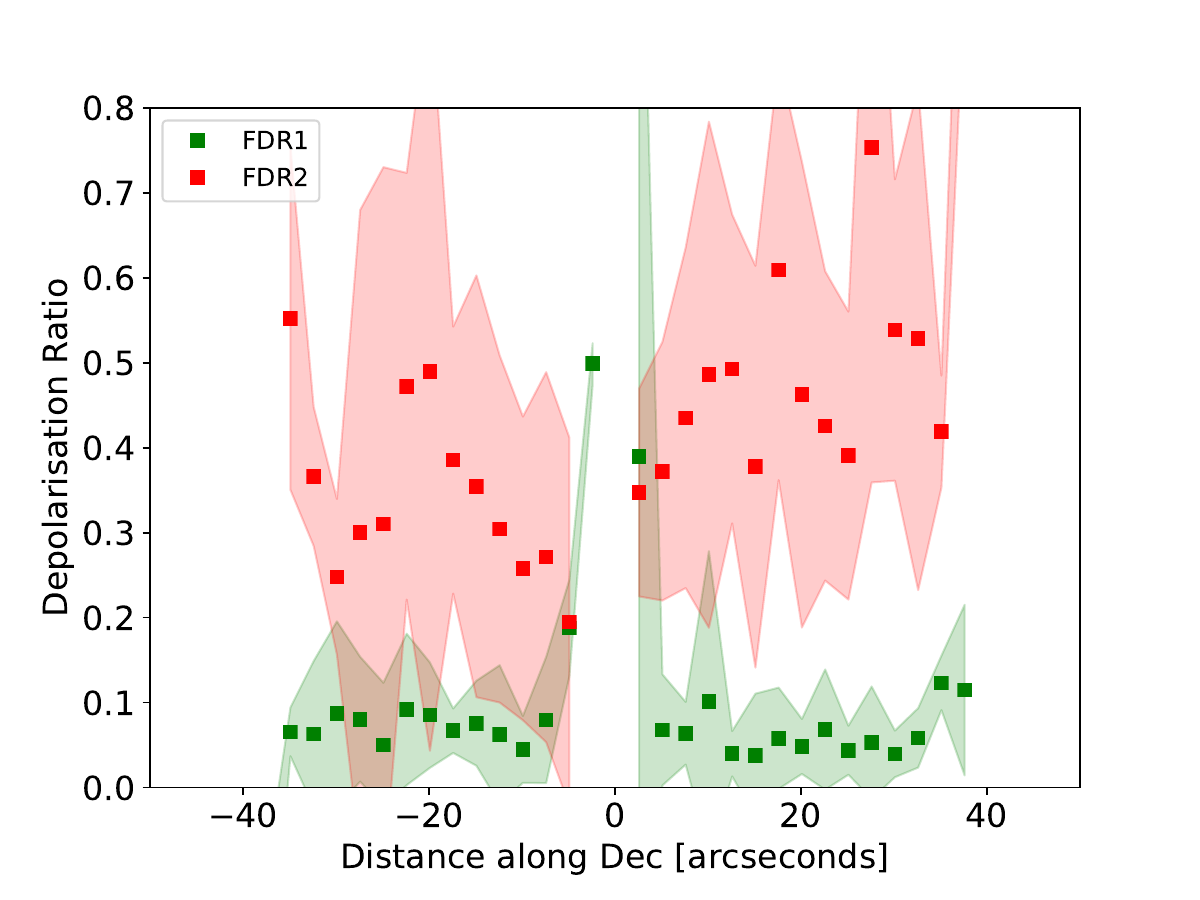}
  \end{minipage}
   \begin{minipage}[b]{1\linewidth}%[3]
    \centering
    \includegraphics[width=1.05\linewidth]{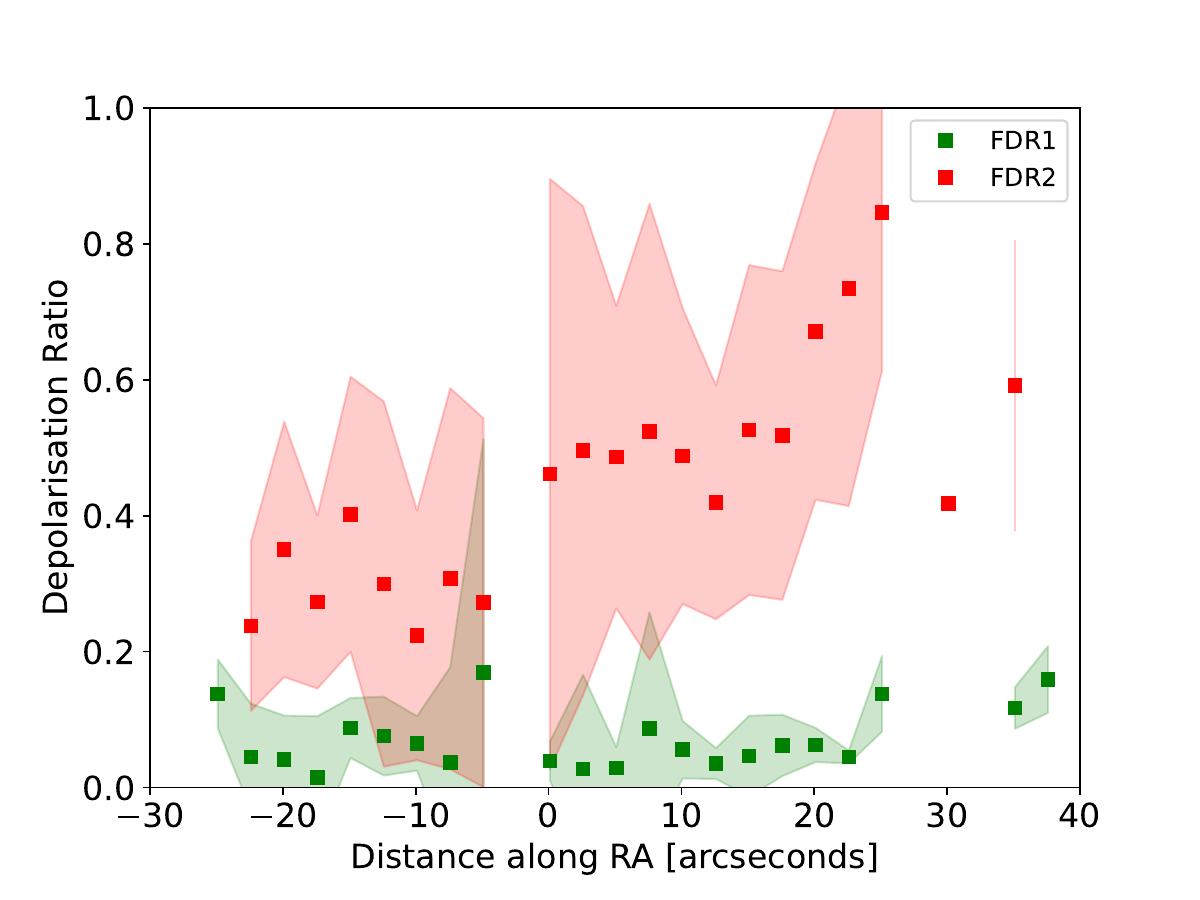}
  \end{minipage}
\caption{Average frequency depolarization ratios in bins of 2.5\arcsec along RA or declination axis. Top: Ratios as a function of declination for all RA. Bottom: Ratios as a function RA for all declination. The southern tail is to the left of RA/Dec $=0''$, and
the northern tail to the right. Red: Ratio of 6 and 10
  GHz denoted as FDR2. Green: Ratio of 2 and 10 GHz (FDR1). Only
  pixels with fractional error $< 80\%$ were used for computing the
  averages and dispersions. The spread shown corresponds to the  standard deviation of the depolarization ratios within bins along declination/RA. The tails
  depolarize to a similar value for FDR1 ranging between
  $0.01-0.1$. The distribution of FDR2 shows radial dependence particularly along RA, with a large spread suggesting
  varying level of depolarization in this higher frequency ratio.
  \label{fig:freqDepol}}
\end{figure}

In order to determine the degree by which the tails depolarize, we
computed the depolarization ratio by dividing the $1.5\arcsec \times 1.0 \arcsec$ resolution fractional polarization map at 2 GHz by the map at 10 GHz
(hereinafter FDR1) and 6 GHz by the same 10 GHz map (hereinafter
FDR2). The spatial distribution of FDR1
  and FDR2 across the tails is shown in Figure  \ref{fig:fdr-distribuion}. Lower values mean stronger depolarization at low frequencies.
The resulting depolarization radial profiles are presented in Figure
\ref{fig:freqDepol}. Due to the shape of the source, it is difficult to orient the source specifically along its extension, so to compute the profiles, we binned the image in slices along right ascension (RA) and declination axis and considered ratios within each bin. The plotted points in the top panel are an average of the depolarization ratios within bins of $2.5''$ sizes as a function of declination for all RA within a bin, with zero defined
as the location of the nucleus. The bottom plot is the depolarization ratio as a function of right ascension for all declination within a bin. The southern tail is to the left and
the northern tail to the right of the AGN in the plot. The spread
shown by the shading corresponds to the standard deviation of the
ratios within each bin at a given declination/right ascension. The depolarization ratio is prone to
large errors since it is a ratio of ratios, so in an attempt to reduce
spurious emission we included only those pixels with fractional error in the depolarization ratio less than 80\%.

The FDR1 (lower frequency ratio) shows significantly stronger
depolarization than the higher frequency ratio, FDR2.  The
depolarization for FDR1 is $\lesssim 0.1$, and
ranges between $0.2-0.8$ for FDR2. This implies that the emission at
10 GHz is depolarized by more than $90\%$ at 2 GHz at this resolution
across the tails. In the case of FDR2, there is no significant trend in
either depolarization ratio with offset from the nucleus except for the radial increase in the ratio across the northern tail as shown in the bottom plot. In general, the northern tail shows large depolarization
values than the southern (i.e., the depolarization is less)
-- implying a more rapid depolarization across the
  southern tail relative to the northern tail. Moreover,  based on FDR2, the
different regions of the tail depolarize widely as suggested by the
spread in the distribution.
 
\subsubsection{Depolarization vs Wavelength Behavior}
We now look at the polarization behavior as a function of $\lambda^2$
for different lines-of-sight across our wideband, high-spectral
resolution data. This is a lengthy process, due to the large number
of lines-of-sight available.  For the purpose of this paper, we will
show a few `representative' example lines-of-sight.

We first define a relative coordinate system, centered on the galaxy
nucleus, with units in tens of milliarcseconds. In this system, north
and west are positive, and east and south are negative.  Thus, a pixel
with coordinate (1585, -3100) is located 15.85$\arcsec$ west, and
31.00$\arcsec$ south of the nucleus. To select our lines-of-sight, we
considered pixels separated by 0.7$\arcsec$ with fractional
polarization above $0.1$ at $8$ GHz. These choices were motivated by obtaining lines-of-sight that are of good signal-to-noise and are usable for scientific analysis. With these restrictions, we
obtain a total of 696 lines-of-sight. However, only 553 of these are
considered for further analysis, as the remaining 134 lines-of-sight
are too noisy to be used for meaningful analysis. The majority of the
excluded lines-of-sight are from the southern tail, and the outermost
regions of the tail.
 
\begin{figure*}
 \center
   \begin{minipage}[b]{1\linewidth}%[3]
    \centering
    \includegraphics[width=0.65\linewidth]{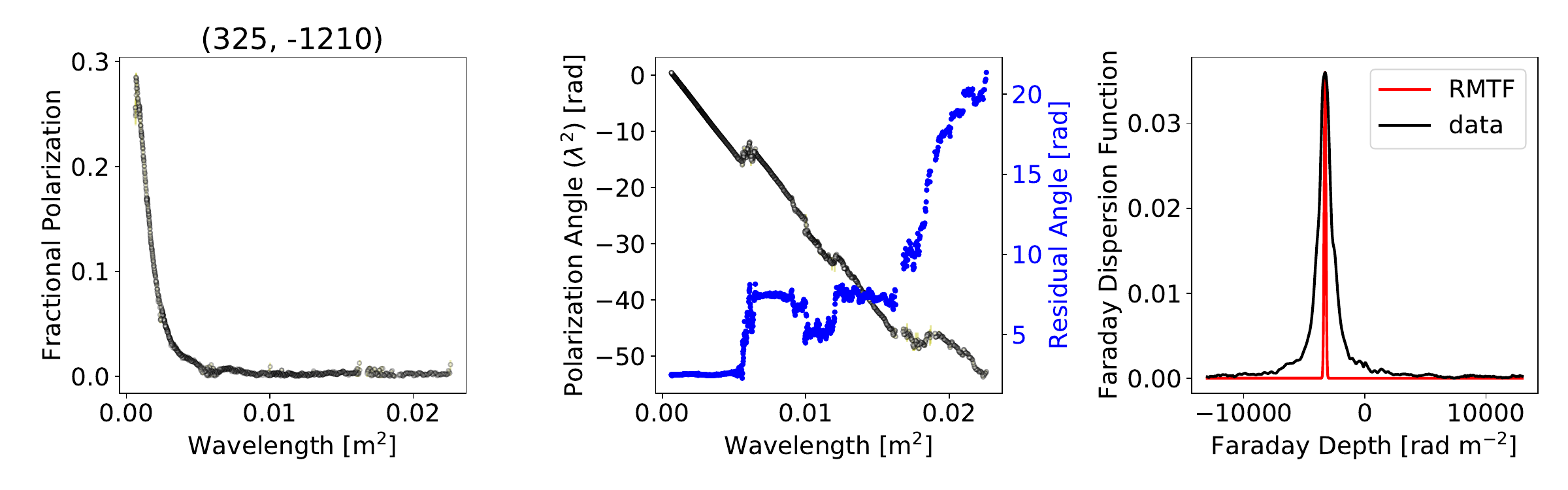}
  \end{minipage}
     \begin{minipage}[b]{1\linewidth}%[5]
    \centering
    \includegraphics[width=0.65\linewidth]{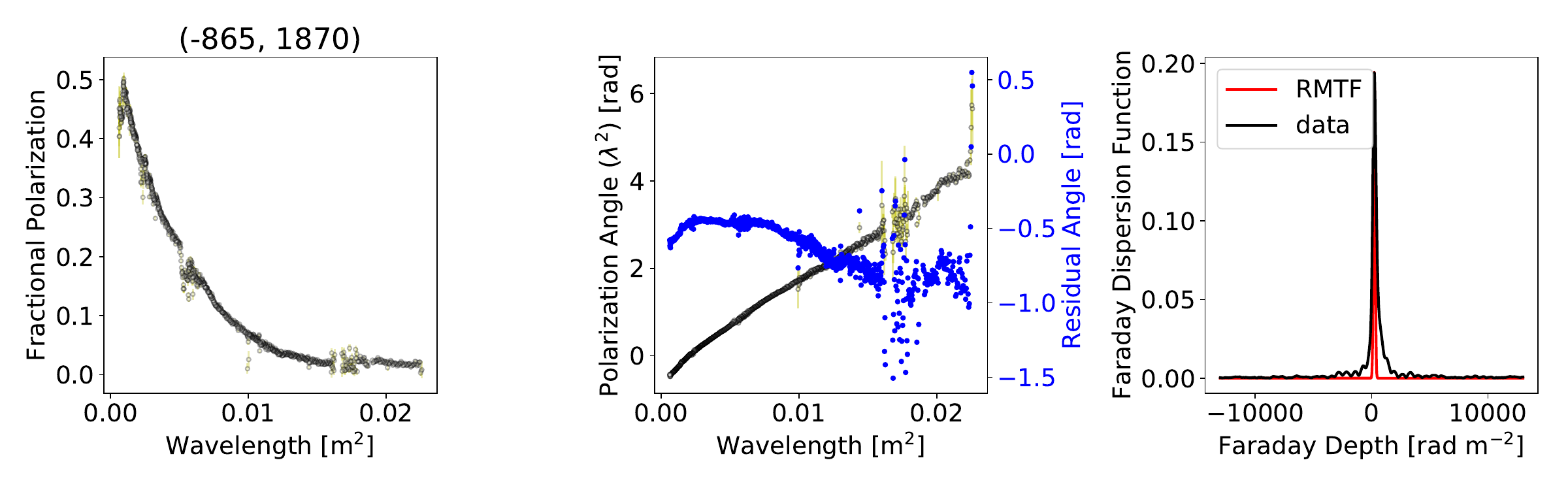}
  \end{minipage}
     \begin{minipage}[b]{1\linewidth}%[3]
    \centering
    \includegraphics[width=0.65\linewidth]{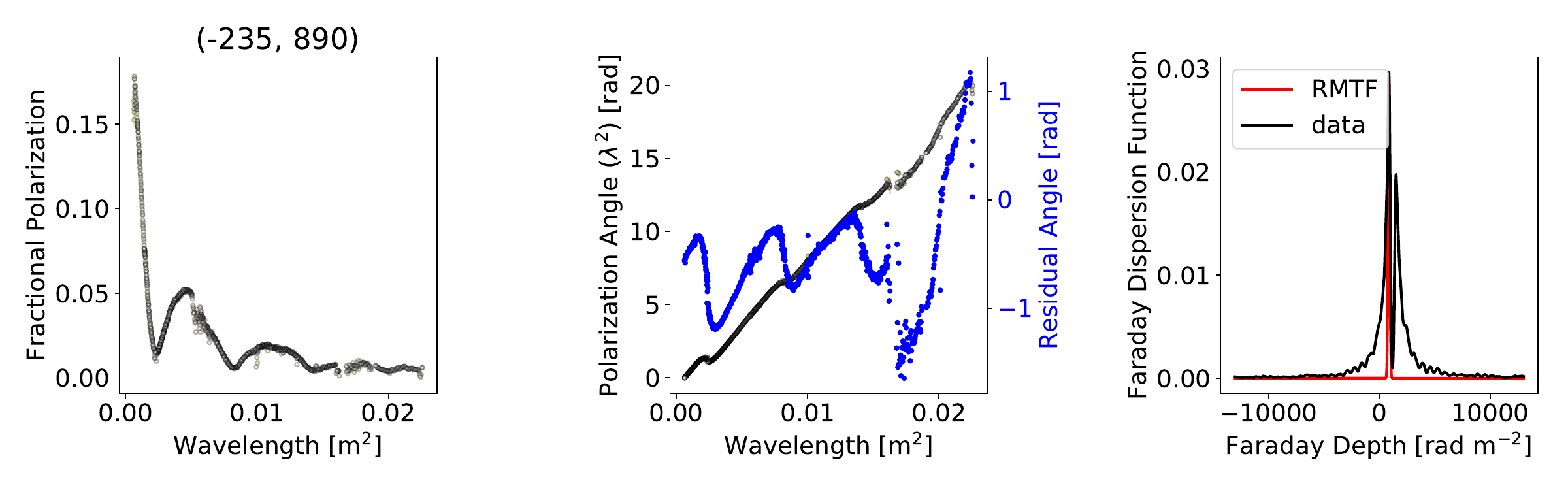}
  \end{minipage}
       \begin{minipage}[b]{1\linewidth}%[3]
    \centering
    \includegraphics[width=0.65\linewidth]{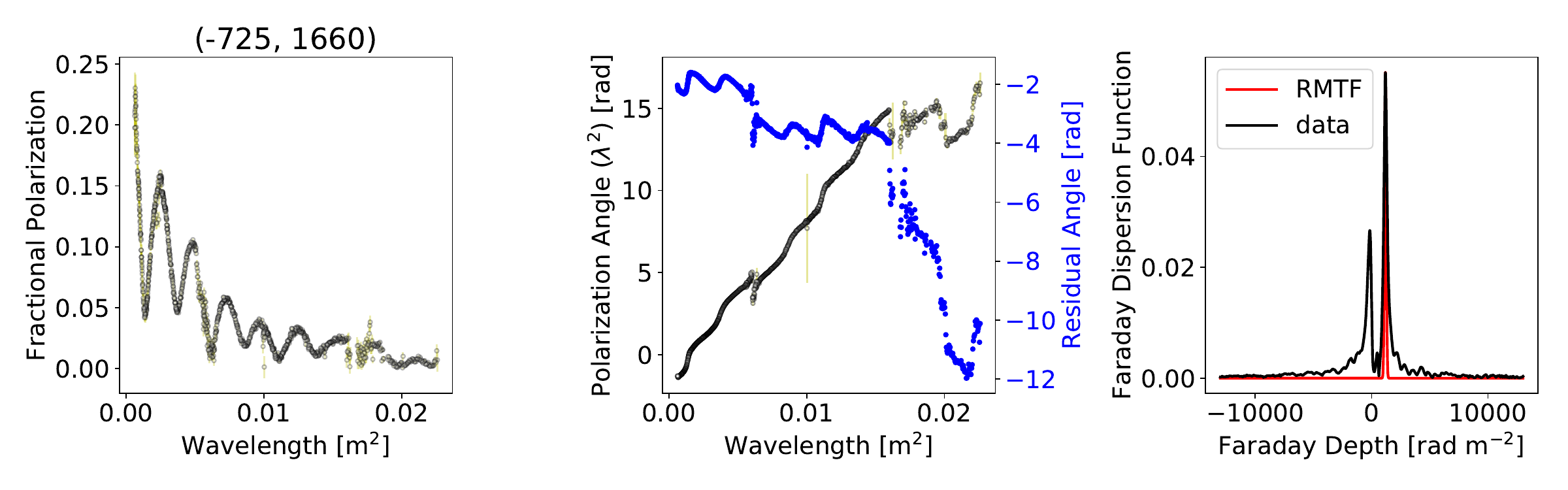}
  \end{minipage}
         \begin{minipage}[b]{1\linewidth}%[3]
    \centering
    \includegraphics[width=0.65\linewidth]{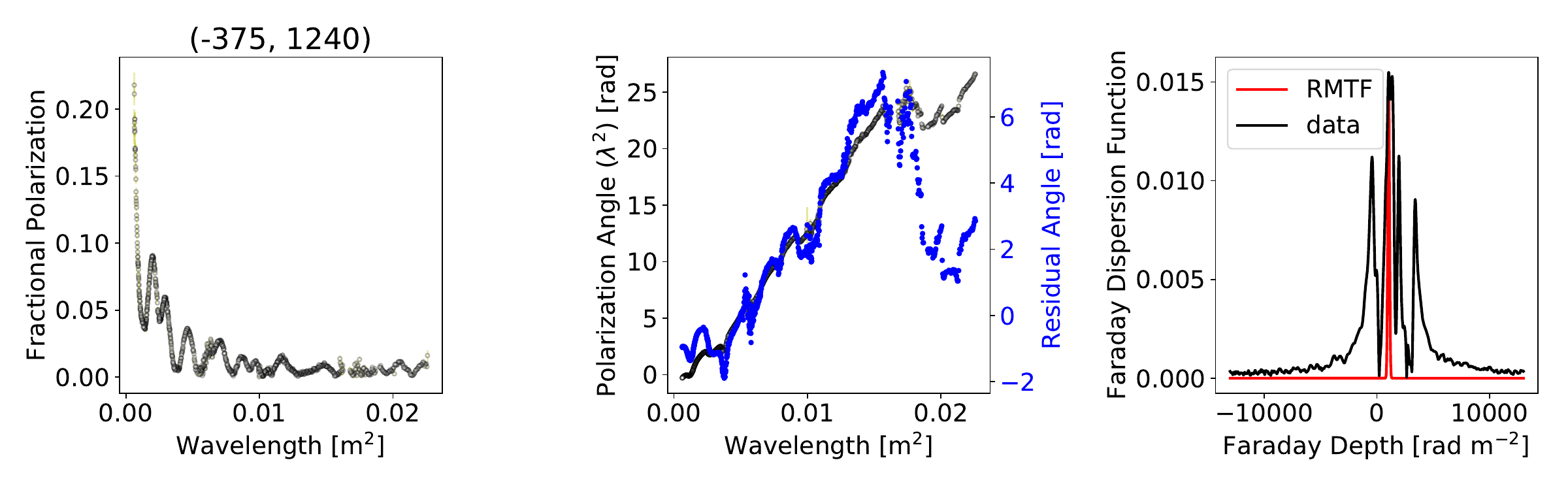}
  \end{minipage}
         \begin{minipage}[b]{1\linewidth}%[3]
    \centering
    \includegraphics[width=0.65\linewidth]{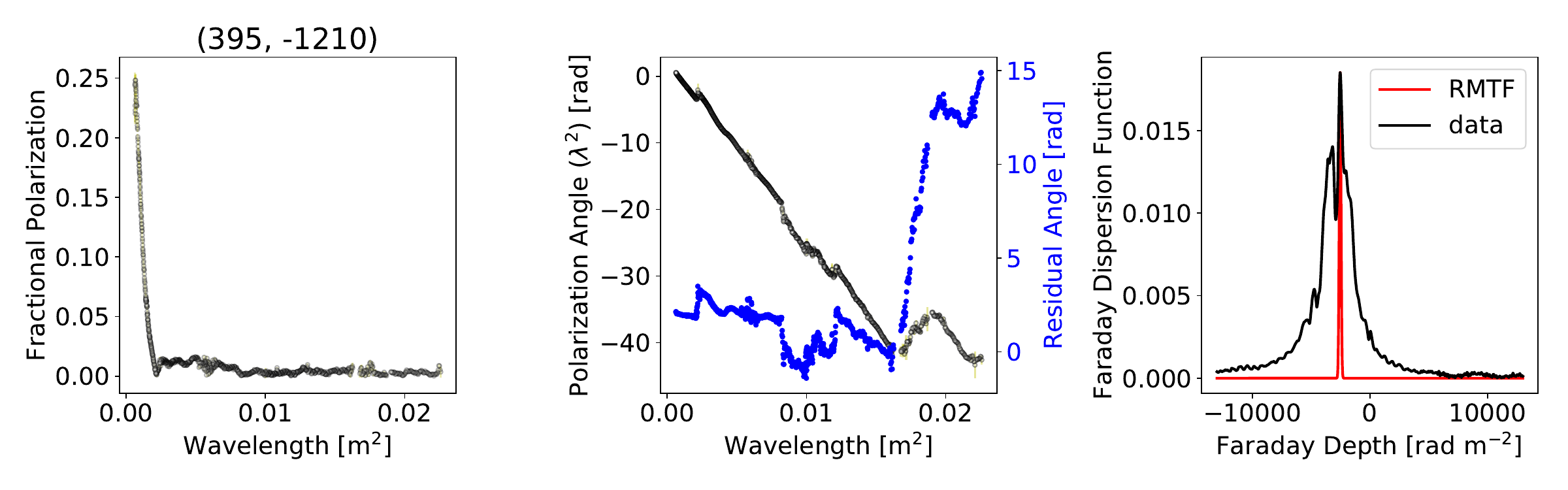}
  \end{minipage}  
  \caption{Example lines-of-sight showing fractional polarization
    vs. $\lambda^2$ (left column), the observed polarization angle
    vs. $\lambda^2$ in black and the residual angles in blue (middle
    column), and Faraday spectra superimposed with the Gaussian RMTF
    in red (right) all at $1.50\arcsec \times 1 \arcsec$. The residual
    polarization angles were obtained by removing the dominant peak in
    the Faraday spectrum (subtracting $RM_{\text{peak}}
    \lambda^2$). The top two rows are from the `smoothely decaying'
    class, the middle two rows from the `sinc-like' class, and the
    bottom two from the `complex class'. \label{fig:freqLoS}}
  \end{figure*}

Figure \ref{fig:freqLoS} shows the depolarization functions for six
representative lines-of-sight. For each line-of-sight (each row) we
display fractional polarization as a function of $\lambda^2$ (left),
polarization angle as a function of $\lambda^2$ (middle) and the
amplitude of the deconvolved Faraday spectrum (right) superimposed
with a real-valued Gaussian of width equal to the full width half
maximum of the rotation measure transfer function (in red). To match
the Gaussian function to the data, we shifted and scaled its location
and amplitude to that of the peak in the Faraday spectrum. Each
line-of-sight is labelled using the derived coordinate system.

The fractional polarizations for all 553 lines-of-sight decrease
significantly with increasing $\lambda^2$. The decline in fractional
polarization is commonly non-monotomic, with many lines-of-sight
showing sinc-like, or more complicated behavior, as shown in the
figure (based on visual inspection).  Example lines-of-sight with a relatively smooth decay are
shown in the top two rows, those with well-defined oscillations in the
third and fourth rows, and those with an intermediate,or more
complicated decay, in the last two rows. Hereon we refer to these
decay behaviors as ``smooth'', ``sinc-like'', and ``complex'' decay,
respectively. We find that roughly 22\% of the lines-of-sight show
smooth decay, 11\% sinc-like, and 67\% are complex. The sinc-like and more complex decays in fractional polarization generally arise from having more than one $\mathrm{RM}$ value within a single observing beam. For instance, the sinc-like decays might occur due to two similarly strong polarized patches with different $\mathrm{RM}$s within the beam,  if each patch itself carries a range of RM values (perhaps due to internal variations in the structure causing the RM patch). A detailed treatment will be presented in Baidoo, Eilek \& Perley (in preparation).

The middle column shows the observed polarization angles as a function
of $\lambda^2$ in black, and the residual polarization angle in
blue. Note the different vertical scales for these.  The residual
angles were obtained by removing the dominant (peak) component in the
Faraday spectrum (right panel) as $\chi - RM_{\text{peak}} \lambda^2$.
There are significant deviations from linearity for most
lines-of-sight. These deviations are most prominent at low frequencies
-- as expected, since the depolarization effects are dominant at this
frequency-regime, while linearity is generally observed at high
frequencies.  

The observed Faraday spectra often reveal very interesting structures
(right panel of Figure \ref{fig:freqLoS}).  In general, the Faraday
spectra of the smoothly decaying lines-of-sight are relatively less
complicated -- a single dominant peak, and some cases with broadened
spectra with respect to the RMTF, while the sinc-like and complex
decaying lines-of-sight have complicated spectra.  In particular, the
latter two classes have broad Faraday spectra that generally consist
of isolated, well resolved peaks (that is, structures separated by
more than our $180$ rad m$^{-2}$ resolution). The peak separations, or
broadening, ranges from a few $\sim 500$ rad m$^{-2}$ up to $\lesssim$
5000 rad m$^{-2}$ -- with the majority of the large peak separations
spanning between 1000 rad m$^{-2}$ and 3000 rad
m$^{-2}$. Complex Faraday structures are also found in
  wideband polarization studies of other sources, such as
  \citet{2012OSULLIVAN, 2016ANDERSON, 2019MA,
    2020RISELEY,2020STUARDI}, however, those presented in this study
  and in \citet{ 2020SEBOKOLODI} have been observed over a wider range
  of frequency, allowing a more detailed analysis.

\subsection{Polarization as a Function of Resolution} \label{sec:newdata2}
The depolarization can be a result of differential rotation along the
line-of-sight (synchrotron-emitting gas mixed with magnetized thermal
gas) or across the synthesized beam (beam depolarization, due to
unresolved transverse structures in the emission or a foreground
screen). These two forms of depolarization have completely different
physical implications, but often similar depolarization
characteristics.  The only definitive way to distinguish between the
two is to eliminate beam depolarization by increasing our resolution
until all transverse variations are resolved out. Any remaining
depolarization can then be attributed to the line-of-sight effect.  In
this section, we investigate whether our data are limited by the
observation's resolution.

\subsubsection{Fractional Polarization Images as a Function of Resolution}
\begin{figure*}
 \center
   \begin{minipage}[b]{0.45\linewidth}%[3]
    \centering
    \includegraphics[width=0.95\linewidth]{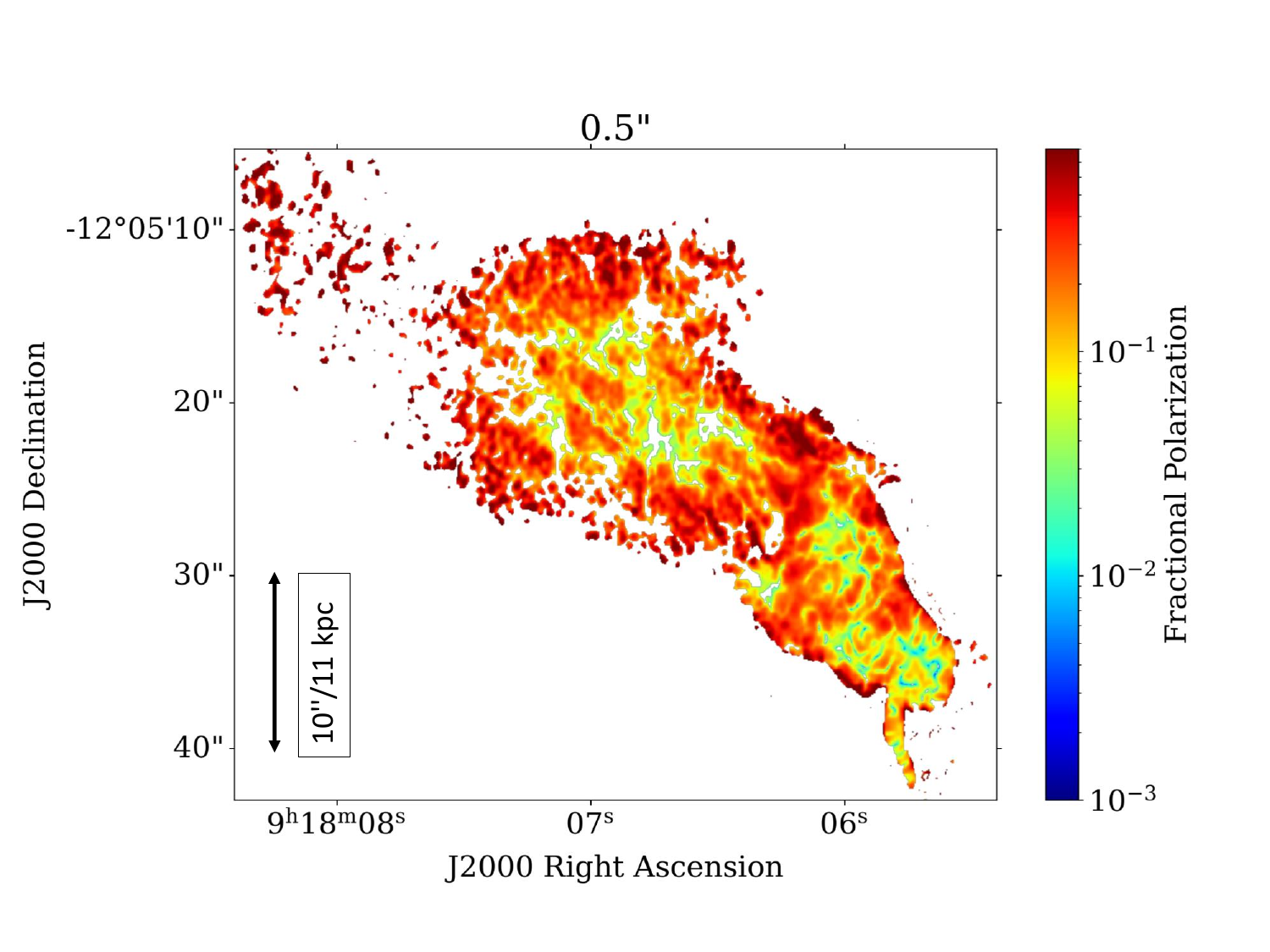}
  \end{minipage}
     \begin{minipage}[b]{0.45\linewidth}%[5]
    \centering
    \includegraphics[width=0.95\linewidth]{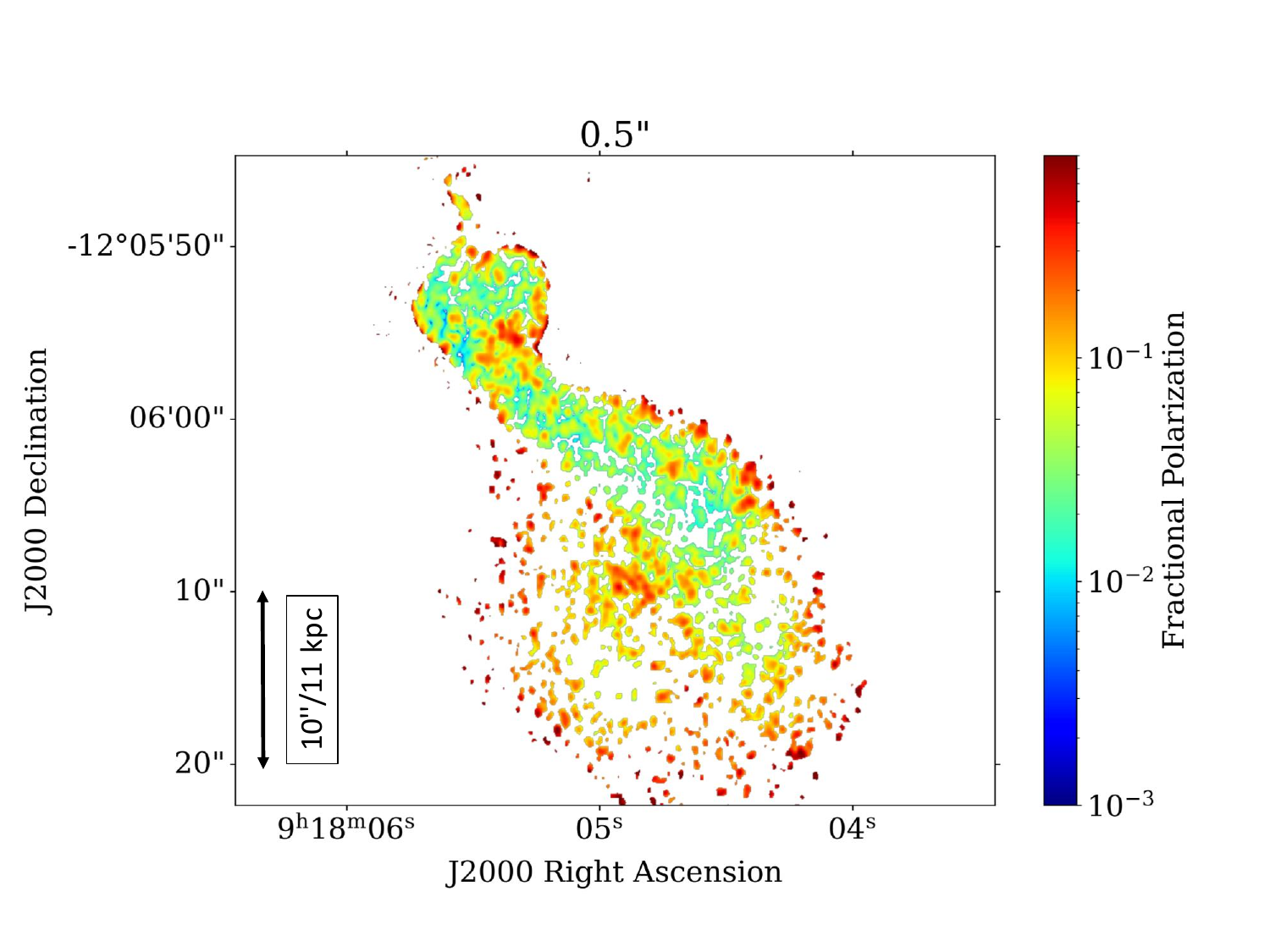}
  \end{minipage}
     \begin{minipage}[b]{0.45\linewidth}%[3]
    \centering
    \includegraphics[width=0.95\linewidth]{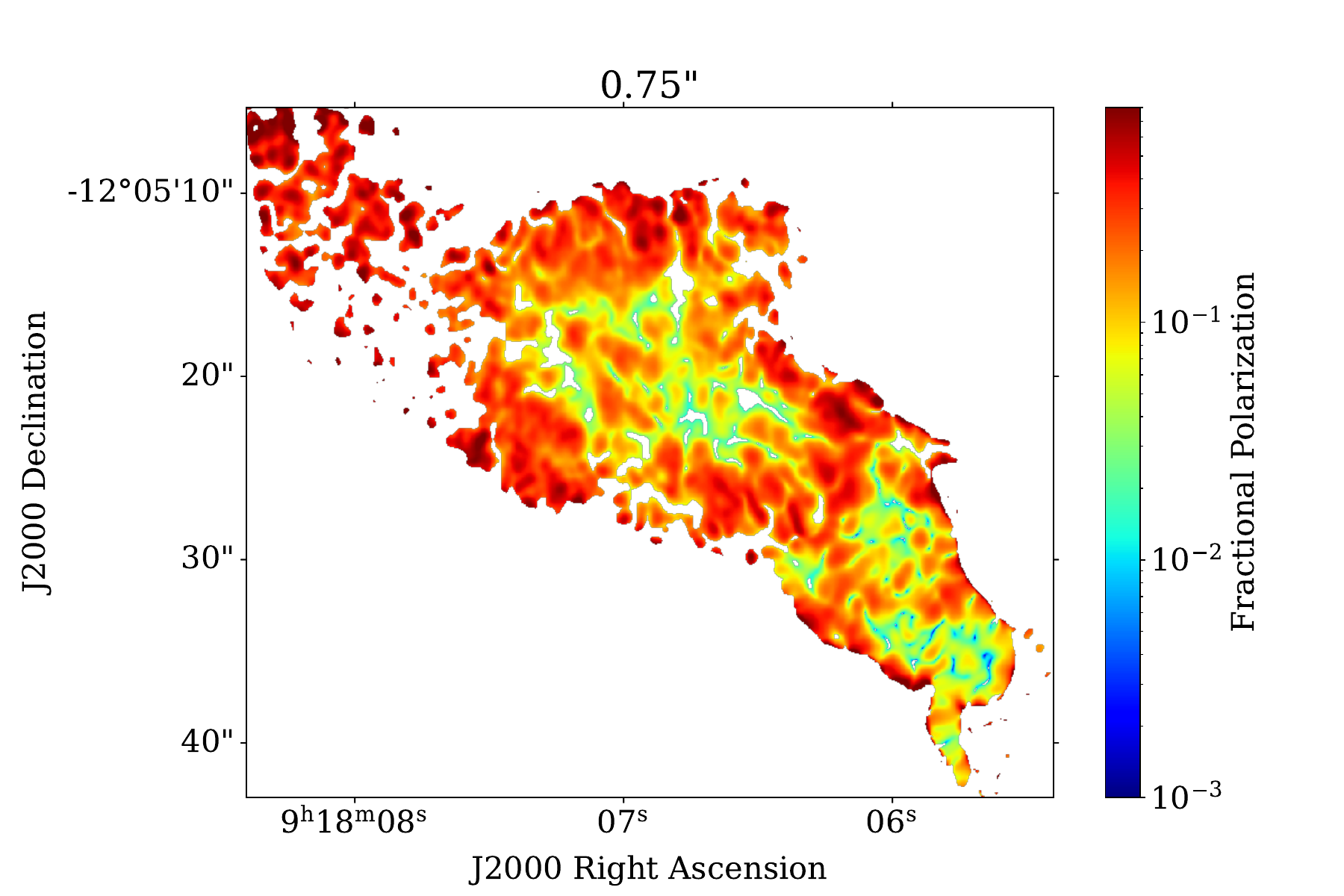}
  \end{minipage}
       \begin{minipage}[b]{0.45\linewidth}%[3]
    \centering
    \includegraphics[width=0.95\linewidth]{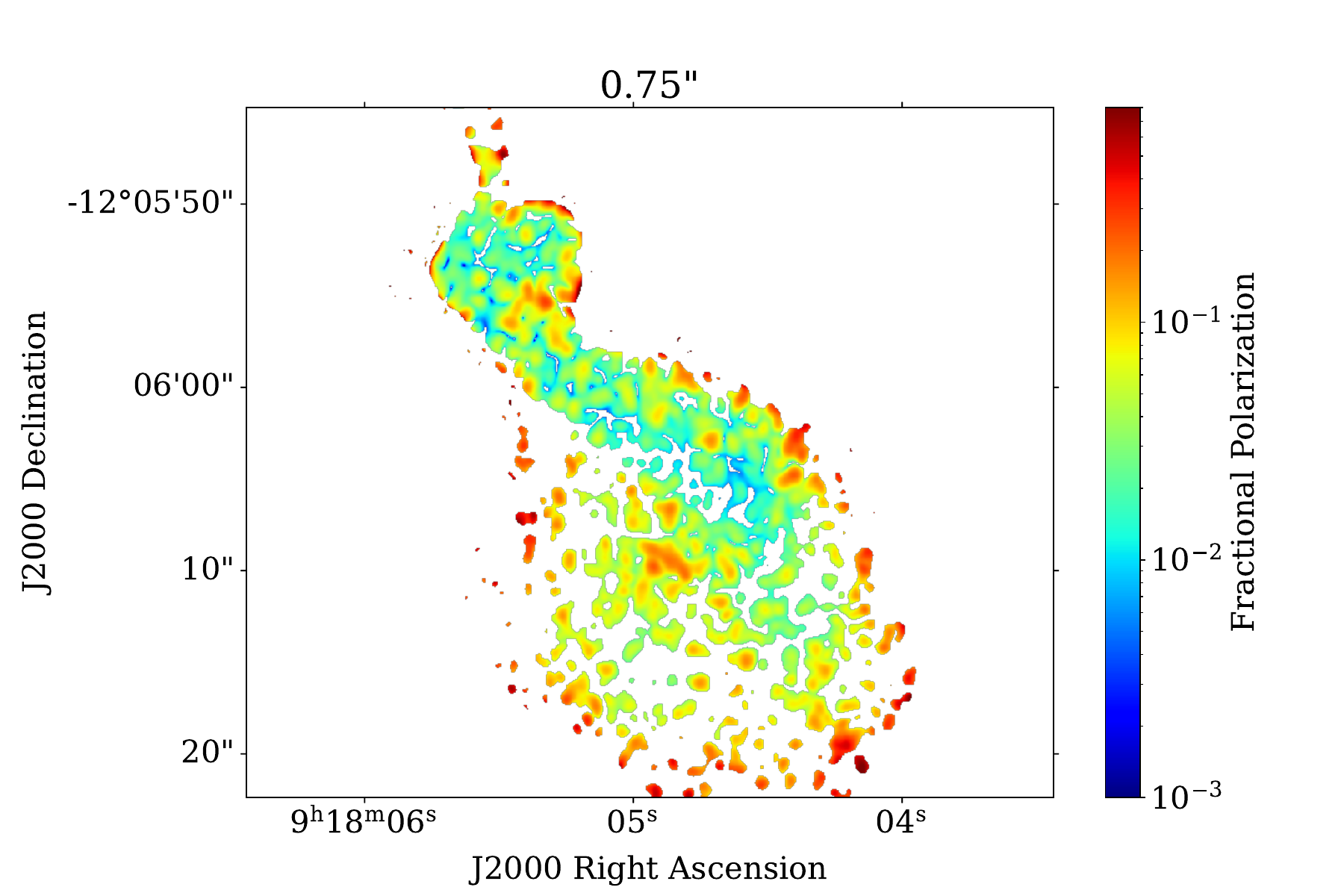}
  \end{minipage}
       \begin{minipage}[b]{0.45\linewidth}%[3]
    \centering
    \includegraphics[width=0.95\linewidth]{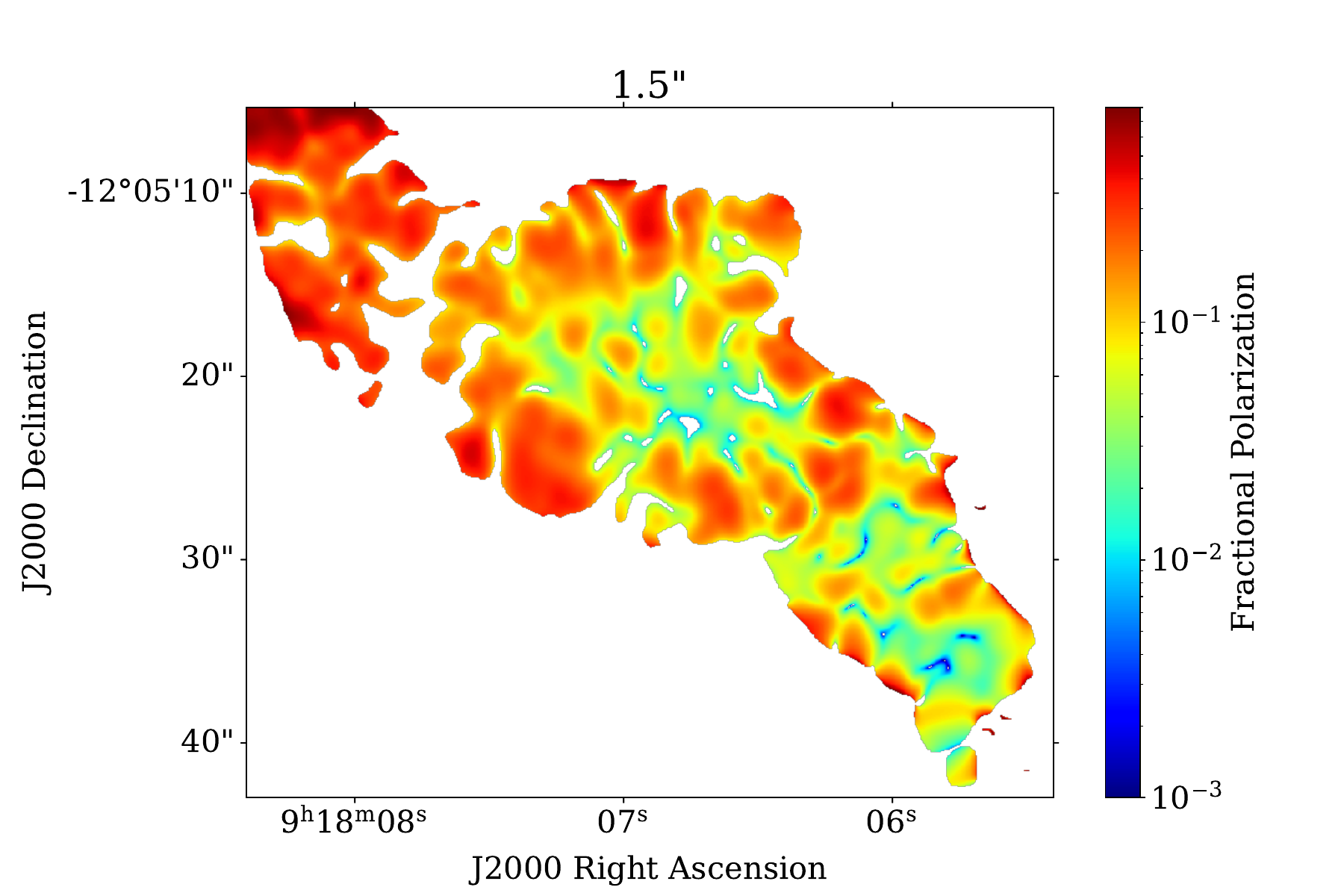}
  \end{minipage}
       \begin{minipage}[b]{0.45\linewidth}%[3]
    \centering
    \includegraphics[width=0.95\linewidth]{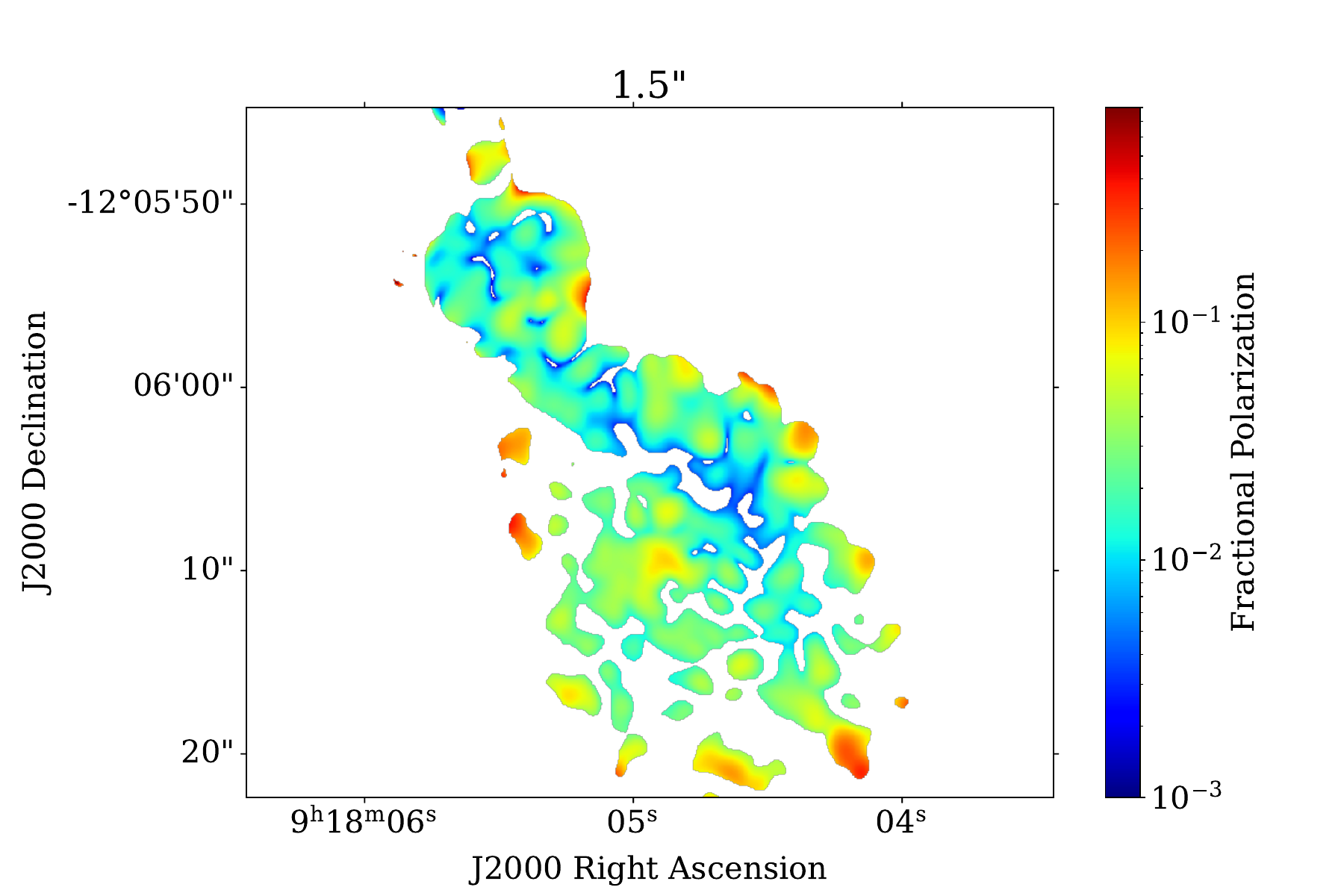}
  \end{minipage}
       \begin{minipage}[b]{0.45\linewidth}%[3]
    \centering
    \includegraphics[width=0.95\linewidth]{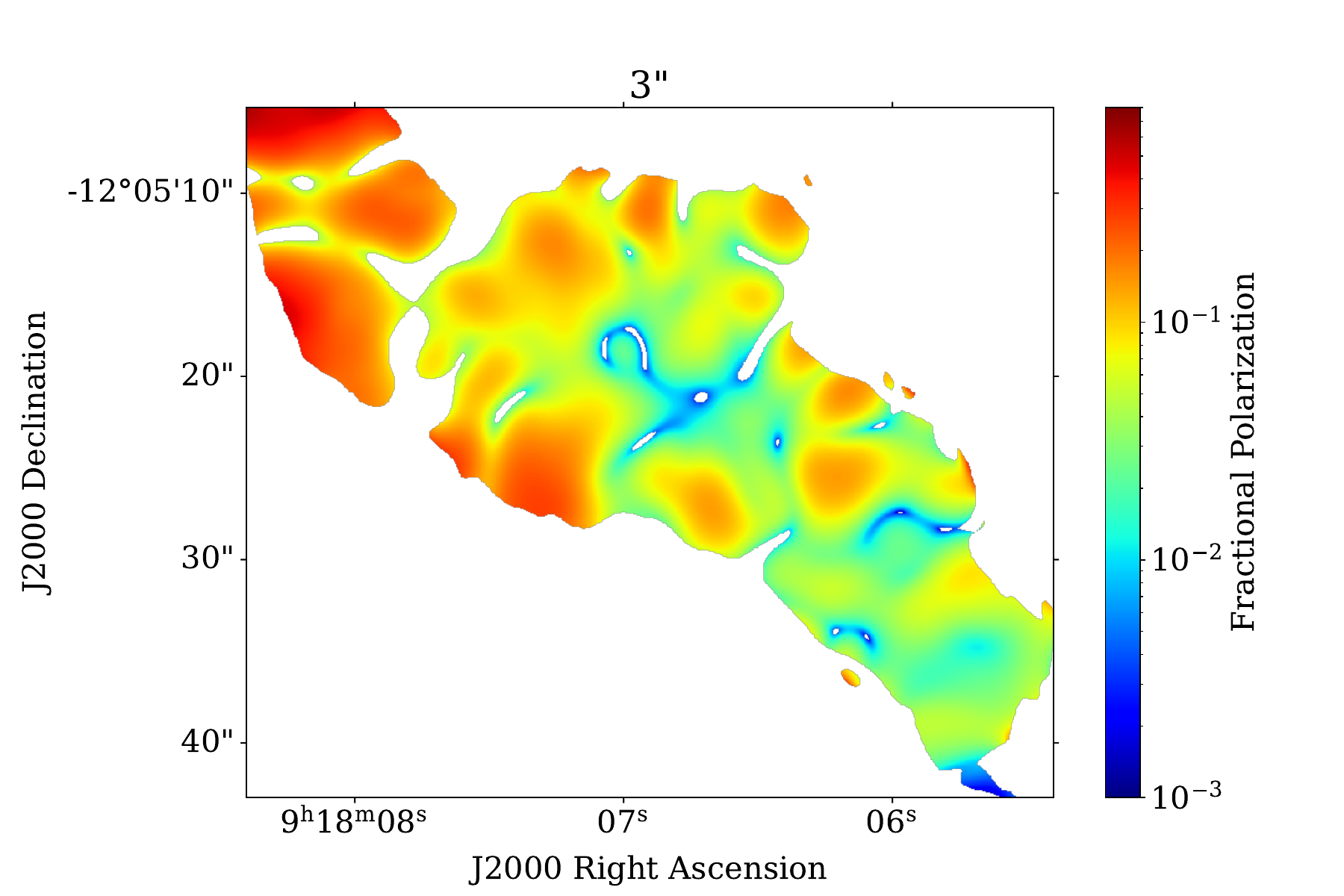}
  \end{minipage}
       \begin{minipage}[b]{0.45\linewidth}%[3]
    \centering
    \includegraphics[width=0.95\linewidth]{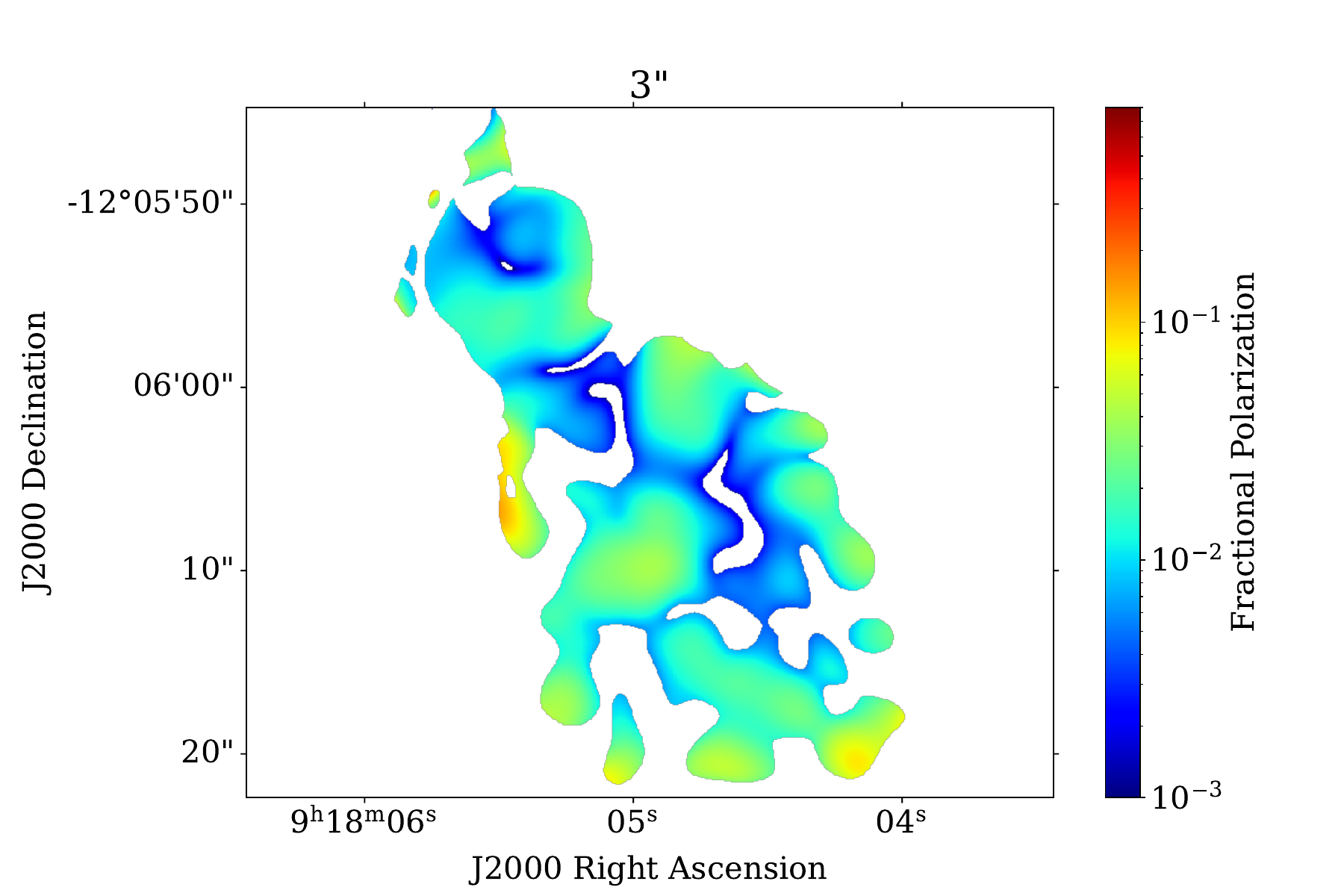}
  \end{minipage}
  \caption{Fractional polarization at $6$ GHz at four resolutions. Left: northern tail. Right: southern tail. Resolution from top to bottom:
    0.5\arcsec, 0.75\arcsec, 1.5\arcsec, and 3\arcsec. Only pixels with SNR
    $p/\sigma_p > 60\%$ are shown. In nearly all locations, the higher
    resolution images are more highly
    polarized. The physical scale shown in the top panel is the same for the rest of the panels. \label{fig:resolmaps}}
  \end{figure*}

Figure \ref{fig:resolmaps} shows a 6 GHz fractional polarization map
at four different resolutions between 3$''$ and 0.5$\arcsec$ (the latter being the highest we
can obtain at this frequency).  The tails depolarize with lower
resolution, with the largest depolarization occurring in the inner
regions of the tails close to the center (most evident across the
northern tail).

\subsubsection{Resolution-Dependent Depolarization Ratio (RDR)}
To see the degree by which the tails depolarize as a function of
resolution, we computed a depolarization ratio map by dividing the 10
GHz fractional polarization maps at 3$\arcsec$ with the maps of same
frequency at 0.3$\arcsec$ (hereinafter RDR1), and 6 GHz fractional
polarization maps at 3$\arcsec$ with 0.5$\arcsec$ (hereinafter
RDR2).

\begin{figure}[!ht]
 \center
   \begin{minipage}[b]{1\linewidth}%[3]
    \centering
    \includegraphics[width=1\linewidth]{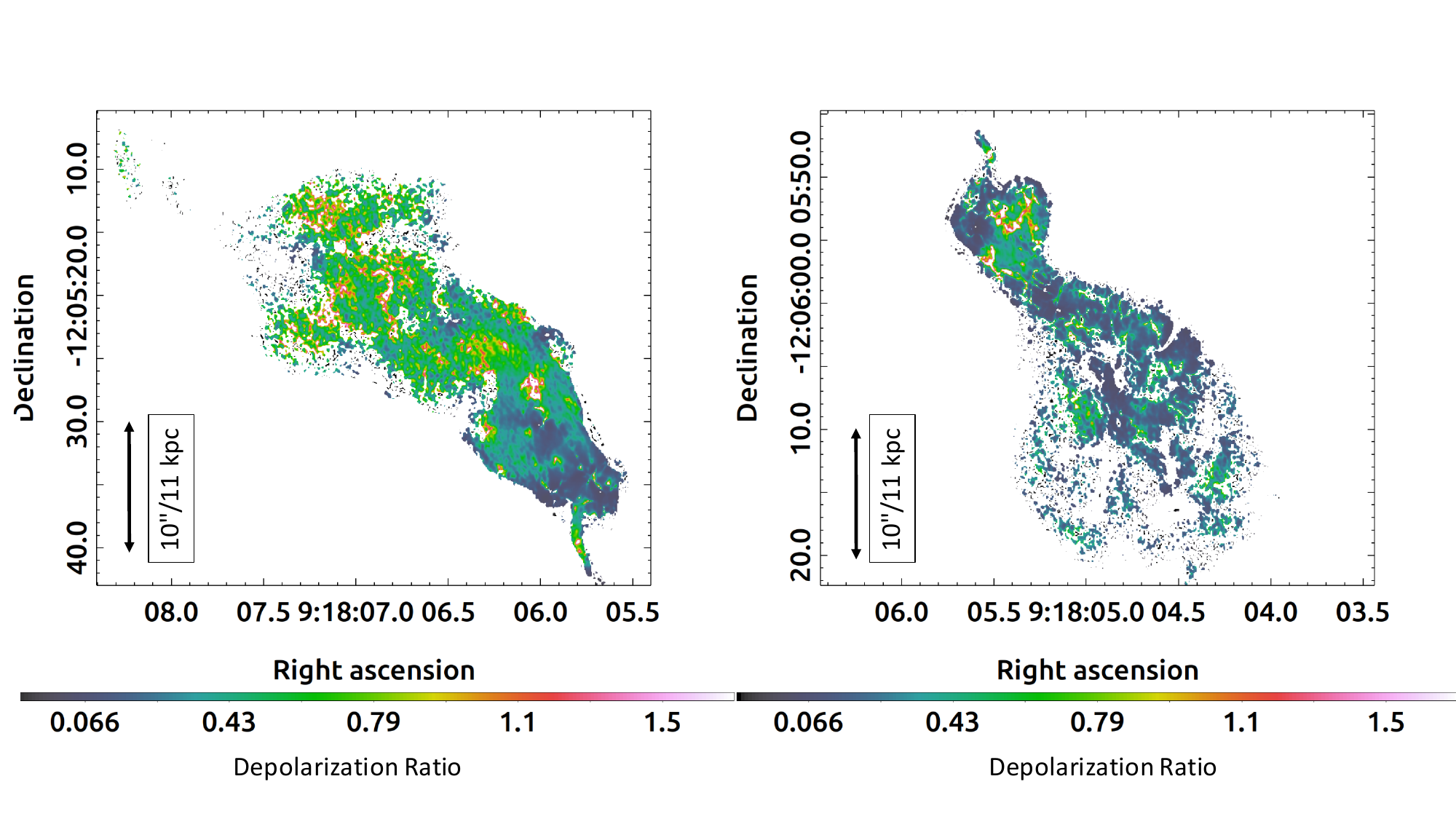}
  \end{minipage}
     \begin{minipage}[b]{1\linewidth}%[5]
    \centering
    \includegraphics[width=1\linewidth]{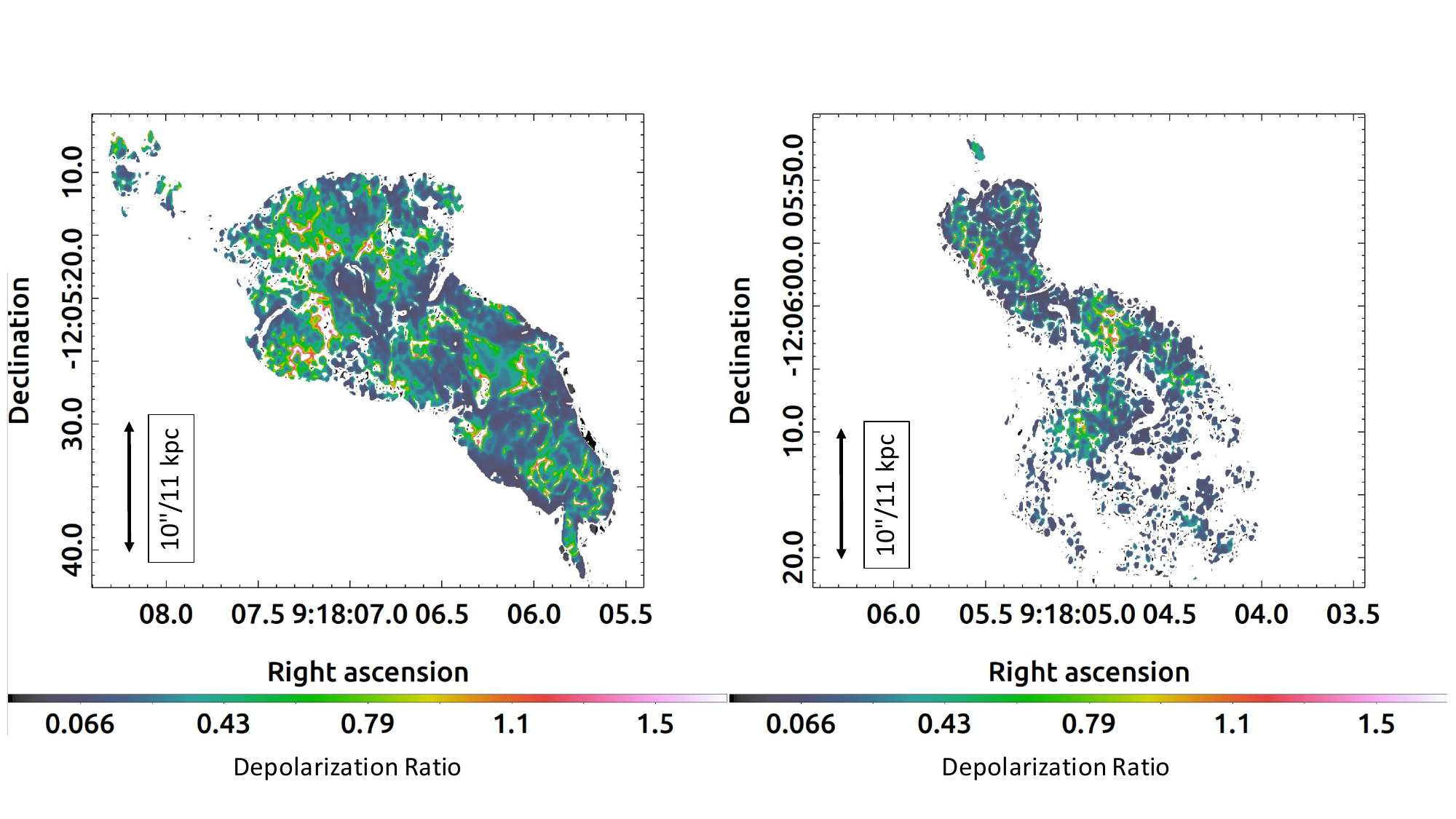}
  \end{minipage}
\caption{Depolarization ratio maps across the tails. Top panel: RDR1
  -- the ratio of the 3$\arcsec$ resolution fractional polarization
  image to the 0.3$\arcsec$ image at 10 GHz. Bottom panel: RDR2 --
  the ratio of the 3$\arcsec$ fractional polarization to the 0.5$\arcsec$ polarization image at 6 GHz. Only pixels with fractional
error of less than $80\%$ are shown. \label{fig:rdr-distribuion}}
\end{figure}

Figure \ref{fig:rdr-distribuion} shows the spatial distribution of RDR across the tails.

\begin{figure}
 \center
    \includegraphics[width=1.05\linewidth]{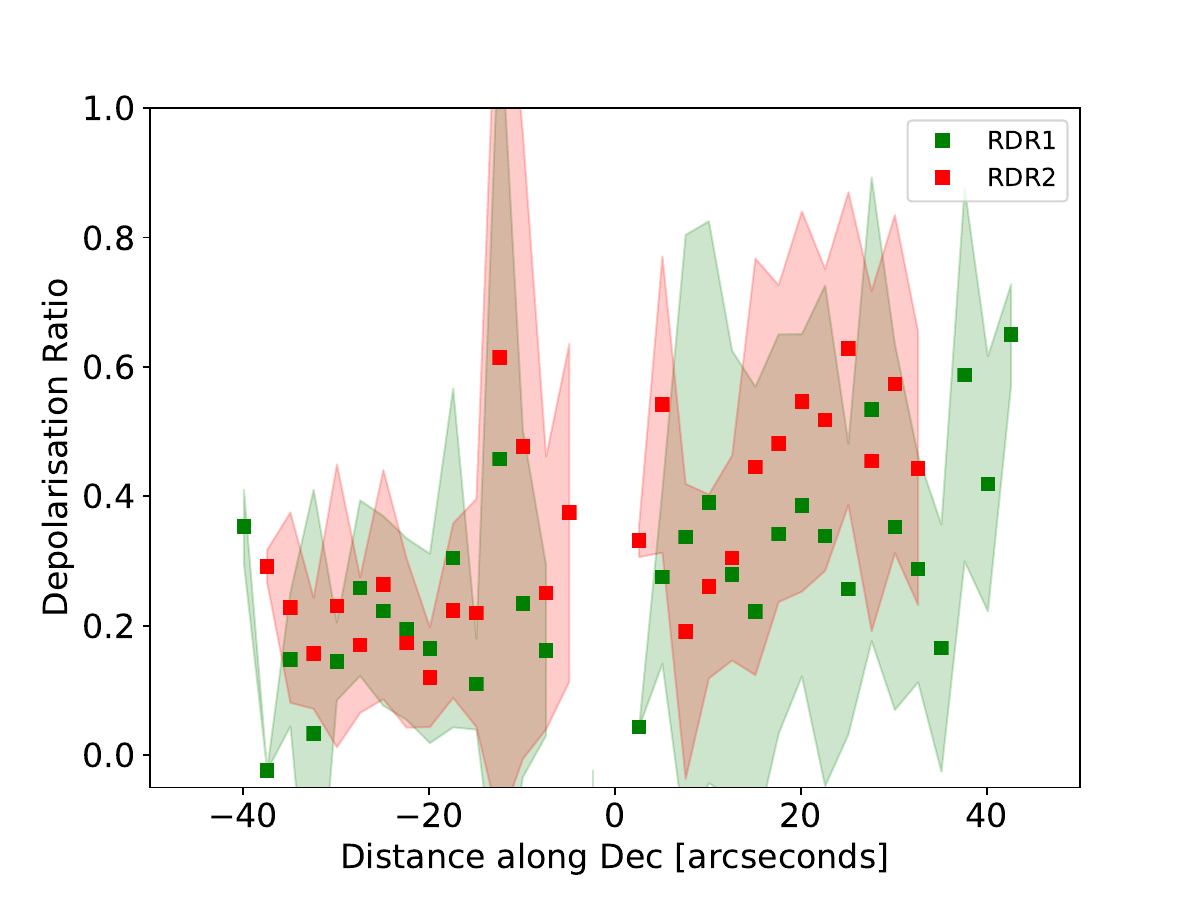}
\caption{Resolution-dependent depolarization ratio. RDR1 is
  $3\arcsec/0.30\arcsec$ at 10 GHz. RDR2 is $3\arcsec/0.50\arcsec$ at
  6 GHz.  The left side of declination zero is the southern tail, and the right side is the northern tail. Pixels shown have fractional error $< 80\%$. The general
  depolarization ratio behavior is similar at the two frequencies. The
  depolarization is stronger across the southern tail. The two tails
  depolarize differently with distance from the center: the
  depolarization across the northern tail decreases with distance from the
  AGN, while the depolarization is strongest away from the center
  across the southern tail. We find the same behavior for RDR binned along RA.
\label{fig:resoldepol}}
\end{figure}

The corresponding radial profiles are presented in Figure
\ref{fig:resoldepol} (similar to FDR profiles shown in Figure
\ref{fig:freqDepol}). We utilized only those pixels with fractional
error of less than $80\%$ to compute the profiles. The general
depolarization ratio behavior is similar at the two frequencies. The
two tails depolarize differently with distance from the center: the
depolarization across the northern tail decreases with distance from the
AGN, while the depolarization is strongest away from the center across
the southern tail. Similarly to the frequency-dependent depolarization
ratios, the resolution-dependent depolarization is much stronger in
the southern tail (left side of declination=$0''$). The same behavior is observed for RDR as a function of distance along right ascension.

\subsubsection{Lines-of-Sight Depolarization}
As noted earlier, in the presence of small-scale transverse
fluctuations, the observed fractional polarization should increase
(usually) monotonically with increasing resolution, until reaching a
resolution which resolves the foreground fluctuations in the
depolarizing screen.  At this resolution, the observed fractional
polarization will be that of the source itself. If this intrinsic
value can be established, any fractional polarization changes with
frequency must then be due to intermixed thermal and synchrotron gas
of the source, or an intermixed boundary layer, allowing an
estimate of the thermal gas content.  This is the only means of
definitively separating beam-depolarization effects from the more
physically relevant depolarization mechanisms.

\begin{figure}
 \centering
   \begin{minipage}[b]{0.45\linewidth}%[3]
  %  \centering
    \includegraphics[width=1.15\linewidth]{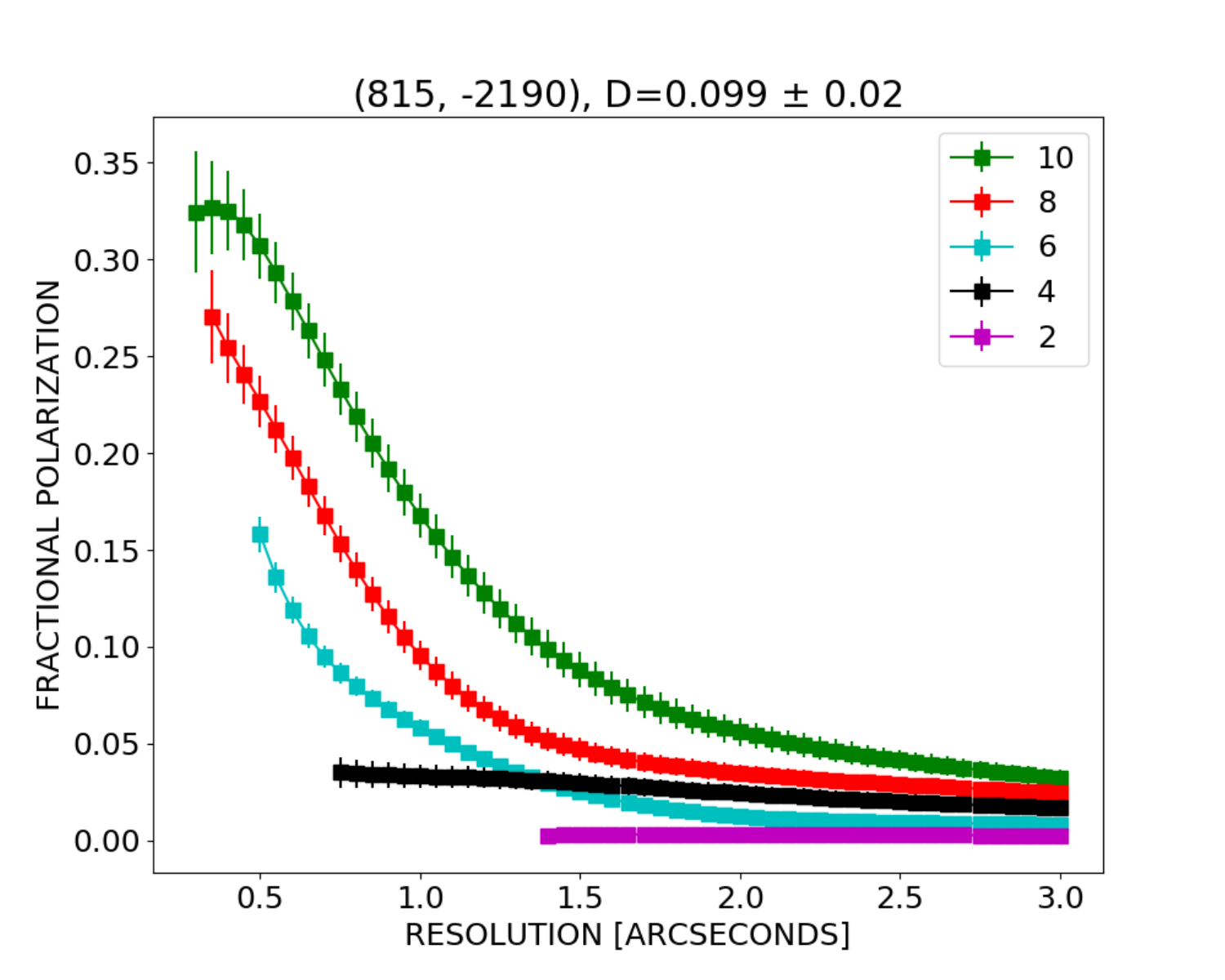}
  \end{minipage}
     \begin{minipage}[b]{0.45\linewidth}%[5]
  %  \centering
    \includegraphics[width=1.15\linewidth]{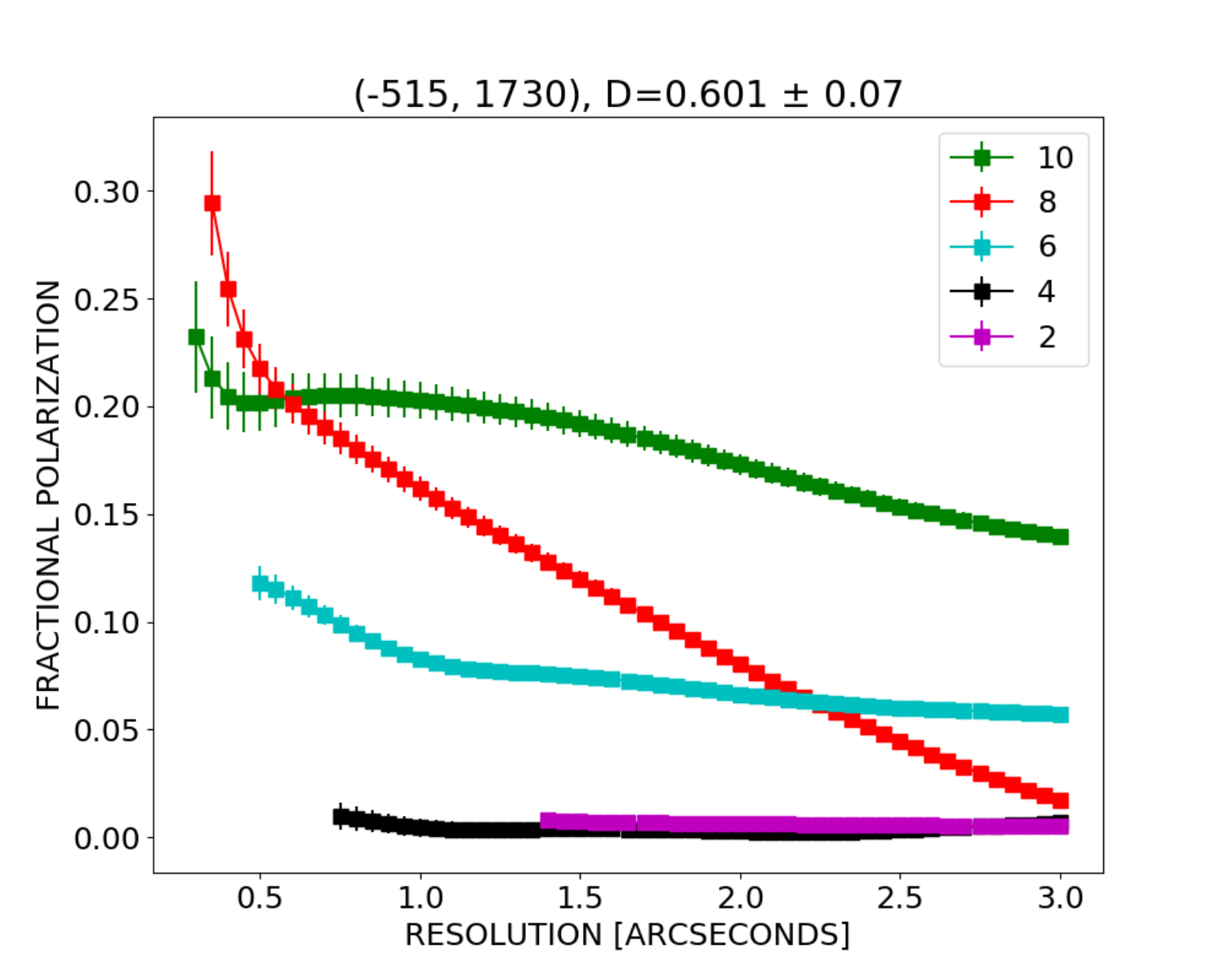}
  \end{minipage} 
     \begin{minipage}[b]{0.45\linewidth}%[3]
  %  \centering
    \includegraphics[width=1.15\linewidth]{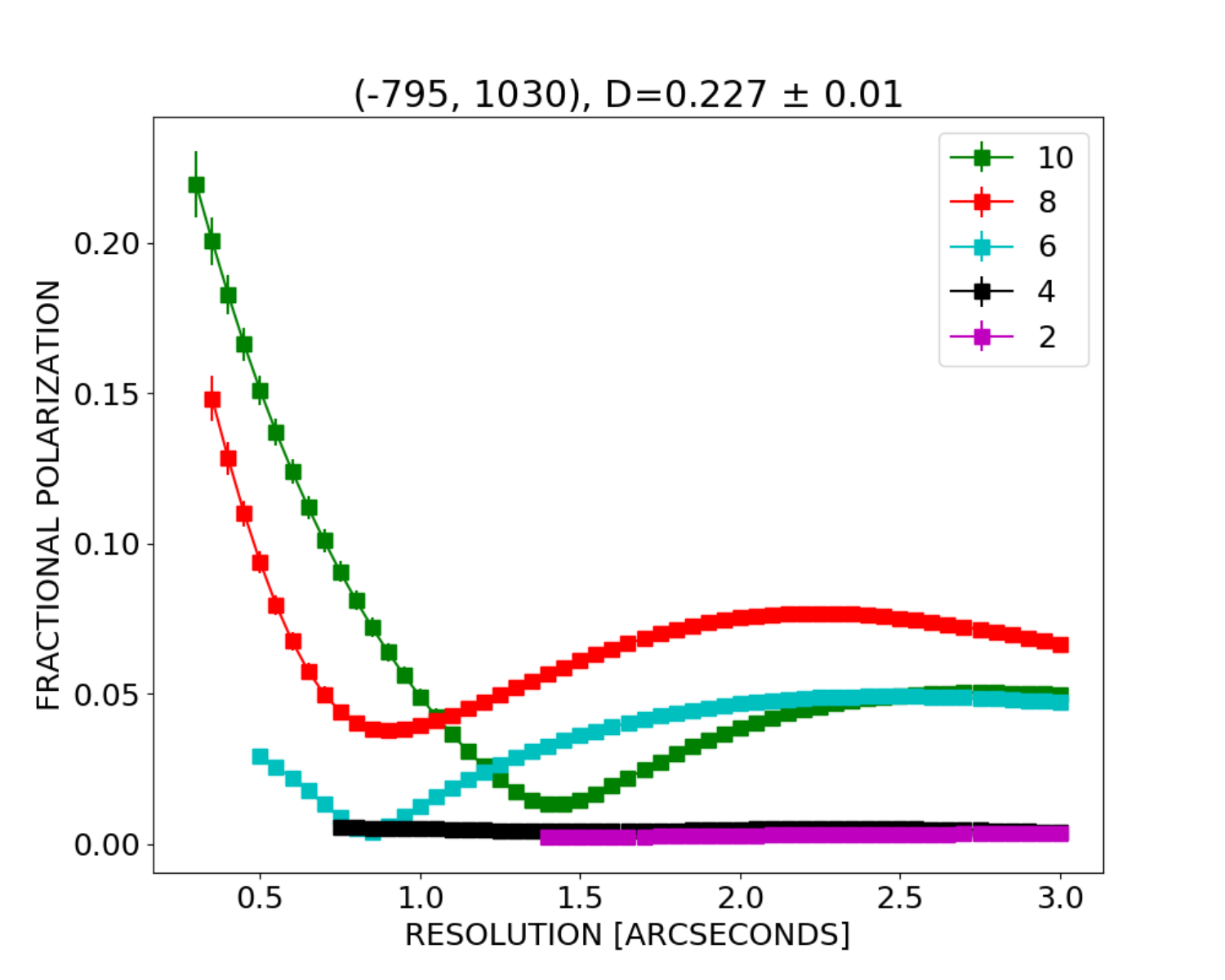}
  \end{minipage}
       \begin{minipage}[b]{0.45\linewidth}%[3]
  %  \centering
    \includegraphics[width=1.15\linewidth]{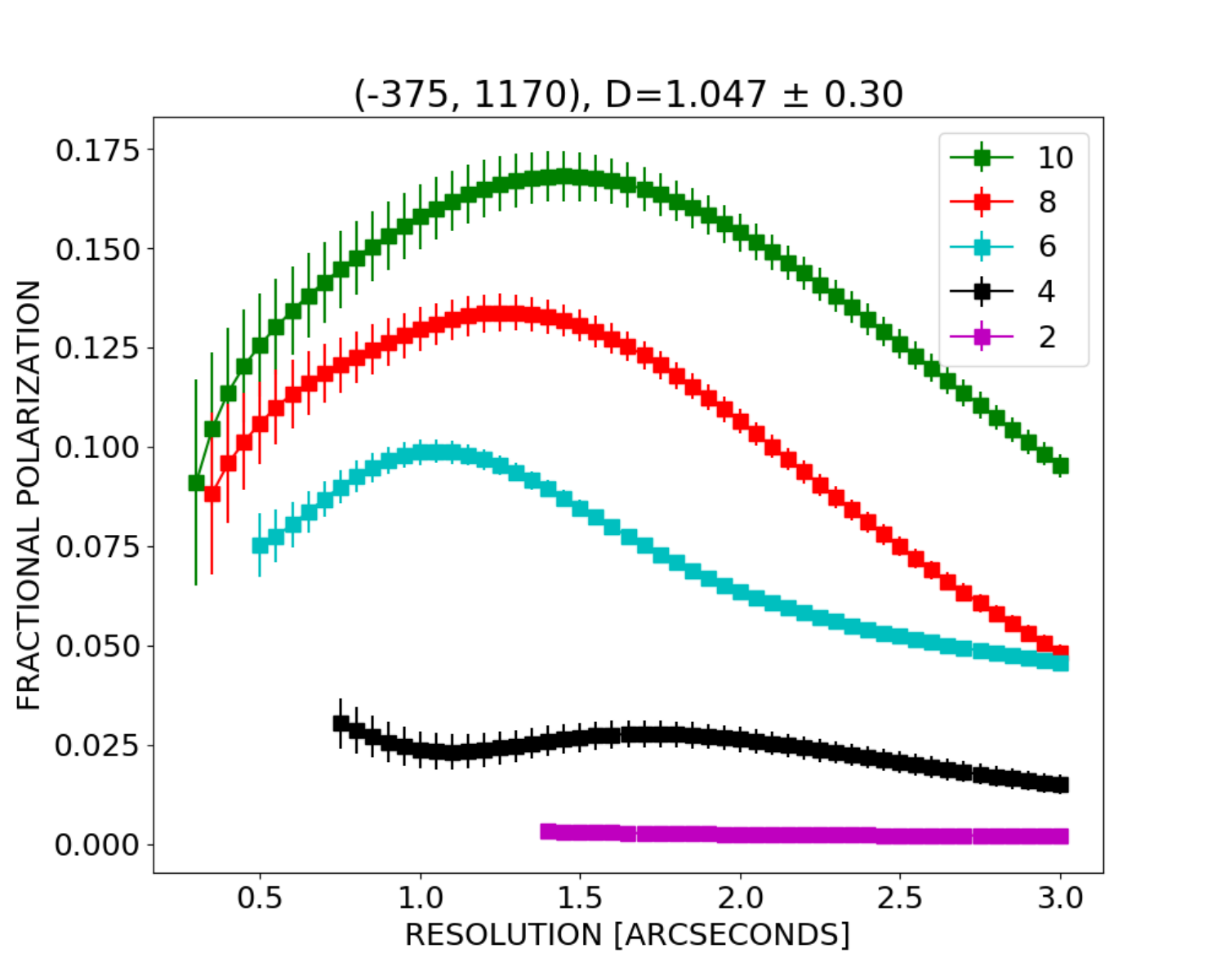}
  \end{minipage}
     \begin{minipage}[b]{0.45\linewidth}%[3]
  %  \centering
    \includegraphics[width=1.15\linewidth]{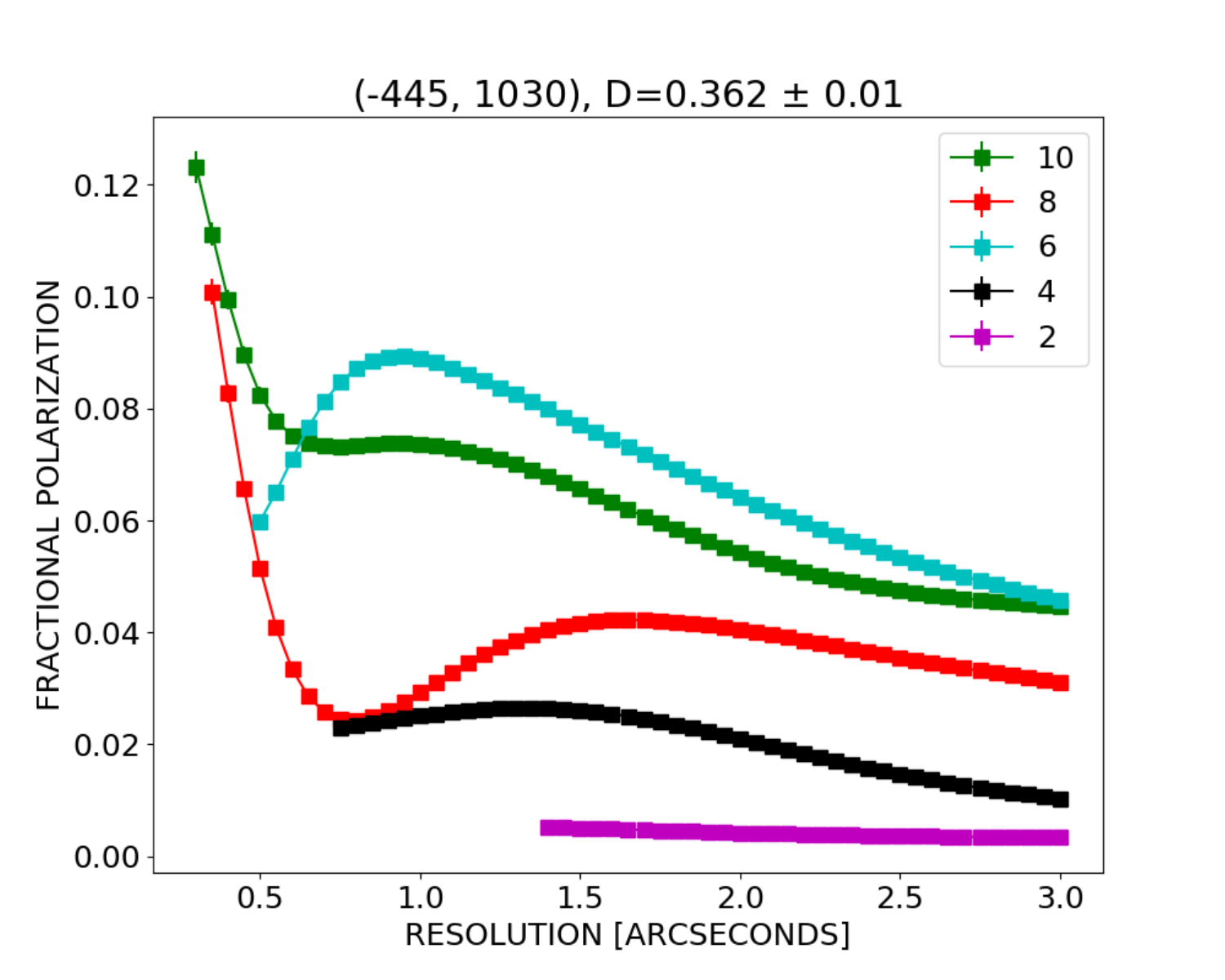}
  \end{minipage}
     \begin{minipage}[b]{0.45\linewidth}%[5]
  %  \centering
    \includegraphics[width=1.15\linewidth]{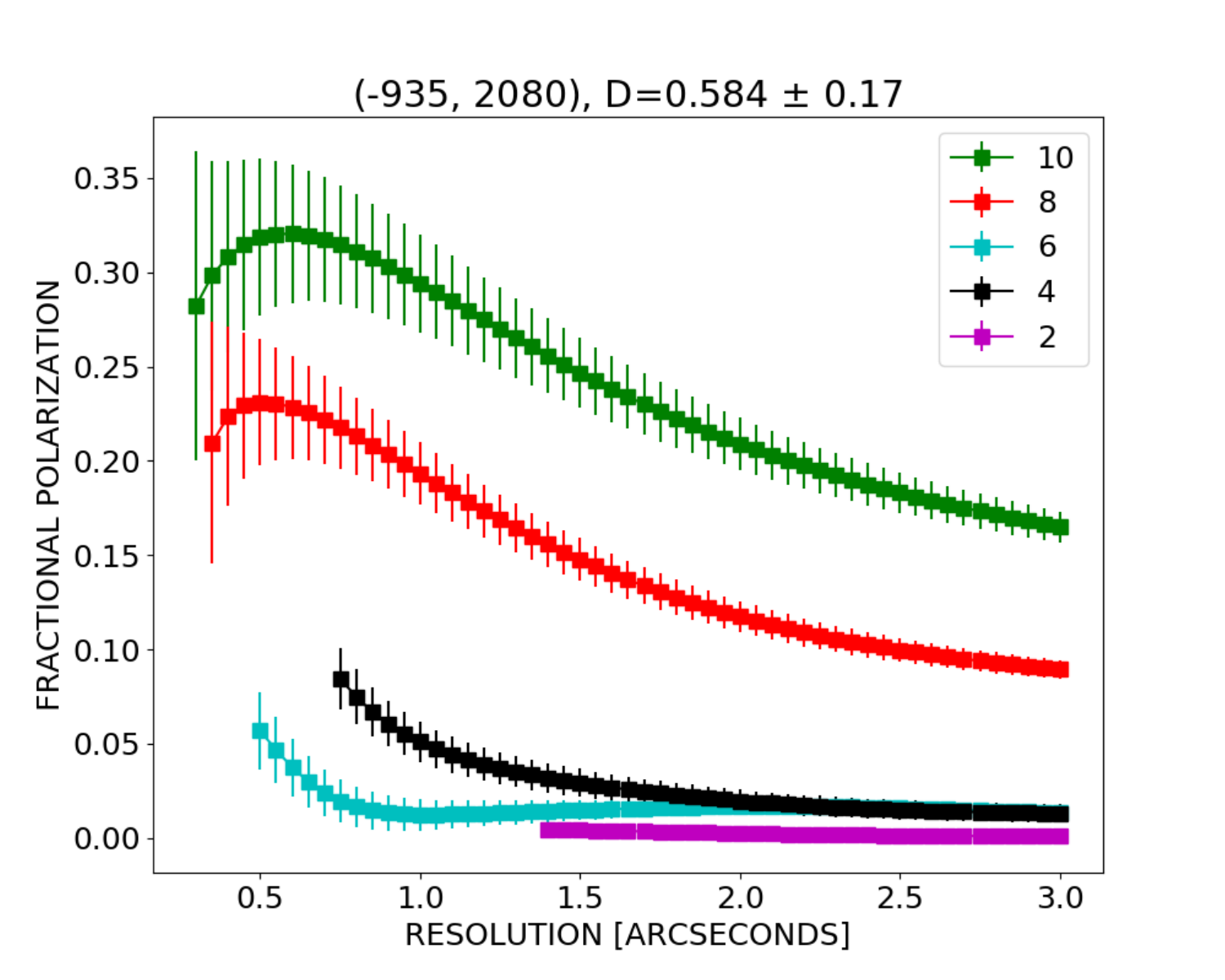}
  \end{minipage}
\caption{Lines-of-sight fractional polarization as a function of
  resolution. Green: 10 GHz. Red: 8 GHz. Cyan: 6 GHz. Black: 4
  GHz. Magenta: 2 GHz. Plot titles: Lines-of-sight position
  coordinates and the RDR value at 3$''/0.3''$ (10 GHz). Some lines-of-sight (top row) show the ``expected'' increase in fractional polarization with increasing resolution (these make up roughly 59\% of the lines-of-sight). In some cases, we find lines-of-sight with fractional polarization changing unexpectedly with increasing resolution (two bottom rows, roughly 41\% of the lines-of-sight behave this way).\label{fig:resolfrac}}
\end{figure}

In Figure \ref{fig:resolfrac}, we show plots of fractional
polarization as a function of resolution for a few selected
lines-of-sight. The fractional polarization changes in various ways
with increasing resolution. Unfortunately, in no cases are we able to
claim that we have resolved out the transverse depolarization
structures at all observing frequencies.

\section{Faraday Rotation}\label{sec:faradayrotation}
In the previous section, we demonstrated that Hydra A is experiencing
both wavelength-dependent and resolution-related depolarization. These
results imply the presence of a turbulent magneto-ionic medium on
scales less than 1.5 kpc. At higher frequencies, we can achieve high
resolution -- which then minimizes beam-related effects. Moreover,
the wavelength-dependent depolarization structures such as the
sinc-like and complex decays are mostly concentrated at longer
wavelengths. Thus, by utilizing the high frequency data, we can derive
the `true' emission properties of the source, without dealing with
complicated depolarization structures.

We can then use these high-frequency, high-resolution approximations
of the `true' emission properties of the tails to predict the
polarization characteristics at lower frequencies and lower
resolution. A good match between such predictions with the observed
data would provide strong evidence that a foreground, turbulent
Faraday rotating medium is primarily responsible for the
majority of the depolarization through beam-related effects.

We thus utilized the 6 - 12 GHz frequency data -- which gives the
highest resolution of $0.5\arcsec \times 0.35\arcsec$. By looking
through the 553 lines-of-sight with suitable signal strength, we find
that the structures in the depolarization functions are negligible
within this frequency range. We thus fit to the fractional $Q$ and $U$ of
our images the real and imaginary parts of the following model,
respectively:
\begin{equation}\label{eqn:tofit}
 p = p_0e^{2i\chi_0}e^{2i RM \lambda^2 - 2\sigma_t^2\lambda^4},
\end{equation}
where $p_0$ and $\chi_0$ are the zero-wavelength fractional
polarization and polarization angle of the tails, respectively, $\mathrm{RM}$
is the rotation measure of the ambient medium with electron density
$n_e$ [cm$^{-3}$], uniform magnetic field component, $\mathbf{B}_u$
[$\mu$G], located a distance $L$ [kpc] from us in $z$-axis direction:
\begin{equation}
 RM = 812 \int\limits_{L}^{0}n_e\mathbf{B}_u\cdot\mathrm{dz}
  \quad \quad [\mathrm{rad\, m}^{-2}],
\end{equation}
and where $\sigma_t$ quantifies the rate of depolarization with
$\lambda^2$. It is commonly interpreted to be due to unresolved
fluctuations in Faraday depths along lines of sight within the
resolution beam.  For a simple random turbulent screen, $\sigma_t$ is
given by

\begin{equation}
 \sigma_t = 812 n_t B_t d\sqrt{N} \quad \text{rad m$^{-2}$},
\end{equation}
and where $N=L/d$ is the number of turbulent cells of size $d$ along
the total path length $L$, $n_t$ is the electron density in the cell,
and $B_t$ is the magnetic field strength of the cell -- representing a
turbulent (assuming a random walk model) magnetic field strength
\citep{1966BURN,1998SOKOLLOF}.

We use a simple non-linear least squares fitting algorithm for fitting
this model to the data. This was done using minimization tools
provided through the software package {\tt LMFIT}\footnote{https://lmfit.github.io/lmfit-py/}. We wrote a
specific code for this particular problem, which can be provided upon
request. The fitting to fractional $Q$ and $U$ was performed
simultaneously and the best-fitting parameters were those that
minimize the difference in the data and model in both $Q$ and $U$.

\begin{figure}[!ht]
 \center
   \begin{minipage}[b]{1\linewidth}%[3]
    \centering
    \includegraphics[width=0.95\linewidth]{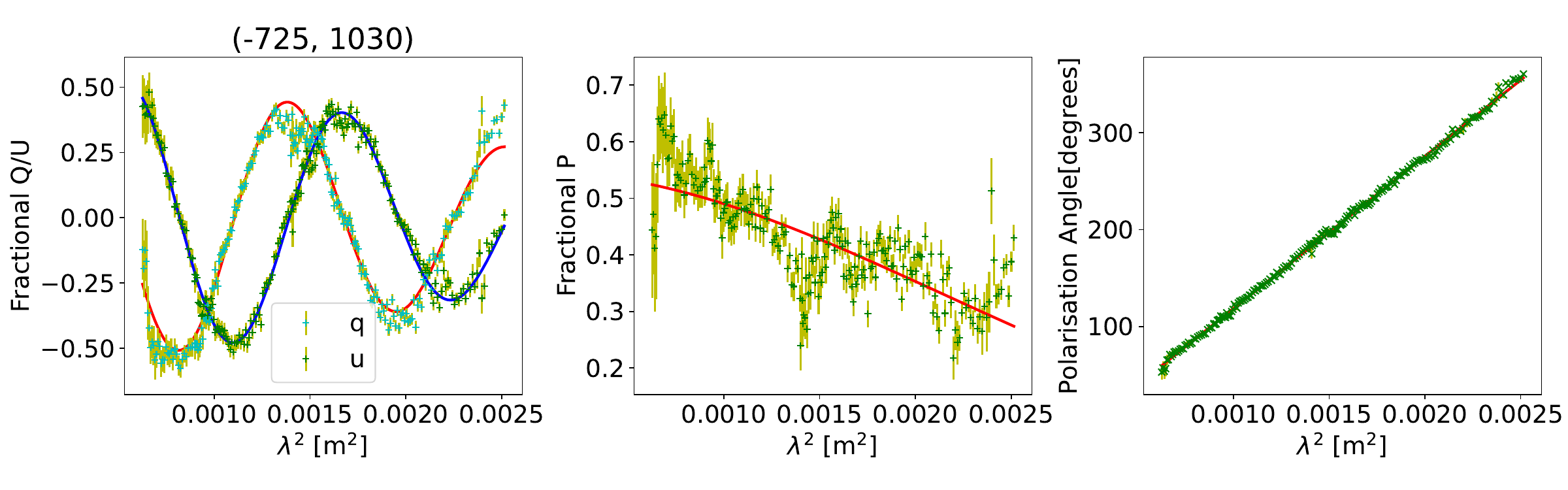}
  \end{minipage}
     \begin{minipage}[b]{1\linewidth}%[5]
    \centering
    \includegraphics[width=0.95\linewidth]{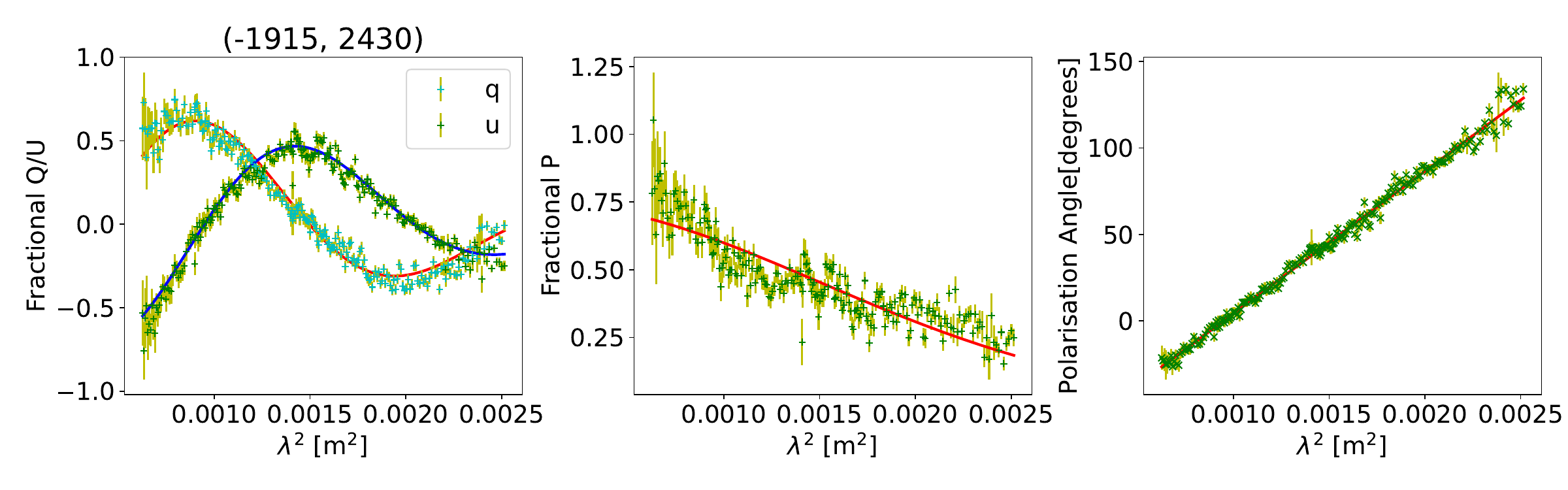}
  \end{minipage}  
     \begin{minipage}[b]{1\linewidth}%[3]
    \centering
    \includegraphics[width=0.95\linewidth]{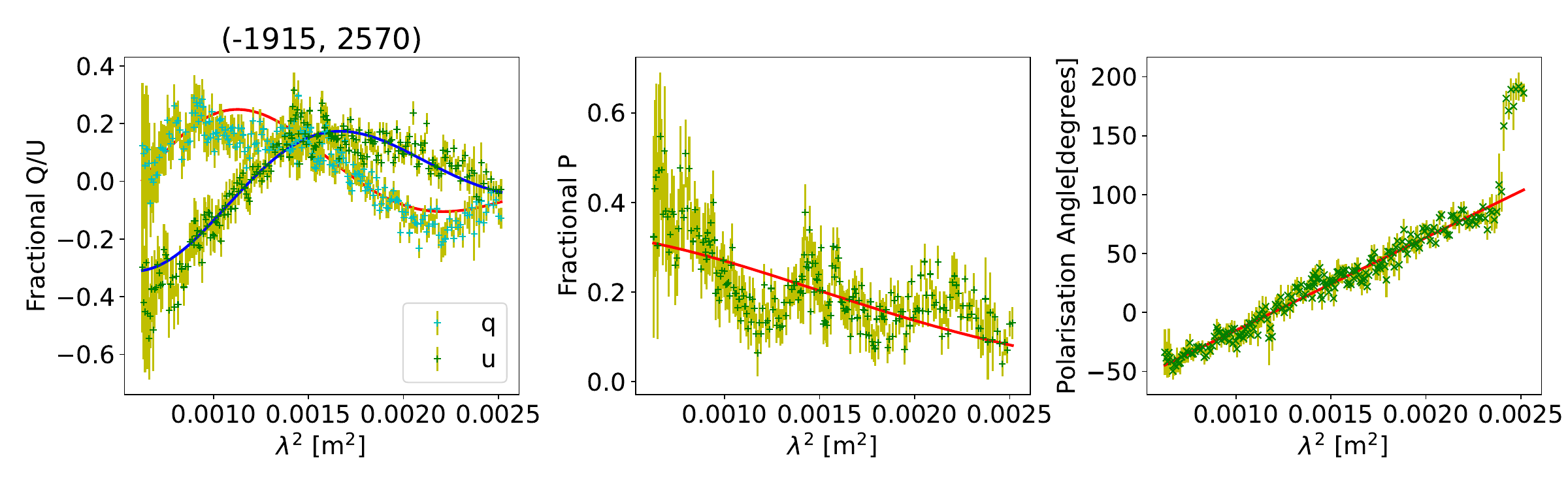}
  \end{minipage}
       \begin{minipage}[b]{1\linewidth}%[3]
    \centering
    \includegraphics[width=0.95\linewidth]{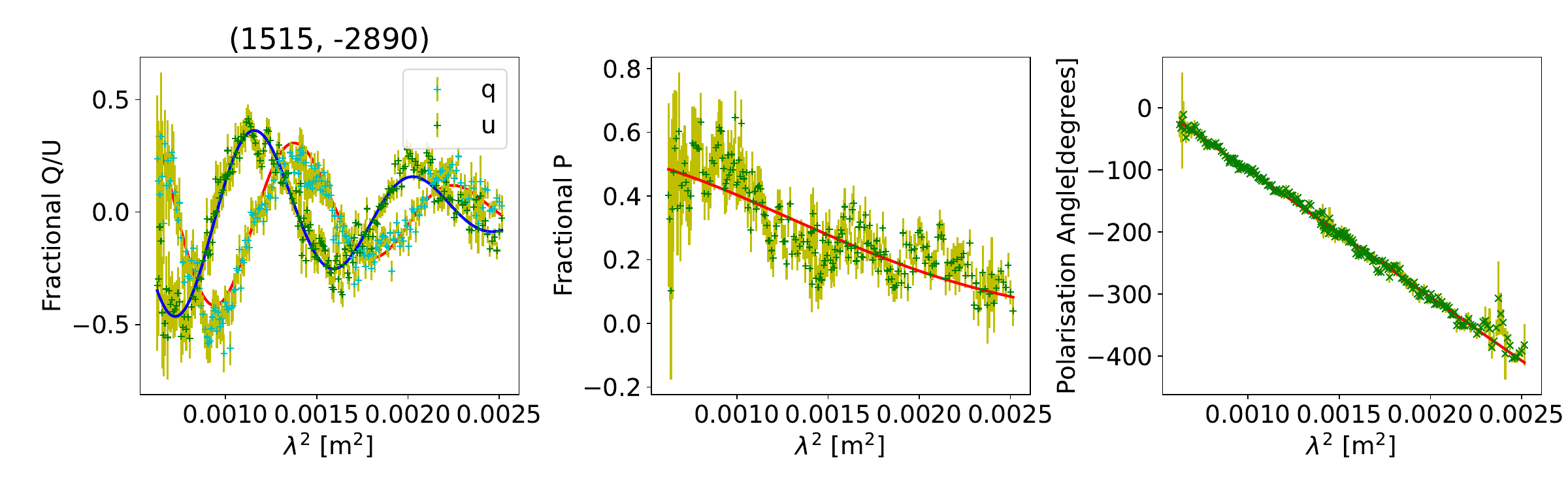}
  \end{minipage}
     \begin{minipage}[b]{1\linewidth}%[5]
    \centering
    \includegraphics[width=0.95\linewidth]{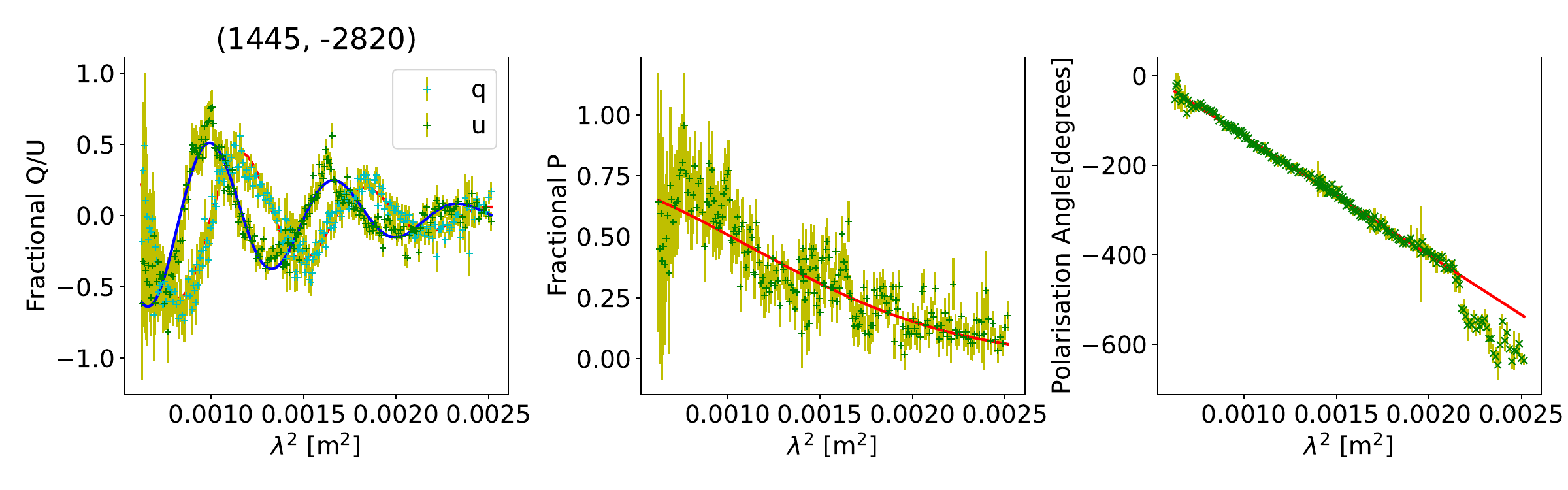}
  \end{minipage}
   \begin{minipage}[b]{1\linewidth}%[5]
    \centering
    \includegraphics[width=0.95\linewidth]{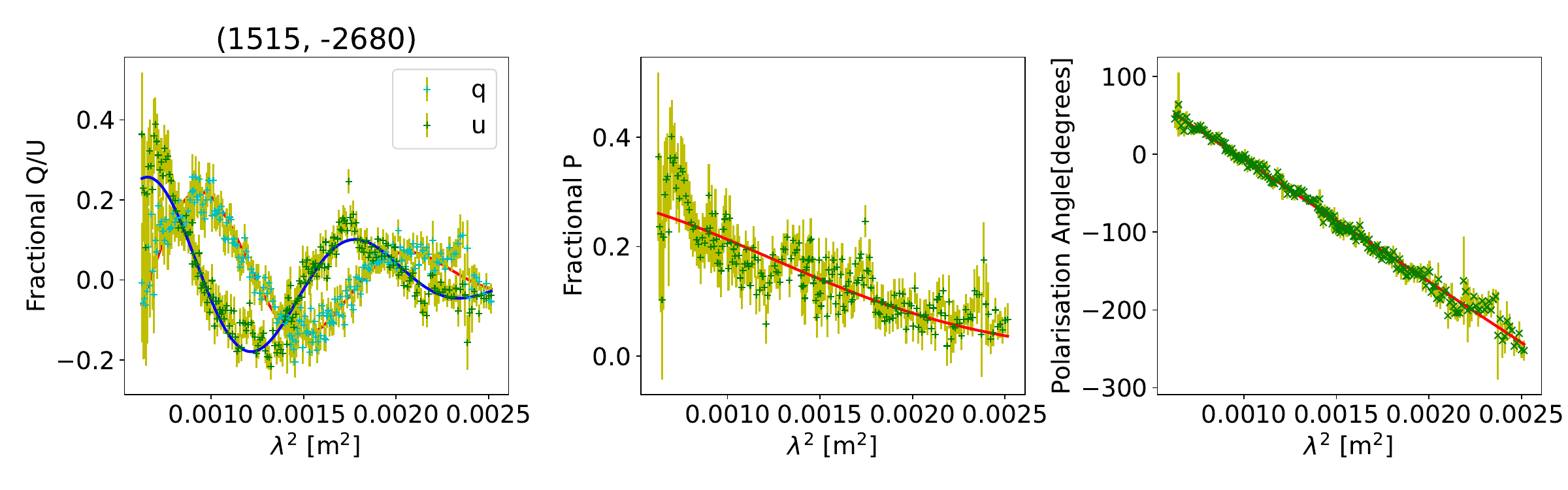}
  \end{minipage}
  \caption{Example of fitting a simple random depolarization screen
    defined in Eq. \ref{eqn:tofit} to the high resolution 6 - 12 GHz
    data. Left column shows fractional $Q$ and $U$, middle column
    shows fractional polarization, $p$, and right column show
    polarization angle all three as a function of
    $\lambda^2$. \label{fig:fitexample}}
  \end{figure}

We confined our search space between $0.0001$ and $1$ for
$p_0$, $\pm \pi/2$ for $\chi_0$, $\pm 12500$ rad m$^{-2}$ for $\mathrm{RM}$ and
[0, 2500] rad m$^{-2}$ for $\sigma$. We only considered pixels with
flux density $> 5\times$ the off-source noise of a 1 GHz Stokes $I$
image. This is so that we avoid evaluating spurious/noisy pixels, and
also to reduce computational time. Figure \ref{fig:fitexample} shows
the example fits, indicating a reasonable fit of this model to the
data.

\begin{figure*}
 \center
   \begin{minipage}[b]{0.45\linewidth}%[3]
    \centering
    \includegraphics[width=0.95\linewidth]{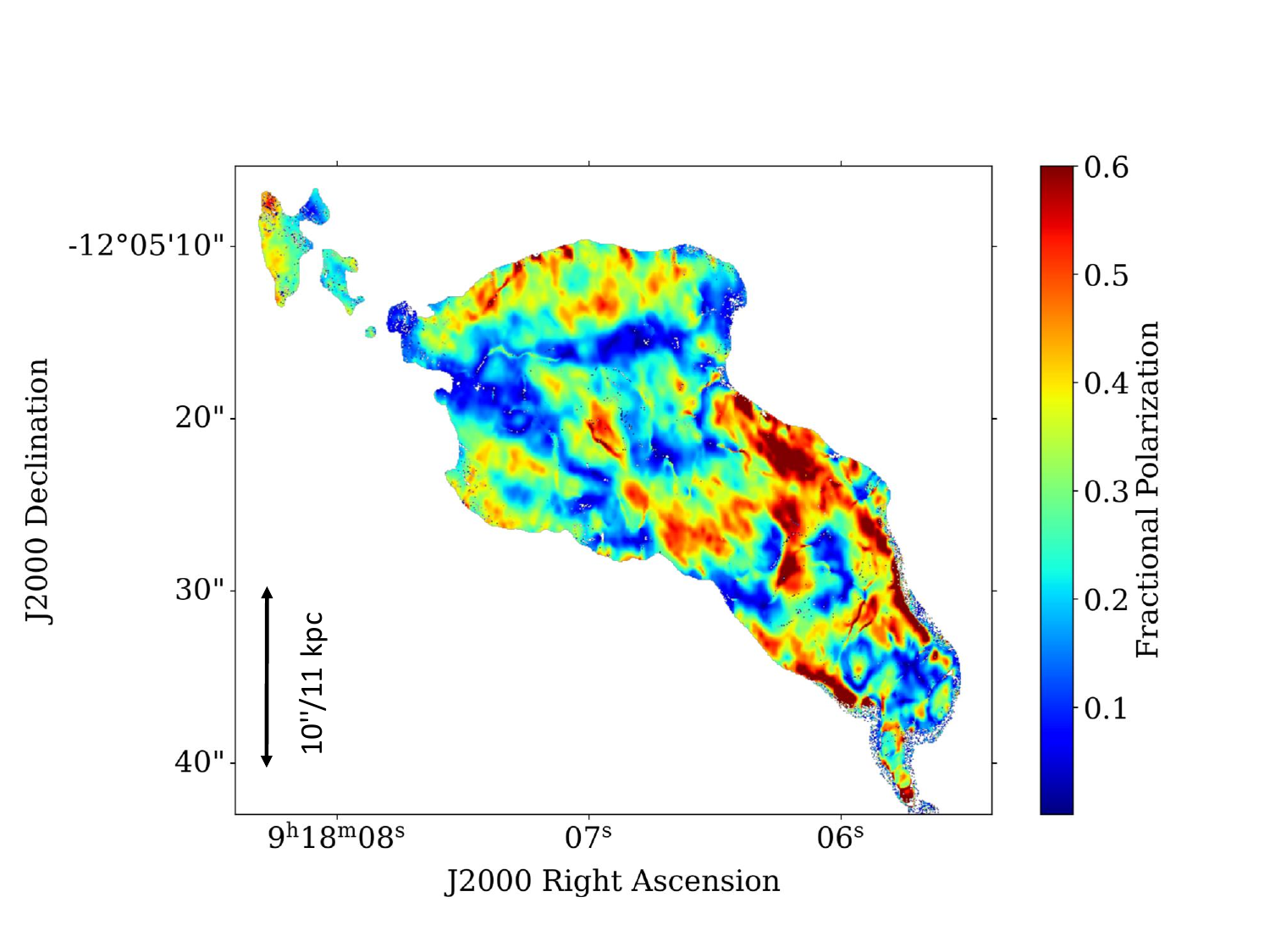}
  \end{minipage}
     \begin{minipage}[b]{0.45\linewidth}%[5]
    \centering
    \includegraphics[width=0.95\linewidth]{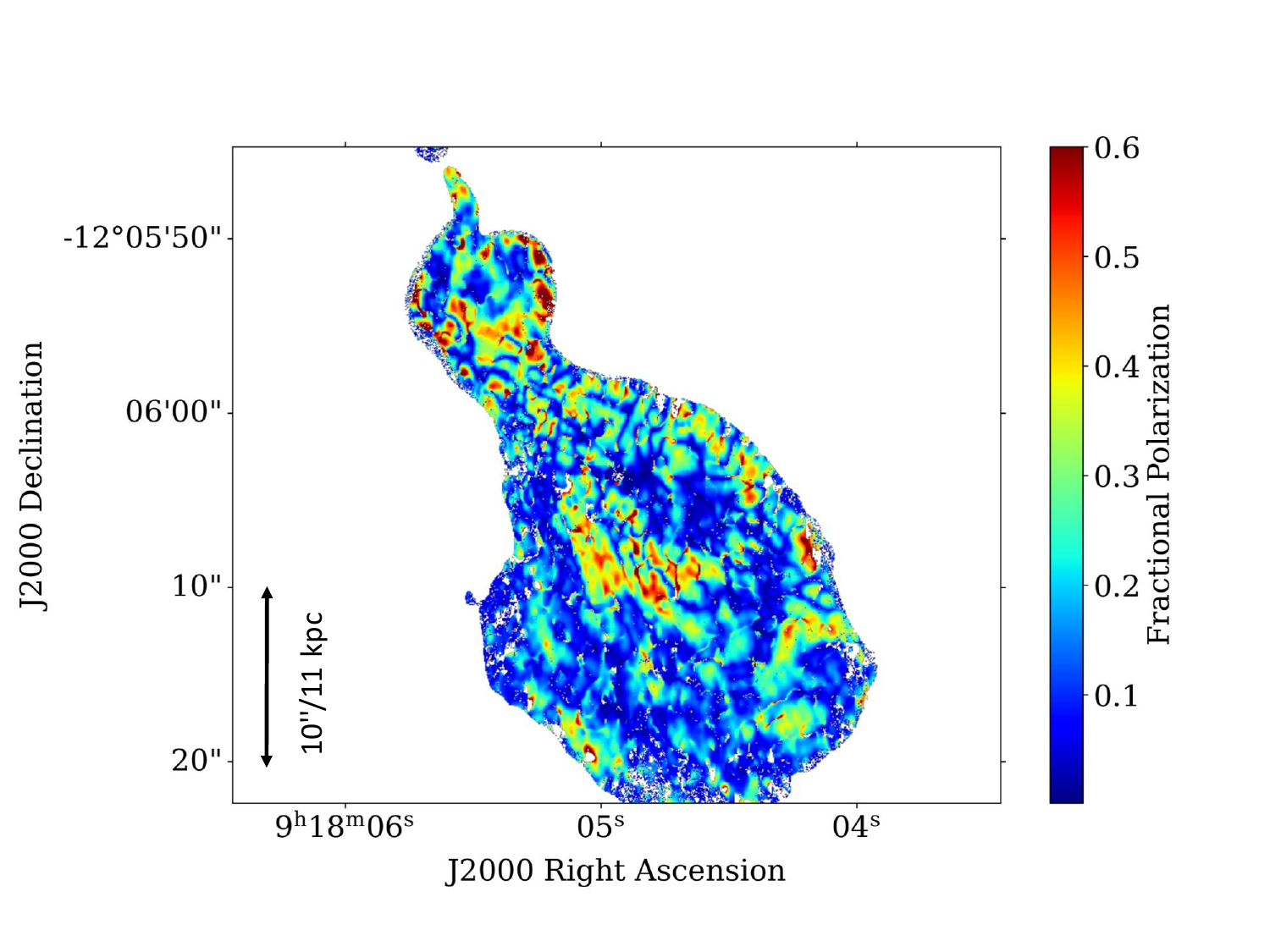}
  \end{minipage}  
     \begin{minipage}[b]{0.45\linewidth}%[3]
    \centering
    \includegraphics[width=0.95\linewidth]{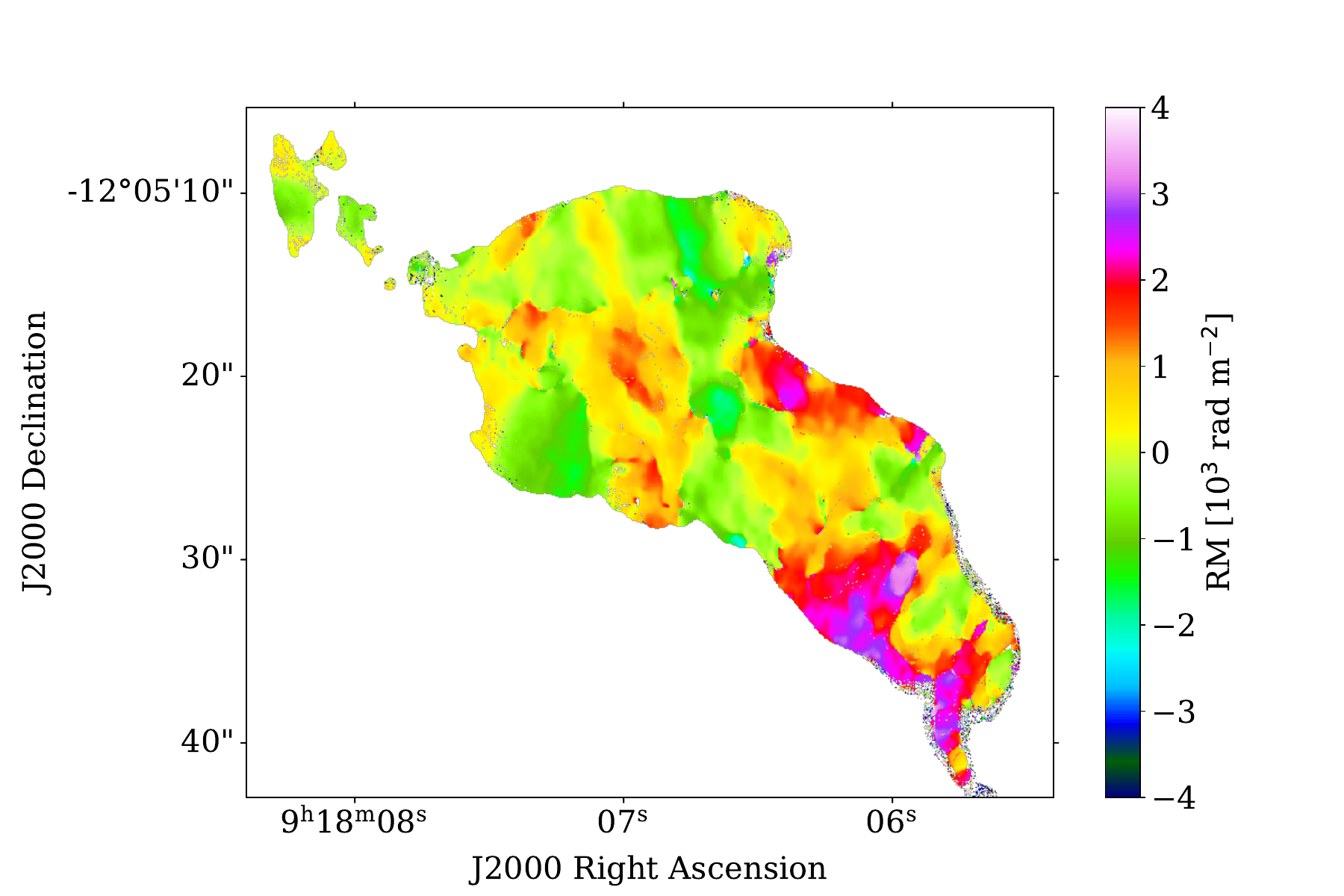}
  \end{minipage}
       \begin{minipage}[b]{0.45\linewidth}%[3]
    \centering
    \includegraphics[width=0.95\linewidth]{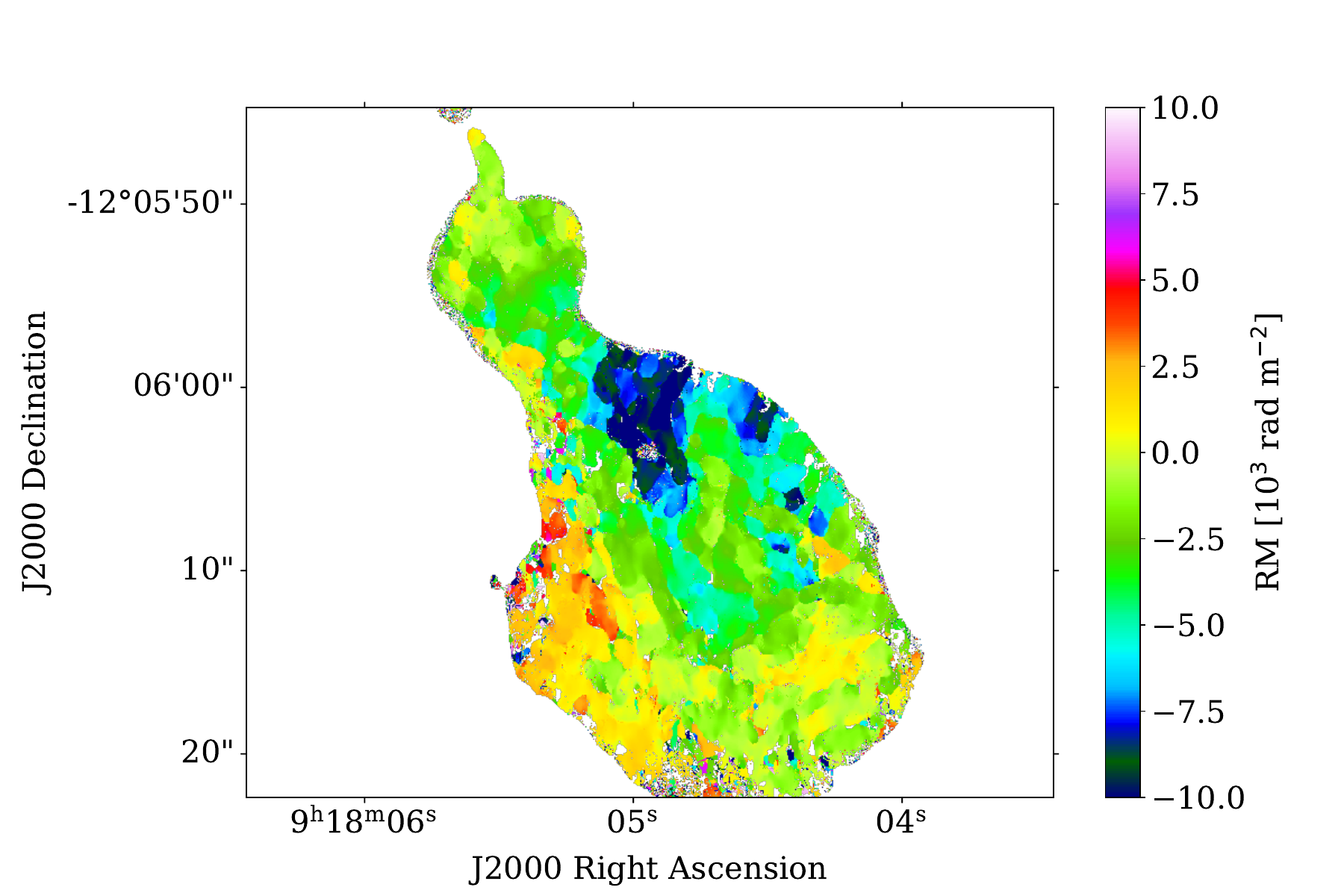}
  \end{minipage}
     \begin{minipage}[b]{0.45\linewidth}%[5]
    \centering
    \includegraphics[width=0.95\linewidth]{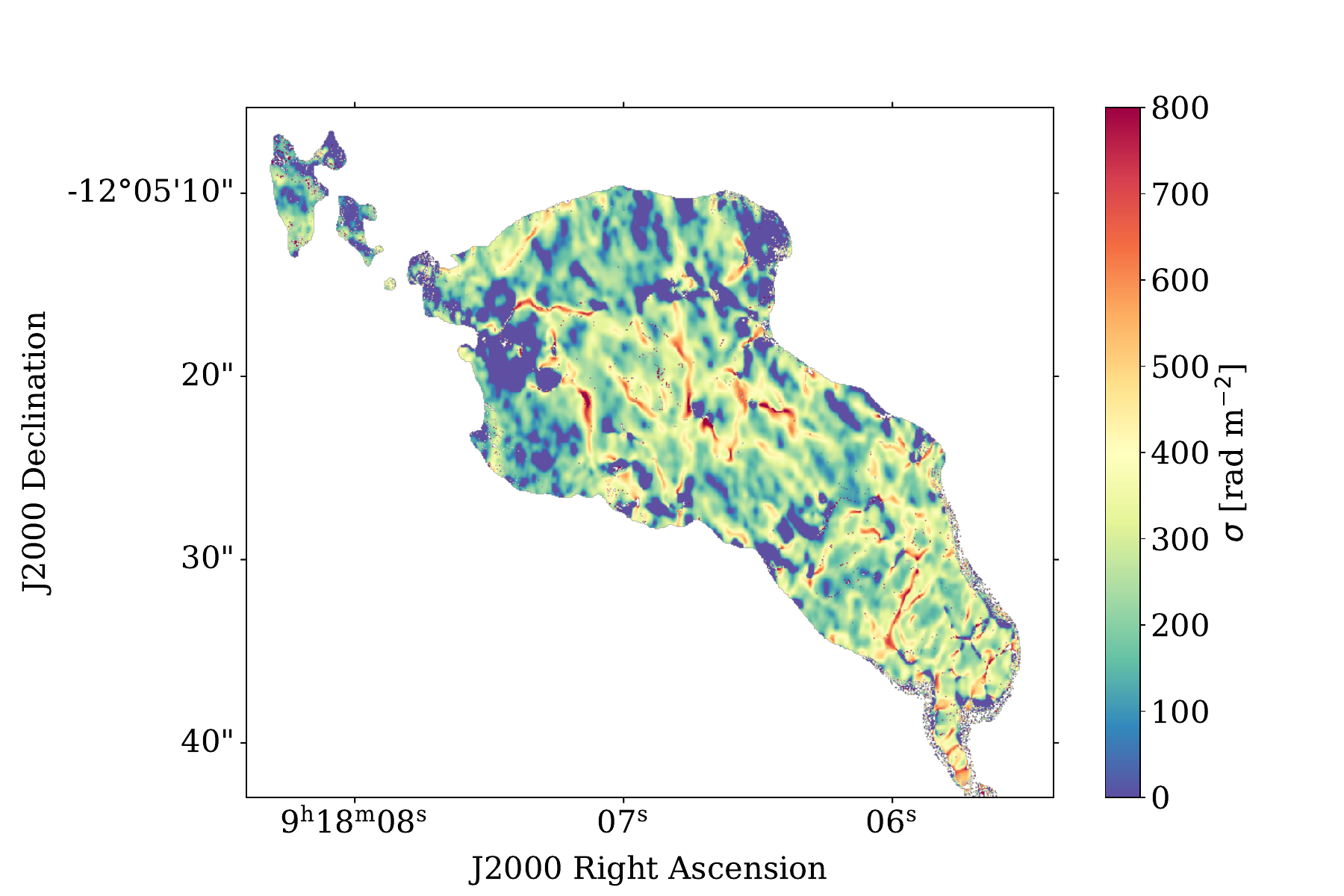}
  \end{minipage}
   \begin{minipage}[b]{0.45\linewidth}%[5]
    \centering
    \includegraphics[width=0.95\linewidth]{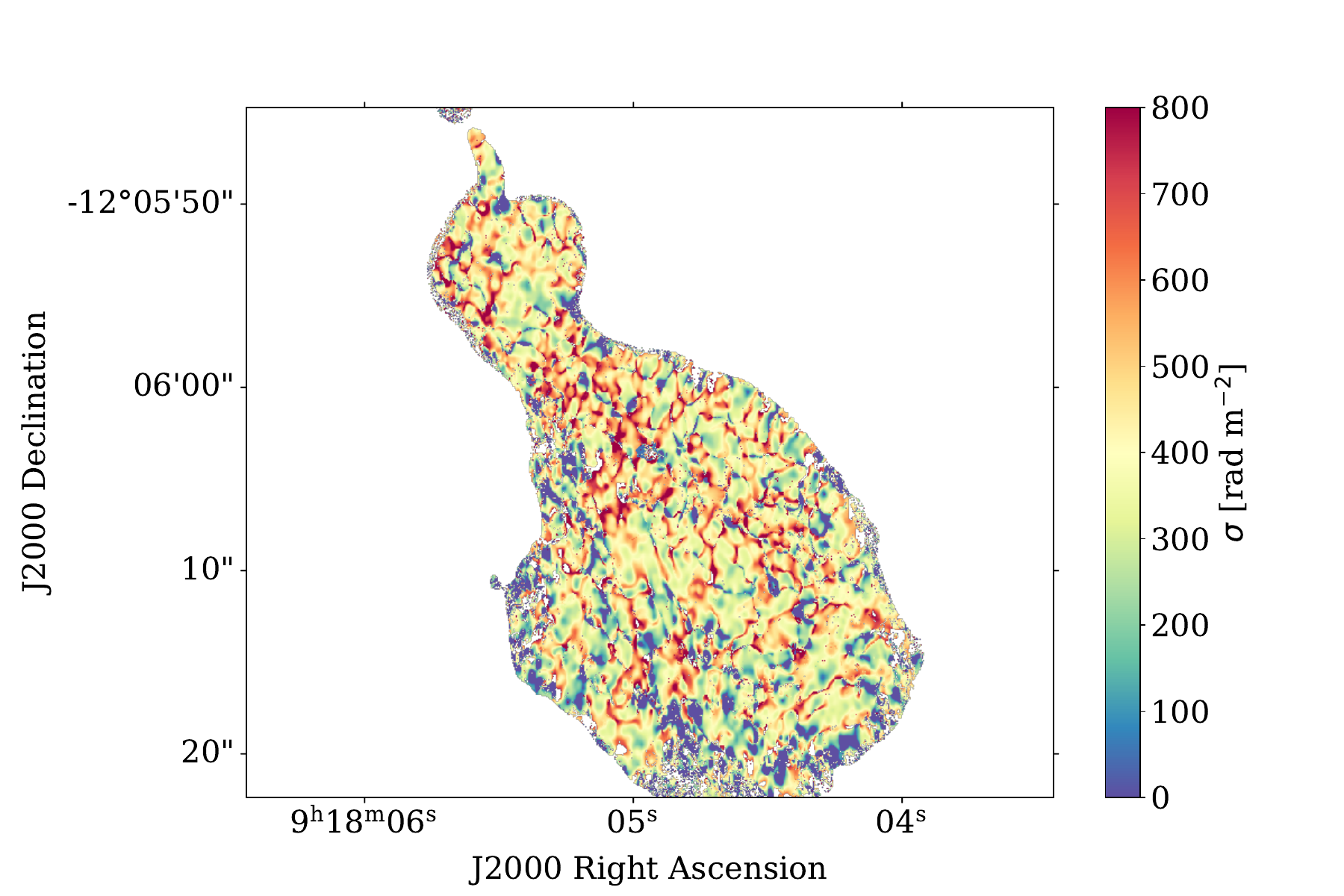}
  \end{minipage}
  \caption{Parameters derived from fitting Eq. 1 to high frequency
    ($6-12$ GHz) high resolution $0.50\arcsec \times 0.35\arcsec$
    polarization data. The left column shows the northern tail, and the
    right column shows the southern tail. The top row shows the intrinsic
    fractional polarization (extrapolated to zero $\lambda$), the middle
    row shows the rotation measure maps, and the third row is the $\mathrm{RM}$
    dispersion. Shown are pixels with error in fractional polarization $<0.1$. The physical scale shown in the top panel is the true for the rest of the panels. \label{fig:fitmaps}}
  \end{figure*}

Figure \ref{fig:fitmaps} shows maps of derived zero-wavelength
(intrinsic) fractional polarization (top panel), rotation measures
(middle panel), and dispersions (bottom panel). The left panel shows
the northern tail, and right panel the southern tail. We display
pixels with fitting error in $p_0$ less than 0.1.

We discuss each of these derived images in the following sections.    
  
\subsection{Intrinsic Fractional Polarization}
The derived zero-wavelength fractional polarizations indicate that
Hydra A is intrinsically highly linearly polarized in both tails, with values ranging between 5\% and 65\% in the northern tail, and up to
75\% at the edges of the tail. The southern tail is less
polarized and extremely patchy -- with polarization values as
low as $2\%$, and as high as 55\% in the inner regions of the tail, and up to 70\% at the edges. The small-scale
patchiness in the fractional polarization of this tail likely
indicates that our resolution is still not sufficient to allow probing
of the source properties, or that there is an interesting phenomenon
occurring inside this tail, which is not present in the northern
tail. We believe the former is probably what is taking place in this
tail, especially since we have seen in Section \ref{sec:newdata} that
the depolarization structures across this tail are mostly complex with
rapid smooth-like decays which indicate the presence of small magnetic
field scales. It should be noted that this does not rule out the
possibility that this tail may be intrinsically different in its
physical properties, but at this point there is no substantial
evidence to support this claim.

\subsection{Rotation Measures}
The rotation measure maps presented here are much more detailed than
those available in the literature
\citep[e.g.,][]{TAYLOR1993,2008LAING}, due to the wide frequency span
and continuous frequency sampling available in the new data.  Rotation
measures range between -2000 rad m$^{-2}$ and 3300 rad m$^{-2}$ across
the northern tail, and -12300 rad m$^{-2}$ and 5000 rad m$^{-2}$
across the southern tail. They are mostly negative across the southern
tail with a small region situated in the south-eastern parts of the
tail with positive rotation measures. The rotation measures associated
with this tail are remarkably patchy on scales of $\sim 1$ kpc. A
region of extremely high rotation measures does not seem to be
associated with any obvious tail features: neither in total intensity,
fractional polarization, or dispersions -- indicative of external (unrelated to the tail) origin.  The rotation measures across
the northern tail, on the other hand, are both negative and positive,
and are relatively more ordered on scales of $\sim$ 2 to 6 kpc. They
also seem to occur in alternating bands of positive and negative
values. Alternating bands in $\mathrm{RM}$ have been observed in few cases in radio sources;
they were observed in Cygnus A \citep{2020SEBOKOLODI}, as well as in
M84, 3C 353, 0206+35, and 3C 270 \citep{2011GUIDETTI}. Similar to the
results of \citet{TAYLOR1993}, the rotation measures across the jets
are consistent with those of the nearby tail, suggesting a common
origin.

\begin{figure}[!ht]
\centering
\includegraphics[width=1\linewidth]{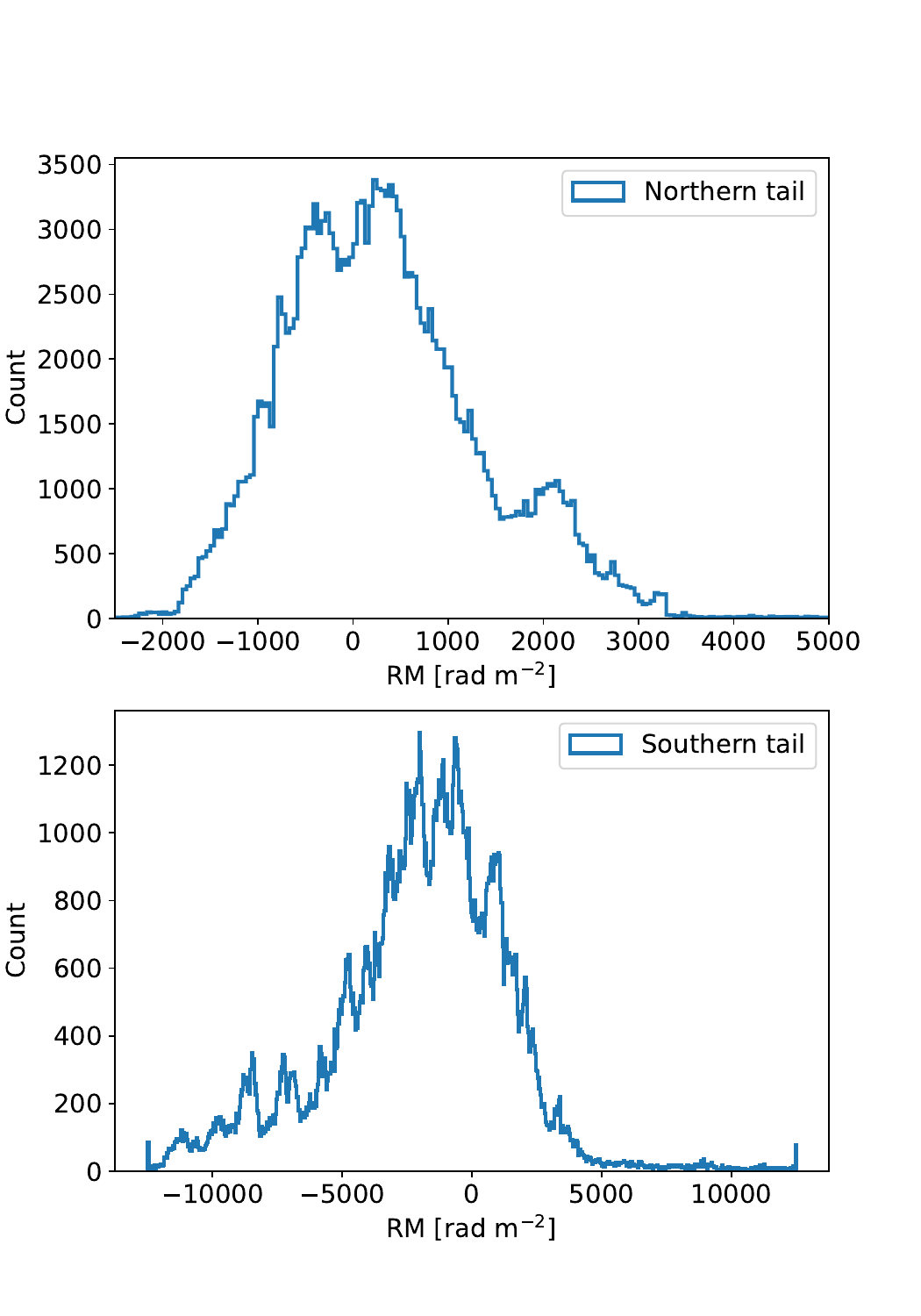}
\caption{A histogram of RM distribution across both tails (in bins of 600). We find a mean of 313 rad m$^{-2}$ and a standard deviation of 1298 rad m$^{-2}$ for the northern tail, and a mean of -2049 rad m$^{-2}$ and a standard deviation of 3396 rad m$^{-2}$  across the southern tail.
\label{fig:RMhist}}
\end{figure}

Figure \ref{fig:RMhist} shows histograms of $\mathrm{RM}$ distribution across
the tails. The distributions are consistent with Figure 4
and 5 of \citet{TAYLOR1993}. We find a mean of 313 rad m$^{-2}$ and a standard deviation of 1298 rad m$^{-2}$ for the
northern tail, a mean of -2049 rad m$^{-2}$ and a standard
deviation of 3396 rad m$^{-2}$ across the southern tail. The
mean $\mathrm{RM}$ across the northern tail is significantly different
from that of \citet{TAYLOR1993}), but the standard deviation is similar.
For the southern tail,  the mean and standard deviation are
different from those of \citet{TAYLOR1993}. This difference
is likely due to
the difference in the $\mathrm{RM}$ range, with our distribution
showing $\mathrm{RM}$s above 2000 rad m$^{-2}$. Our data show bumps
in $\mathrm{RM}$ at 2000 rad m$^{-2}$ in the northern tail
and -8000 rad m$^{-2}$ in the southern tail. The latter was also seen
by \citet{TAYLOR1993}. The bump in the northern
tail is slightly visible in \citet{TAYLOR1993}. The mean and spread in $\mathrm{RM}$ distribution of the two tails is extremely different, suggesting the astrophysical situation is remarkably different for the two tails, either within the tails or in their local environment, on scales of $\sim$30 kpc (the extent of the individual tail).

\begin{figure}[!ht]
\centering
\includegraphics[width=1\linewidth]{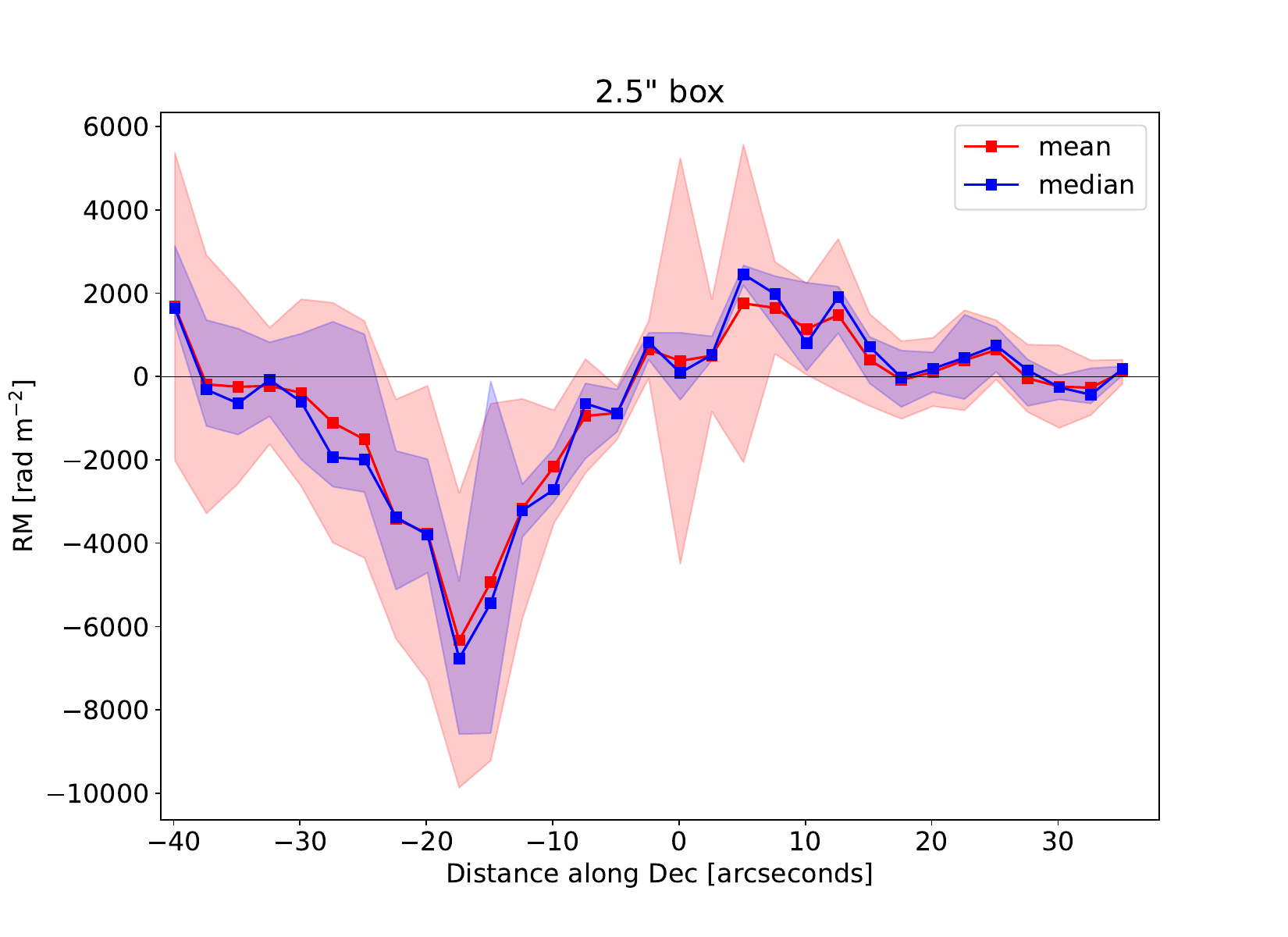}
\caption{The $\mathrm{RM}$ profile as a function of distance from the image center along declination, in bins of 50 pixels (2.5$''$). The mean of $\mathrm{RM}$s computed across RA is shown in red data points  and a standard deviation in a red shade. The median is shown in blue data points, the first and third quartile in blue shade. The $\mathrm{RM}$ profile is consistent with that of \citet{TAYLOR1993}.
   \label{fig:RMprofile}}
\end{figure}

Figure \ref{fig:RMprofile} shows the $\mathrm{RM}$ profile of the tails along
the declination. The profile was obtained by computing the statistics of $\mathrm{RM}$s across all right ascension for a specific declination. The estimates were made across bins of 50 pixels (2.5$''$). In red, we show the mean (data points) and standard
deviation (shade), and in blue is the median (data points) and first and third quartile in shade. The magnitude of $\mathrm{RM}$s across the northern tail reduces radially outward, showing non-uniform oscillations. The magnitudes of the $\mathrm{RM}$s across the southern tail increase radially outward until
reaching a maximum at (RA, Dec) = $(10'', 20'')$, and then decrease
beyond 20$''$ to $\sim 0$ rad m$^{-2}$, and begins to increase and change
sign. The $\mathrm{RM}$ changes smoothly -- there are no large jumps between
successive regions (with the exception of the high $\mathrm{RM}$ region in the
southern tail). Notably, the $\mathrm{RM}$ smoothly reduces close to $0$ rad m$^{-2}$
before changing sign. This profile is consistent with the profile obtained by \citet{TAYLOR1993} (see Figure 7).

\subsection{Faraday Dispersions}
\begin{figure}[!ht]
\centering
\includegraphics[width=0.8\linewidth]{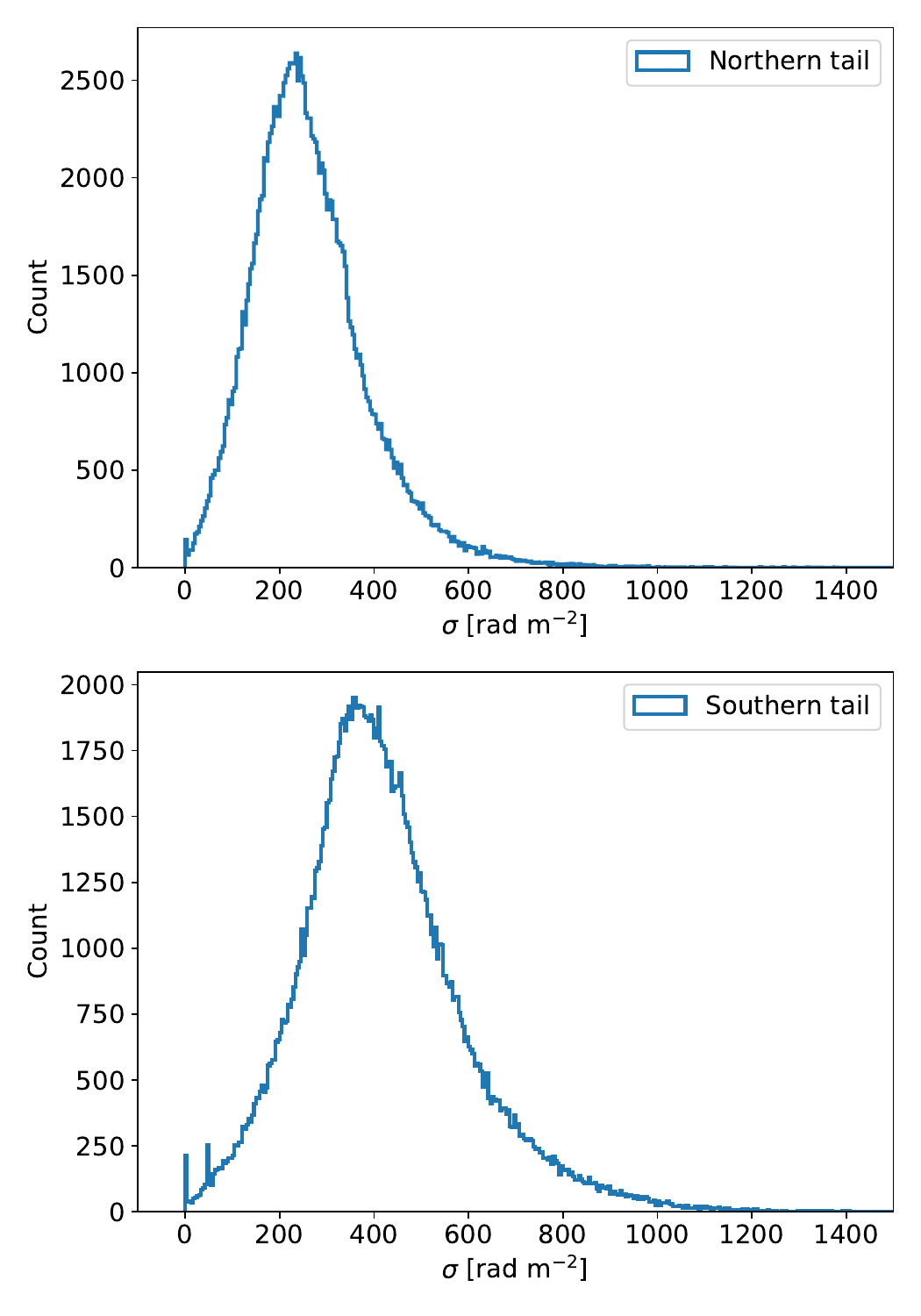}
\caption{A histogram of $\mathrm{RM}$ dispersions distribution across both tails (in bins of 600). We find a mean of 234 rad m$^{-2}$ and a standard deviation of 200 rad m$^{-2}$ for the northern tail and a mean of 380 rad m$^{-2}$ and a standard deviation of 226 rad m$^{-2}$  across the southern tail. The $\sigma$ range roughly between 0 rad m$^{-2}$  and 800 rad m$^{-2}$ for the northern tail, and 0 rad m$^{-2}$  and 1000 rad m$^{-2}$ for the southern tail.
\label{fig:dRMhist}}
\end{figure}

This metric characterizes the rate of depolarization between 12 GHz
and 6 GHz, at a resolution of $0.5 \arcsec \times 0.35 \arcsec$. Figure \ref{fig:dRMhist} shows the distribution of the $\mathrm{RM}$ dispersions more clearly. The dispersions in the northern tail range roughly between 0 and 800 rad
m$^{-2}$, with most regions of the tail $<450$ rad
m$^{-2}$. The dispersions across the southern tail range between 0 and 1000 rad
m$^{-2}$, with the majority concentrated  $\lesssim$ 800 rad m$^{-2}$. The large dispersions are associated with narrow regions. These narrow regions are relatively common across the southern tail. The majority of the dispersions are associated with fitting errors $\lesssim$ 100 rad m$^{-2}$. The dispersions $\le$ 20 rad m$^{-2}$ and $\ge$ 1500 rad m$^{-2}$ are associated with very large errors, and are therefore, not reliable (most of which are failed fits). The large dispersions are removed after applying the masking approach noted in Fig \ref{fig:dRMhist} caption, which masks pixels with fractional error in $p_0 < 0.1$.

\begin{figure}[!ht]
    \centering
\includegraphics[width=1\linewidth]{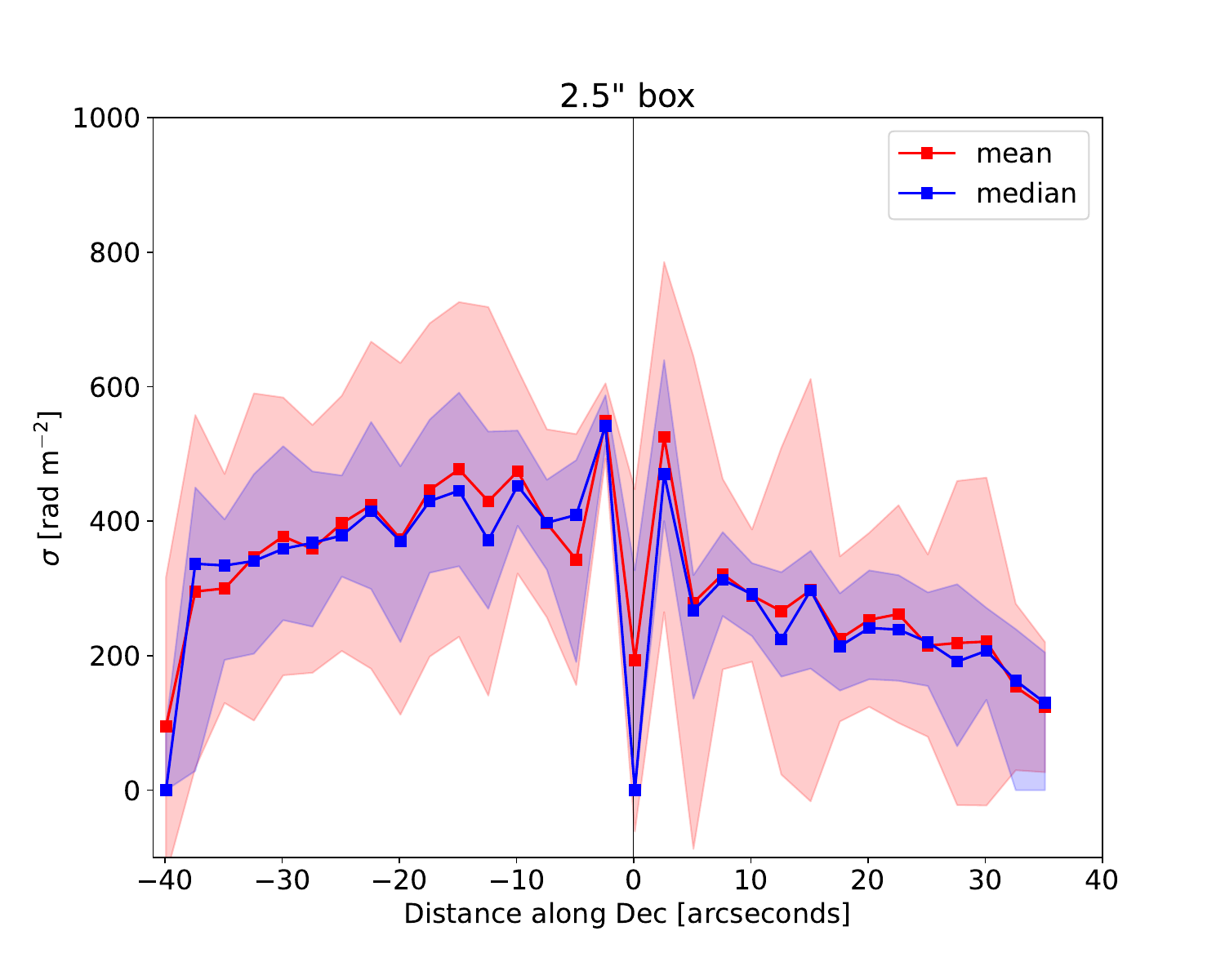}
\caption{$\mathrm{RM}$ dispersion profile as a function of distance from the image center along declination, in bins of 50 pixels (2.5$''$). The mean of dispersions computed across RA is shown in red data points  and standard deviation in a red shade. The median is shown in blue data points, the first and third quartile in a blue shade.\label{fig:dRMprofile}}
\end{figure}

Figure \ref{fig:dRMprofile} shows the $\sigma_t$ profiles along declination. The dispersions across the southern tail decrease steadily with radius, while the dispersions decrease rapidly with radius across the northern tail. There is no correlation between $\mathrm{RM}$ and $\sigma_t$ profile in the southern tail (see Figures \ref{fig:RMprofile} and \ref{fig:dRMprofile}), while both quantities decrease radially in the northern tail.

\subsection{Intrinsic Projected Magnetic Field Orientation}
\begin{figure}
 \center
   \begin{minipage}[b]{1\linewidth}%[3]
    \centering
    \includegraphics[width=1.1\linewidth]{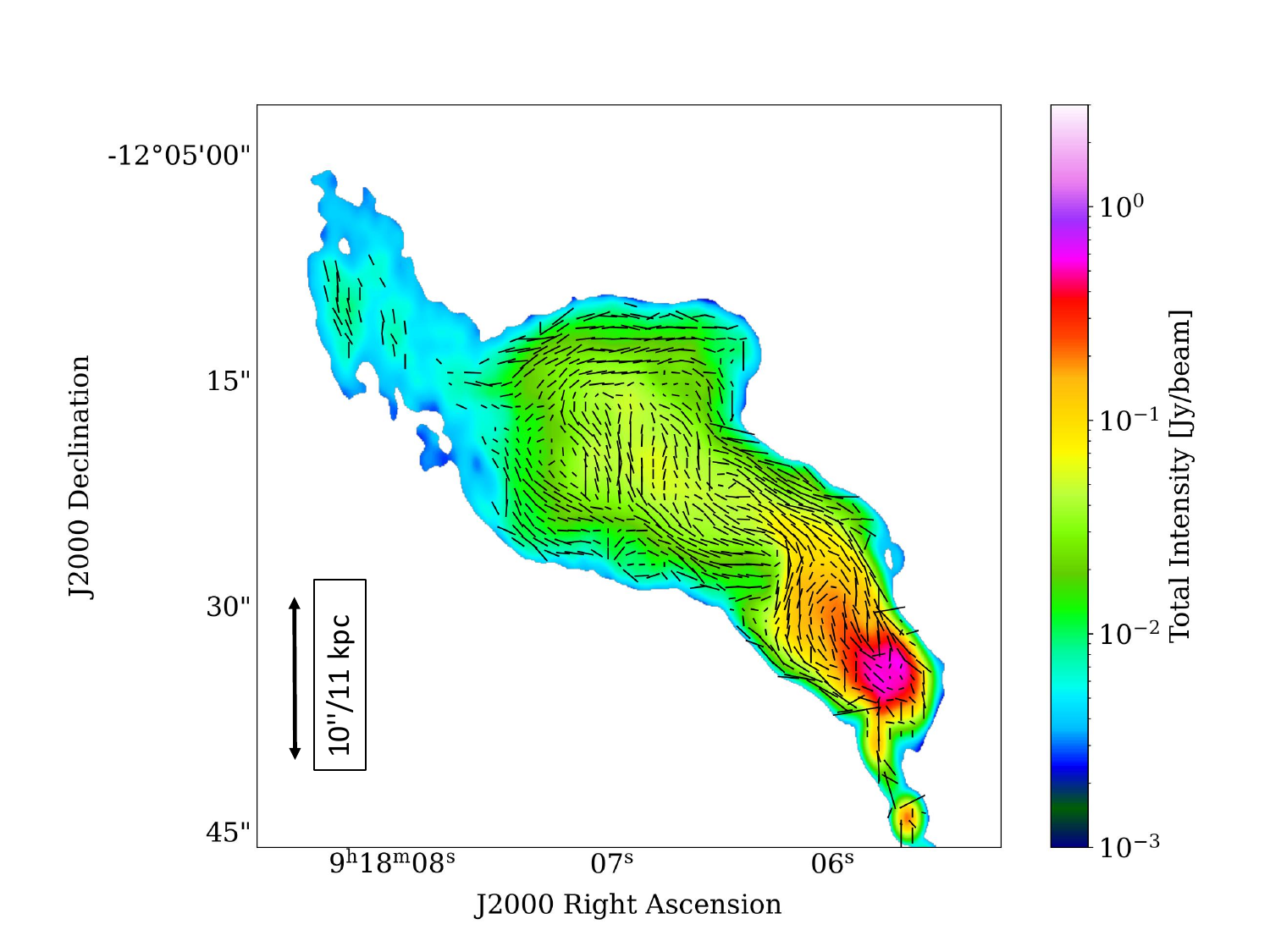}
  \end{minipage}
     \begin{minipage}[b]{1\linewidth}%[5]
    \includegraphics[width=1.15\linewidth]{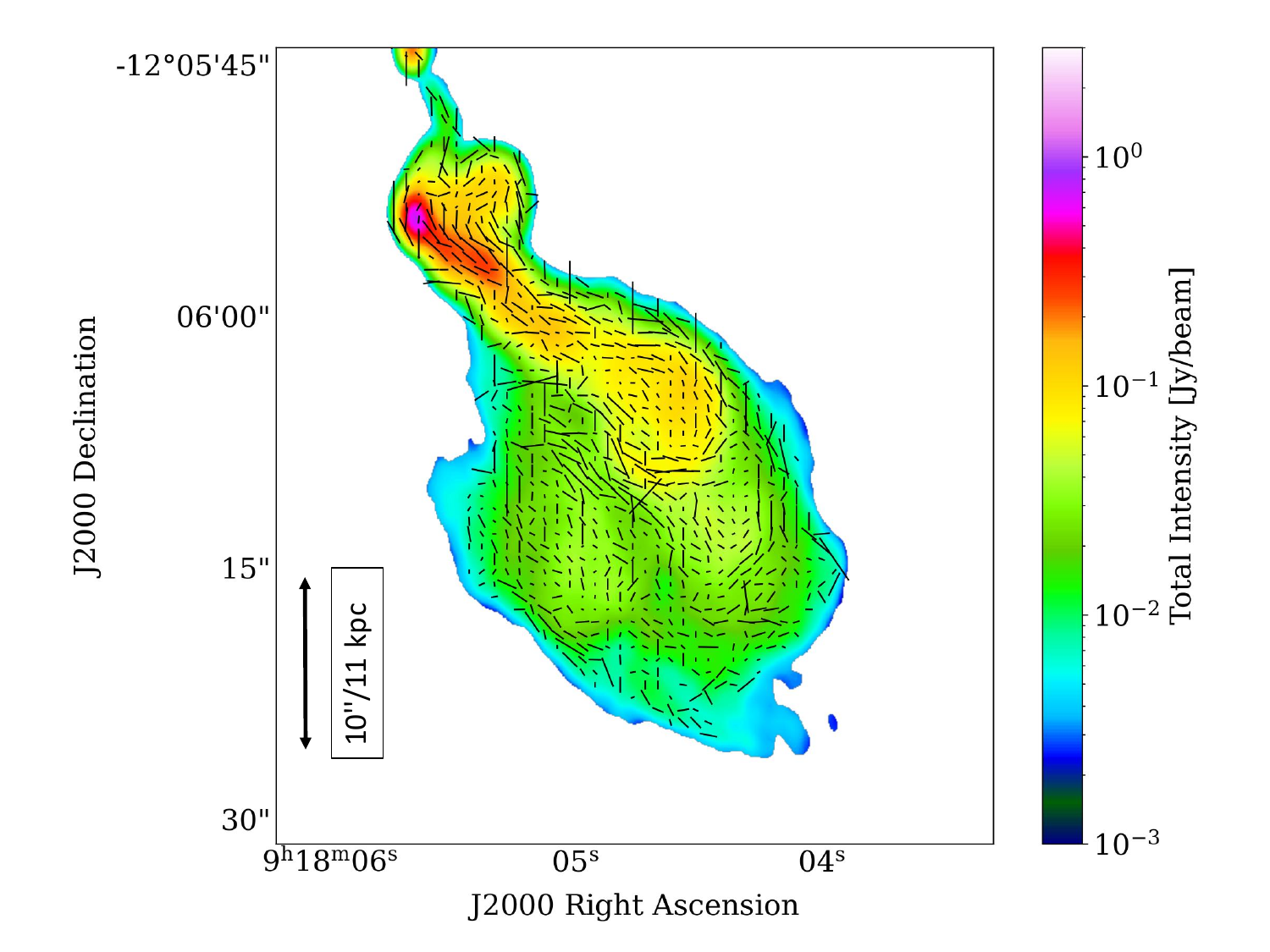}
  \end{minipage}
  \caption{Magnetic field orientations superimposed on a 2 GHz total
    intensity map.  The vector lengths are proportional to the
    fractional polarization (magnified by a factor of 30). Showing only pixels with fitting error in $p_0 <0.1$.  \label{fig:fitPA}}
  \end{figure}

Figure \ref{fig:fitPA} shows the intrinsic magnetic field orientation
across the tails obtained by adding $\pi/2$ to the derived intrinsic
polarization angle, $\chi_0$. The fields follow the boundaries and
filamentary structures of the tail emission. This behavior is quite
common in radio galaxies, for example 3C 465 \citep{2002EILEK}, Cygnus
A \citep{1987DREHER,2020SEBOKOLODI}, and Pictor A \citep{1997PERLEY},
and is generally understood as an effect resulting from shearing (and
compression at outer parts of the tails) of the tangled magnetic
field at the tail boundary, resulting in suppression of field
components normal to the tail boundaries \citep{1980LAING}.  The field
vectors are generally smooth across the northern tail, while slightly
chaotic across the southern tail.  As with the other fitted
parameters, this is likely due to significant structures on scales
less than the $0.5\arcsec$ resolution utilized here.
  
\subsection{A Curious V-shaped Structure in Northern Tail}

\begin{figure}[!ht]
 \centering 
 \includegraphics[width=1.1\linewidth]{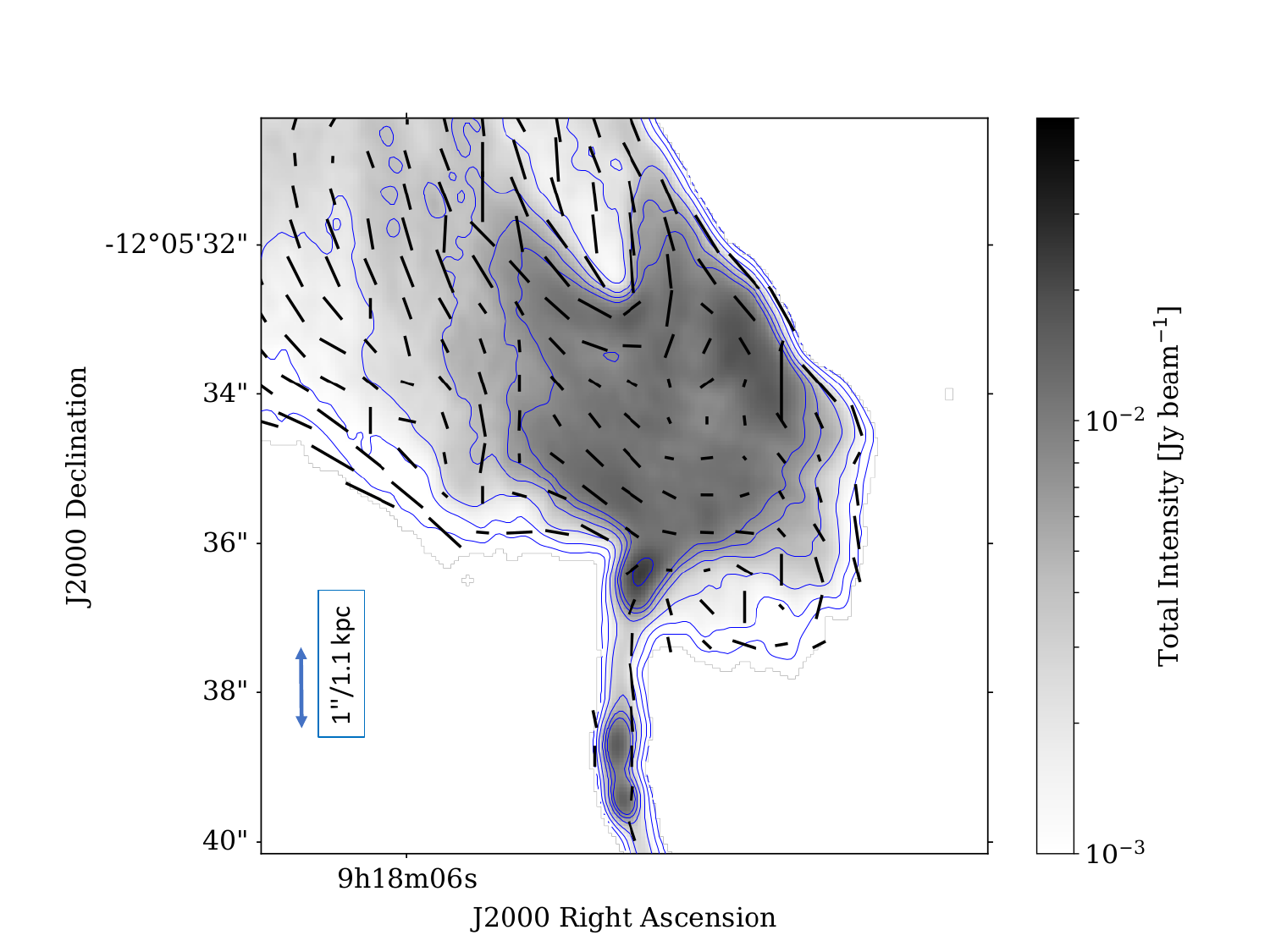}
 \caption{The `V'-shaped structure in brightness and magnetic fields
   in the northern tail of Hydra A. The orientation vectors correspond
   to the magnetic field orientation and the lengths are proportional
   to the fractional polarization. The magnetic field orientation
   traces the `V', while the fractional polarization has a minimal
   effect. \label{fig:bwshock}}
\end{figure}

An examination of the highest resolution intensity image from the
inner part of the northern tail shows a distinct `V'-shaped feature.  It
sits $\sim 12$ kpc north of the galactic core and is oriented roughly
along the local direction of the tail.  It is apparent in total
intensity and is also traced by projected magnetic field
lines. However, there is only a small effect in the fractional
polarization on the `V'-shape. A close-up image is shown in Figure
\ref{fig:bwshock}. The apex of the `V' points toward the galaxy,
which is also the presumed `upstream' direction relative to the systematic
outflow likely to be moving through the northern bright spot.

We do not know the cause of this structure, however, its shape
suggests a bow shock or magnetic draping around an object within the
tail.  If it is a bow shock, its opening angle requires a Mach number
of $\sim 3-4$.  Although large-scale tails in FR I sources are thought
to be subsonic, the jets close to the AGN are likely supersonic.  If
this is the case in Hydra A, the supersonic flow may continue through
the growth of the instability which causes the bright spot and changes
the narrow inner jet to a broad tail.  Alternatively, a slower flow
can ``drape'' a weak magnetic field around an object in the flow
\citep[for example,][]{2006LYUTIKOV,2008DURSI} and can also explain
the ordered magnetic fields along the sides of the structure. A third
possibility suggested by the `V'-shape could be a wake behind some
object in the flow.  However, the ordered magnetic field along the
sides of the `V' seems inconsistent with the turbulence characteristic
of subsonic wakes.

Each of these possibilities depends on the existence of a dense object
-- say a cold gas cloud -- within the flow.  The inner regions of many
galaxies in cool-core clusters contain filaments and clouds of thermal
and molecular gas \citep{2019OLIVARES}. Similar objects might exist in
the Hydra A galaxy, however they have not yet been detected. The Hydra
A galaxy contains a $5$ kpc cool gas disk, rotating around the core of
the central galaxy \citep{2019ROSE} but no $H\alpha$ or molecular
emission has been detected yet outside of this disk.

\section{Predictions of Low-Frequency Data}\label{sec:beamtest}
 In section \ref{sec:newdata2} we showed that interpretation of the
 Hydra A depolarization data is limited by resolution -- our
 observations do not provide enough resolution to properly resolve out
 the variations occurring across the tail. Although it is clear
 that the beam-related effects are important, can we really claim that
 they are the dominant effects responsible for the majority of the
 depolarization? This question can only be fully answered with high
 resolution observations, particularly at low frequencies. Based on
 our current data, we would need resolutions much better than
 0.30$\arcsec$ (perhaps 10 times better), at frequencies down to 2
 GHz, to determine the significance of the beam depolarization to the
 overall depolarization. Instruments with such observing capability
 are currently not available. We have thus developed a method of using
 the high-frequency, high-resolution images of the $\mathrm{RM}$ and polarized
 emission to predict the lower resolution, lower frequency emission
 properties.  Comparison of the observed with the predicted emission
 will then allow a judgement on whether the assumptions inherent in
 the prediction are correct.  The basic assumption here is that these
 high-resolution maps approximate that of the true emission of the
 tails and rotating gas. A close prediction will be a strong
 indication for a foreground turbulent Faraday rotating screen. A
 mis-match will indicate either the presence of much smaller scales
 ($\ll0.3\arcsec$), which we could not accurately model, or that the
 beam-related effects are not a dominant effect, and the
 depolarization is due to a different physical origin.  Both of these
 have an important physical implication, but this is out of scope for
 this paper.

Given the derived $p_0$, $\chi_0$, and $\mathrm{RM}$, from the high-resolution,
high-frequency images, we calculate the model polarized flux as
\begin{equation}\label{eqn:predict}
 P = p_0 I e^{2i \chi_0} e^{2i RM \lambda^2}.
\end{equation}
We obtain $I$ by first determining the spectral index at $0.50\arcsec
\times 0.35\arcsec$ (using 6 -12 GHz data), and using this spectral index map
to predict total intensities across 2 - 12 GHz. The polarized emission in
Eq. \ref{eqn:predict} is computed for $\lambda^2$ between 2 - 12 GHz
-- the resulting polarized cube has a resolution of the input
maps; $0.50\arcsec \times 0.35\arcsec$. We then obtain Stokes $Q$ and
$U$ by taking the real and imaginary part of Eq. \ref{eqn:predict},
then convolve these Stokes maps, including Stokes $I$ to $1.50\arcsec
\times 1.0\arcsec$. The convolution is done using AIPS task CONVL,
with factor input as $0$.  This is the same procedure applied
in the case of Cygnus A \citep{2020SEBOKOLODI}. Our simple modelling does not take into consideration the observing noise.

\begin{figure*}
  \centering
   \begin{minipage}[b]{1\linewidth}%[3]
     \centering
    \includegraphics[width=0.8\linewidth]{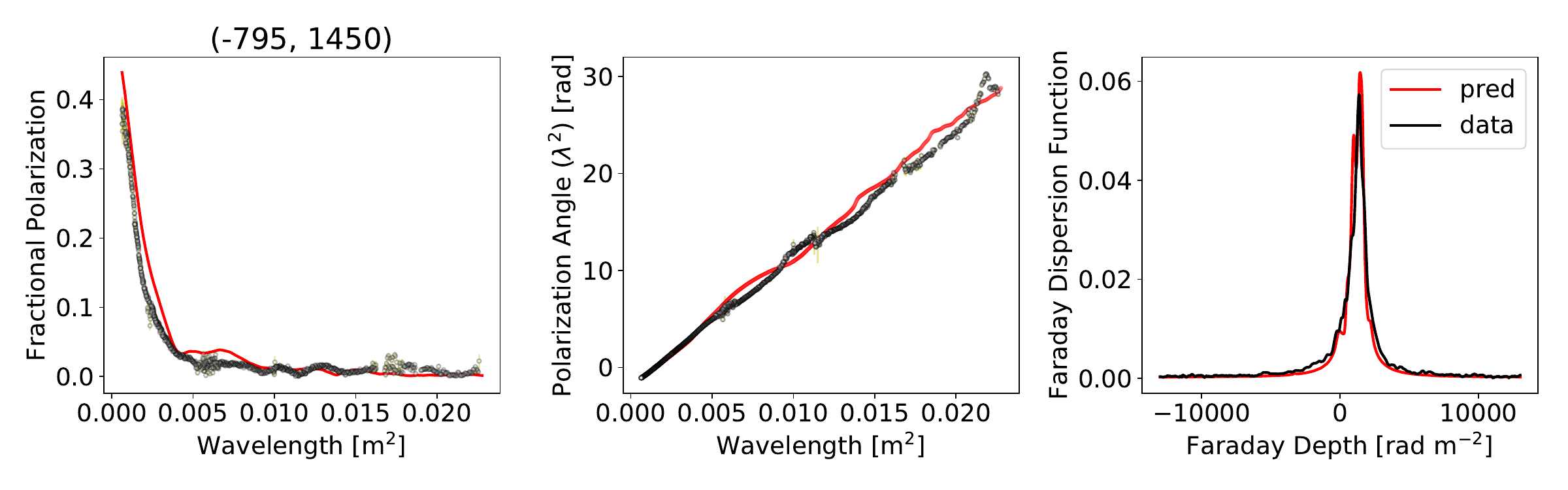}
  \end{minipage}
     \begin{minipage}[b]{1\linewidth}%[5]
       \centering
    \includegraphics[width=0.8\linewidth]{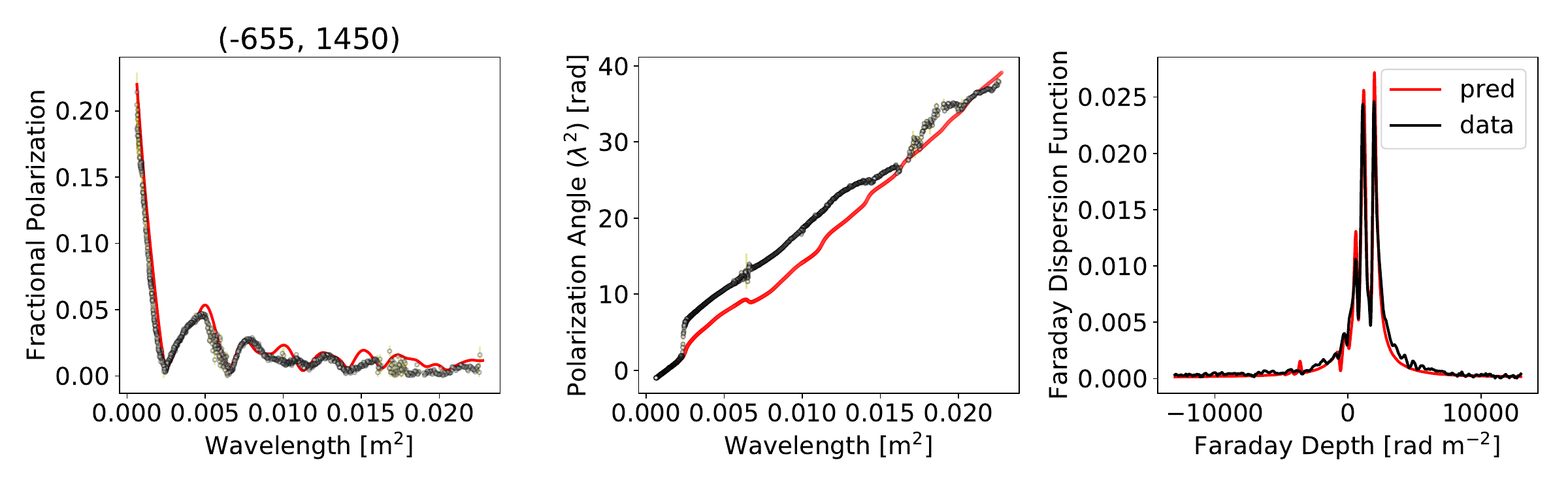}
  \end{minipage}
     \begin{minipage}[b]{1\linewidth}%[5]
       \centering
    \includegraphics[width=0.8\linewidth]{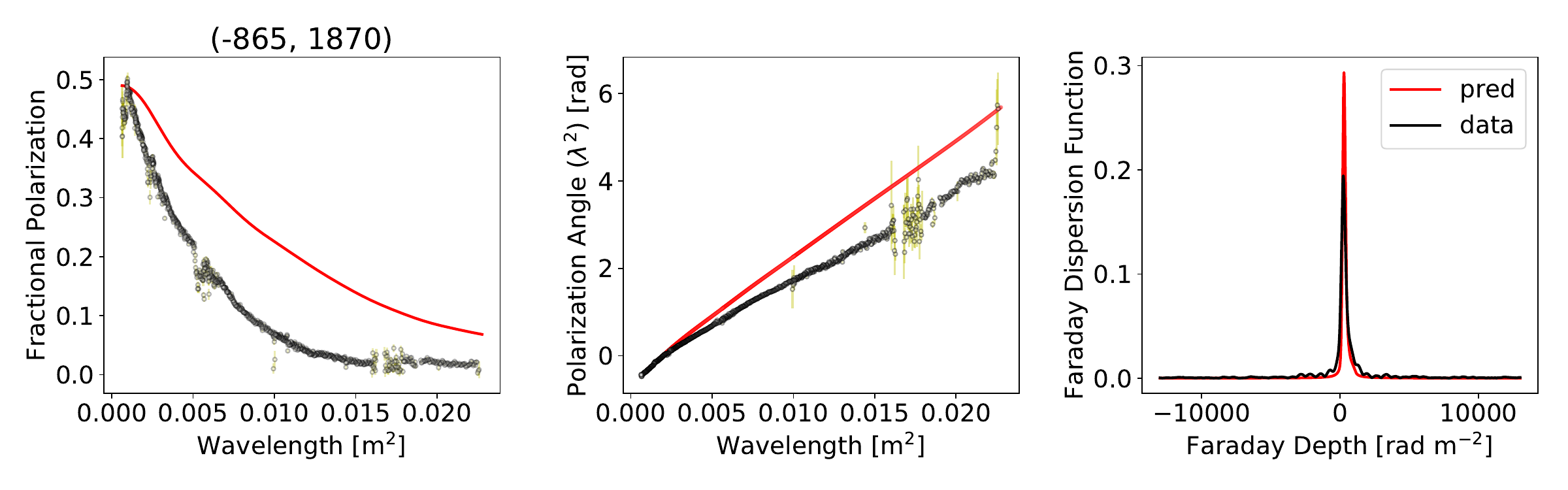}
  \end{minipage}
     \begin{minipage}[b]{1\linewidth}%[5]
       \centering
    \includegraphics[width=0.8\linewidth]{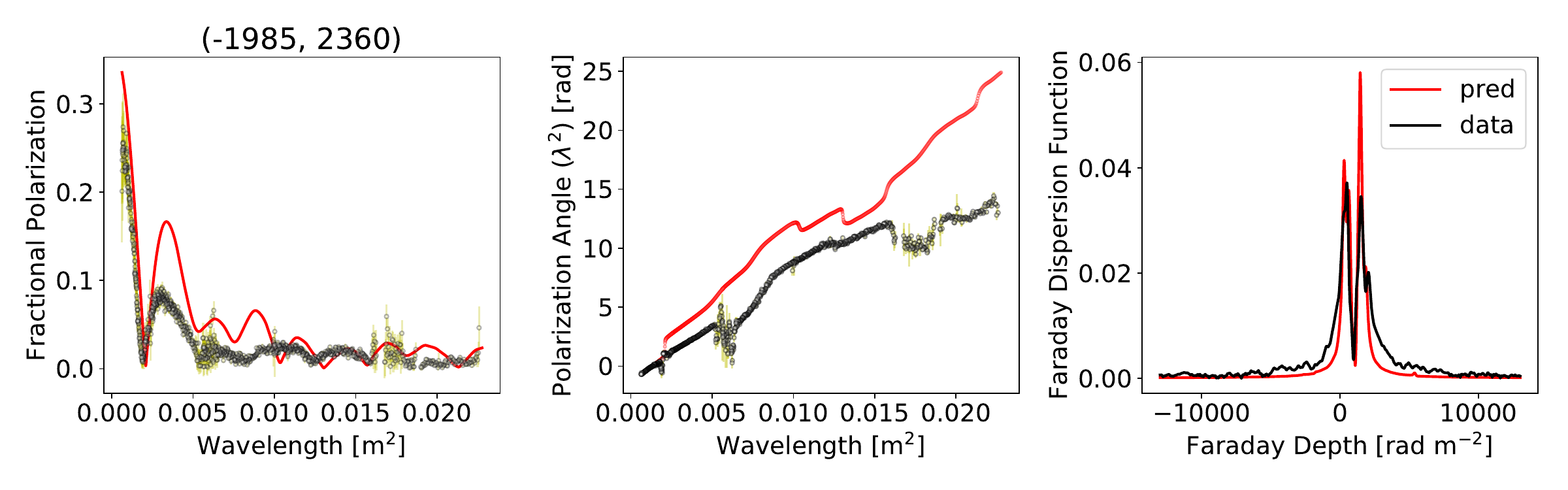}
  \end{minipage}
     \begin{minipage}[b]{1\linewidth}%[3]
       \centering
    \includegraphics[width=0.8\linewidth]{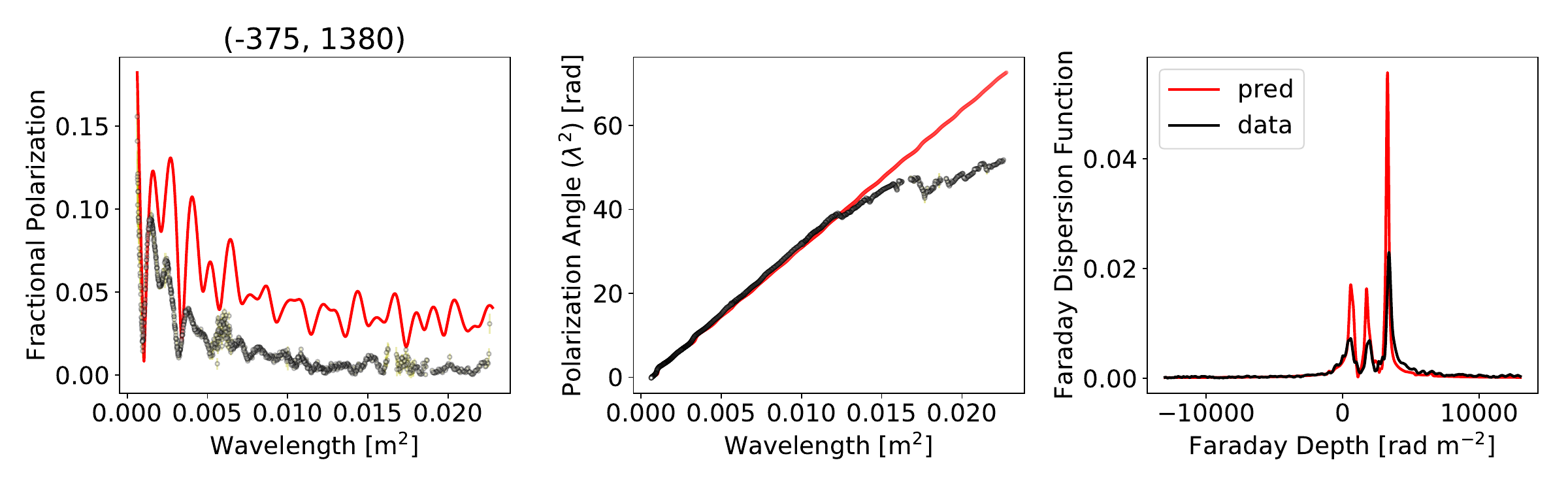}
  \end{minipage}
  \caption{Predictions of low-frequency, low-resolution depolarization
    using high-resolution polarization and $\mathrm{RM}$ maps derived from
    high-resolution, high-frequency data. Left column: Fractional
    polarization vs $\lambda^2$. Middle column: Polarization angle vs
    $\lambda^2$. Right column: Faraday spectra.  Black: Observed
    data. Red: Predictions. Two top rows: Reasonably good predictions
    (roughly $4\%$ of the lines-of-sight). Last three rows: Partially
    reproduced ($71\%$).
  \label{fig:pred1}}
  \end{figure*}

Figure \ref{fig:pred1} shows the predictions in red, and the actual
data in black. The plots show fractional polarization as a function of
$\lambda^2$ in the left panel, polarization as a function of
$\lambda^2$ in the middle column, and Faraday spectra in the right
panel. We compared the data and the predicted data based on fractional
polarization (left panel). The top two rows show examples of the
lines-of-sight whose data are predicted well to within measurement
errors -- with roughly 4.1\% of the lines-of-sight reproduced
reasonably well. We find that these lines-of-sight are not in any way special: they neither occupy a special spatial location, nor do they occupy a special region in the parameter spaces (e.g. Stokes $I$, fitted $p_0$, $\mathrm{RM}$ and $\sigma_t$). The remaining rows show those whose general
depolarization pattern/structure is being reproduced but with slight
differences due to either the underestimation of the depolarization,
and/or the shift in the nulls of the oscillations. We find that
roughly 70.5\% of the lines-of-sight are partially reproduced. The
remaining 25.4 \% are poorly predicted, or are too noisy to make any
accurate judgment.

Although this approach is simple and naive, we find that the
depolarization structures that are seen in the data are, overall,
reproduced remarkably well. This has an important implication -- that
the structures in the rotation measure map are responsible for the
observed depolarization.  This suggests that beam-related effects
are the main contributor to the observed depolarization. The
misalignment of the nulls in the sinc-like lines-of-sight occur very
rarely, while the underestimation of the depolarization is
common. These underestimations may be a result of unmodelled
small-scale fluctuations and/or systematic errors in the prediction
model. Thus, we emphasize that these predictions should be treated as
indicative, not as a complete proof.

\section{Summary}\label{sec:discussion}
In this paper we have presented initial results from our wideband (2 - 12
GHz), high spectral resolution polarimetry data on Hydra A. We look at
how the source polarization emission changes across both frequency and
resolution. We have also derived high resolution maps of the intrinsic
polarized emission of the tails ($p_0e^{2i\chi_0}$), the rotation
measure ($\mathrm{RM}$) and Faraday dispersion ($\sigma_t$) of a foreground
cluster gas. Further, we used these high resolution maps to predict
the data at low frequencies and low resolution. 

The results are summarized as follows:
\begin{enumerate}
 \item The tails including the jets depolarize globally with
   decreasing frequency, with regions closest to the nucleus
   depolarizing more quickly than those far away.
 \item The fractional polarization across the northern tail is smooth
   at high frequencies while the southern tail is relatively
   patchier. However, both tails become clumpy at lower frequencies.
 \item We find that the tails depolarize by more than 90\% between 10
   GHz and 2 GHz.
 \item Fractional polarization as a function of $\lambda^2$ of the
   different lines-of-sight across the tails reveals very complex
   depolarization behavior, with some lines-of-sight showing smooth
   decaying fractional polarization, some are sinc-like and others are
   complex/intermediate.  The depolarization across the
   southern tail is mostly complex, with a few smooth decays. The
   northern tail consist of the three depolarization structure with
   the smooth decay concentrated at extreme regions of the tail
   (further from the nucleus).
 \item We derived Faraday spectra of the lines-of-sight using
   RM-Synthesis, and we find interesting structures in the spectra. In
   general the spectra for smooth decaying lines-of-sight consist of
   single (or few closely-separated) peak(s), while the spectra of
   sinc-like and complex decays are complicated, showing multiple
   peaks, and large broadening.
 \item Polarization angle as function of $\lambda^2$ show significant
   deviations from linearity ($>1$ rad). Most deviations are
   associated with multiple-peaked Faraday spectra.
 \item We find that the tails depolarize with decreasing resolution,
   with the inner regions depolarizing more rapidly than regions
   further from the nucleus.
 \item The fractional polarization across the tail decreases with
   decreasing resolution for most lines-of-sight. However, for some
   lines-of-sight the fractional polarization changes in an
   unpredictable manner by decreasing at high resolution and changing
   abruptly across resolution -- indicating very complex beam-related
   effects.
 \item The rotation measures, $\mathrm{RM}$, across the northern tail range
   between $2000$ rad m$^{-2}$ and 3300 rad m$^{-2}$, and between $-2000$
   rad m$^{-2}$ and $\sim$ 11900 rad m$^{-2}$ across the southern
   tail. The rotation measures occur on scales of 2 to 6 kpc in the
   northern tail, and very small scales of $\sim$ 1 kpc across the
   southern tail. Rotation measures across the northern tail show
   bands of alternating positive and negative values. Those of the
   southern tail are mostly negative, with a small region of positive
   values situated in the outskirts of the tail (south-east).
 \item The derived intrinsic fractional polarization at
   0.5\arcsec$\times$0.35\arcsec shows polarization of 20\% to 50\%
   across the northern tail, and 0.5\% to 50\% across the southern
   tail. There are also highly polarized region of up to 65\% at the edge
   of the northern tail (west side), and narrow regions of up to 70\%
   polarization across the southern tail. In general, the southern
   tail is relatively less polarized, and patchy. The southern tail
   may be intrinsically different or it could be that we haven't fully
   resolved structures across the tail. We argue that the latter is
   probably the major reason for the observed polarization behavior,
   but our data cannot disprove any intrinsic phenomena.
 \item The magnetic field orientation of the source follows the
   boundary and filamentary structures of the tails -- consistent with
   other radio galaxies. The orientations are slightly chaotic in the
   southern tail.
 \item The rotation measure dispersions range between 150 rad m$^{-2}$
   and 350 rad m$^{-2}$ across the northern tail, and 300 rad m$^{-2}$
   and 950 rad m$^{-2}$. The dispersions show no radial dependence. The
   dispersions are extremely chaotic across the southern tail -- with
   narrow regions of high dispersions. These are associated with large
   gradients.
 \item We used the derived intrinsic fractional polarization, and
   rotation measures at high resolutions
   (0.5\arcsec$\times$0.35\arcsec) to predict low frequency, low
   resolution data (1.50\arcsec$\times$1.0\arcsec). The assumption is
   that these high resolution maps present a close representation of
   the 'true' source properties and foreground Faraday rotating
   medium. We find that the depolarization structure in the data are
   reproduced -- with a few 4.1\% closely reproduced, and 70.5\%
   partially reproduced. For the latter, the depolarizations are mostly
   underestimated, and in some rare cases the nulls in the sinc-like
   decay are shifted to low frequencies. These results suggest that
   the depolarization is  mostly a result of small-scale fluctuations
   across a foreground Faraday rotating medium. This depolarizing
   medium must consist of multiscale magnetic fields ordered on scales 0.30 - 1.5 kpc.
\end{enumerate}

\acknowledgments

\section{acknowledgments}

The financial assistance of the South African Radio Astronomy
Observatory (SARAO) towards this research is hereby acknowledged
(www.ska.ac.za). Additional financial assistance was provided by the
National Radio Astronomy Observatory -- a facility of the National
Science Foundation operated under cooperative agreement by Associated
Universities, Inc.  This work is based upon research supported by the
South African Research Chairs Initiative of the Department of Science
and Technology and National Research Foundation.


\begin{thebibliography}{}

%A
\bibitem[Abell(1958)]{ABELL1958} Abell, G.~O.\ 1958, \apjs, 3, 211
\bibitem[Anderson et al.(2016)]{2016ANDERSON} Anderson, C.~S., Gaensler, B.~M., \& Feain, I.~J.\ 2016, \apj, 825, 59.

%B
 \bibitem[Baum et al.(1988)]{1988BAUM} Baum, S.~A., Heckman, T.~M., Bridle, A., et al.\ 1988, \apjs, 68, 643
\bibitem[Blanton et al.(2001)]{2001BLANTON} Blanton, E.~L., Sarazin, C.~L., McNamara, B.~R., et al.\ 2001, \apjl, 558, L15
\bibitem[Boehringer et al.(1993)]{1993BOEHRINGER} Boehringer, H., Voges, W., Fabian, A.~C., et al.\ 1993, \mnras, 264, L25
\bibitem[Burn(1966)]{1966BURN} Burn, B.~J.\ 1966, \mnras, 133, 67 
\bibitem[Brentjens \& de Bruyn(2005)]{2005BRENTJENS} Brentjens, M.~A., \& de Bruyn, A.~G.\ 2005, \aap, 441, 1217 

%C
\bibitem[Carilli et al.(1994)]{1994CARILLI} Carilli, C.~L., Perley, R.~A., \& Harris, D.~E.\ 1994, \mnras, 270, 173
\bibitem[Cotton et al.(2009)]{2009Cotton} Cotton, W.~D., Mason, B.~S., Dicker, S.~R., et al.\ 2009, \apj, 701, 1872. %doi:10.1088/0004-637X/701/2/1872

%D 
\bibitem[David et al.(1990)]{DAVID1990} David, L.~P., Arnaud, K.~A., Forman, W., et al.\ 1990, \apj, 356, 32
\bibitem[David et al.(2001)]{2001DAVID} David, L.~P., Nulsen, P.~E.~J., McNamara, B.~R., et al.\ 2001, \apj, 557, 546
\bibitem[Dreher et al.(1987)]{1987DREHER} Dreher, J.~W., Carilli, C.~L., \& Perley, R.~A.\ 1987, \apj, 316, 611
\bibitem[Dwarakanath et al.(1995)]{DWARAKANATH1995} Dwarakanath, K.~S., Owen, F.~N., \& van Gorkom, J.~H.\ 1995, \apjl, 442, L1
\bibitem[Dursi \& Pfrommer(2008)]{2008DURSI} Dursi, L.~J. \& Pfrommer, C.\ 2008, \apj, 677, 993.


%E
\bibitem[Eilek, \& Owen(2002)]{2002EILEK} Eilek, J.~A., \& Owen, F.~N.\ 2002, \apj, 567, 202

%F
\bibitem[Fabian et al.(2000)]{2000FABIAN} Fabian, A.~C., Sanders, J.~S., Ettori, S., et al.\ 2000, \mnras, 318, L65
\bibitem[Fabian et al.(2003)]{2003FABIAN} Fabian, A.~C., Sanders, J.~S., Allen, S.~W., et al.\ 2003, \mnras, 344, L43


%G
\bibitem[Garrington et al.(1991)]{1991GARRINGTON} Garrington, S.~T., Conway, R.~G., \& Leahy, J.~P.\ 1991, \mnras, 250, 171
\bibitem[Guidetti et al.(2011)]{2011GUIDETTI} Guidetti, D., Laing, R.~A., Bridle, A.~H., et al.\ 2011, \mnras, 413, 2525
\bibitem[Greisen(1990)]{GREISEN1990} Greisen, E.~W.\ 1990, Acquisition, Processing and Archiving of Astronomical Images, 125


%H
\bibitem[Heinz et al.(2002)]{2002HEINZ} Heinz, S., Choi, Y.-Y., Reynolds, C.~S., et al.\ 2002, \apjl, 569, L79
\bibitem[Hutschenreuter et al.(2022)]{2022Hutschenreuter} Hutschenreuter, S., Anderson, C.~S., Betti, S., et al.\ 2022, \aap, 657, A43%. doi:10.1051/0004-6361/202140486

%I


%J


%K
\bibitem[Kuchar \& En{\ss}lin(2011)]{2011Kuchar} Kuchar, P. \& En{\ss}lin, T.~A.\ 2011, \aap, 529, A13. %doi:10.1051/0004-6361/200913918

%L
\bibitem[Laing(1980)]{1980LAING} Laing, R.~A.\ 1980, \mnras, 193, 439
\bibitem[Laing(1988)]{1988LAING} Laing, R.~A.\ 1988, \nat, 331, 149
\bibitem[Laing et al.(2008)]{2008LAING} Laing, R.~A., Bridle, A.~H., Parma, P., et al.\ 2008, \mnras, 391, 521
\bibitem[Lane et al.(2004)]{LANE2004} Lane, W.~M., Clarke, T.~E., Taylor, G.~B., et al.\ 2004, \aj, 127, 48
\bibitem[Lyutikov(2006)]{2006LYUTIKOV} Lyutikov, M.\ 2006, \mnras, 373, 73.


%M
\bibitem[McNamara et al.(2000)]{MCNAMARA2000} McNamara, B.~R., Wise, M., Nulsen, P.~E.~J., et al.\ 2000, \apjl, 534, L135

\bibitem[Ma et al.(2019)]{2019MA} Ma, Y.~K., Mao, S.~A., Stil, J., et al.\ 2019, \mnras, 487, 3432.


%N

\bibitem[Nulsen et al.(2002)]{NULSEN2002} Nulsen, P.~E.~J., David, L.~P., McNamara, B.~R., et al.\ 2002, \apj, 568, 163
\bibitem[Nulsen et al.(2005)]{2005NULSEN} Nulsen, P.~E.~J., McNamara, B.~R., Wise, M.~W., et al.\ 2005, \apj, 628, 629

%O
\bibitem[Owen et al.(1995)]{OWEN1995} Owen, F.~N., Ledlow, M.~J., \& Keel, W.~C.\ 1995, \aj, 109, 14

\bibitem[Olivares et al.(2019)]{2019OLIVARES} Olivares, V., Salome, P., Combes, F., et al.\ 2019, \aap, 631, A22.

\bibitem[O'Sullivan et al.(2012)]{2012OSULLIVAN} O'Sullivan, S.~P., Brown, S., Robishaw, T., et al.\ 2012, \mnras, 421, 3300.


%P

\bibitem[Perley et al.(1997)]{1997PERLEY} Perley, R.~A., Roser, H.-J., \& Meisenheimer, K.\ 1997, \aap, 328, 12
\bibitem[Perley et al.(2001)]{2001PERLEY} Perley, R.A., Chandler, C.J., Butler, B.J., and Wrobel, J.M.\ 2001  \apj, Lett, 739:L1

%Q

%U


%R
\bibitem[Rose et al.(2019)]{2019ROSE} Rose, T., Edge, A.~C., Combes, F., et al.\ 2019, \mnras, 485, 229.
\bibitem[Riseley et al.(2020)]{2020RISELEY} Riseley, C.~J., Galvin, T.~J., Sobey, C., et al.\ 2020, \pasa, 37, e029.


%S
\bibitem[Sebokolodi et al.(2020)]{2020SEBOKOLODI} Sebokolodi, M.~L., Perley, R., Eilek, J., et al.\ 2020, \apj, 903, 36
\bibitem[Simionescu et al.(2009)]{2009SIMIONESCU} Simionescu, A., Roediger, E., Nulsen, P.~E.~J., et al.\ 2009, \aap, 495, 721
\bibitem[Snios et al.(2018)]{SNIOS2018} Snios, B., Nulsen, P.~E.~J., Wise, M.~W., et al.\ 2018, \apj, 855, 71
\bibitem[Sokoloff et al.(1998)]{1998SOKOLLOF} Sokoloff, D.~D., Bykov, A.~A., Shukurov, A., et al.\ 1998, \mnras, 299, 189
\bibitem[Stuardi et al.(2020)]{2020STUARDI} Stuardi, C., O'Sullivan, S.~P., Bonafede, A., et al.\ 2020, \aap, 638, A48.


%T
\bibitem[Taylor et al.(1990)]{TAYLOR1990} Taylor, G.~B., Perley, R.~A., Inoue, M., et al.\ 1990, \apj, 360, 41
\bibitem[Taylor \& Perley(1993)]{TAYLOR1993} Taylor, G.~B., \& Perley, R.~A.\ 1993, \apj, 416, 554
\bibitem[Taylor(1996)]{TAYLOR1996} Taylor, G.~B.\ 1996, \apj, 470, 394


%U


%V
\bibitem[van Moorsel et al.(1996)]{VANMOORSEL1996} van Moorsel, G., Kemball, A., \& Greisen, E.\ 1996, Astronomical Data Analysis Software and Systems V, 37
\bibitem[Vogt \& En{\ss}lin(2005)]{2005Vogt} Vogt, C. \& En{\ss}lin, T.~A.\ 2005, \aap, 434, 67. %doi:10.1051/0004-6361:20041839



%W

%X

%Y
\bibitem[Young et al.(2002)]{2002YOUNG} Young, A.~J., Wilson, A.~S., \& Mundell, C.~G.\ 2002, \apj, 579, 560 % shocks M87

%Z

\end{thebibliography}
\end{document}